\documentclass[longauth]{aa}

\usepackage{graphicx}
\usepackage{natbib}
\usepackage{scalerel}

\usepackage[table]{xcolor}

\usepackage{txfonts}
\usepackage[pdfencoding=auto,psdextra]{hyperref}
\hypersetup{
    colorlinks=true,
    linkcolor=blue,
    filecolor=magenta,      
    urlcolor=blue,
    citecolor=blue
}
\makeatletter
\renewcommand*\aa@pageof{, page \thepage{} of \pageref*{LastPage}}
\makeatother

\usepackage[utf8]{inputenc}

\usepackage[switch, modulo]{lineno}

\usepackage{euclid}

\usepackage{bm}
\usepackage{tabularx}

\definecolor{amaranth}{rgb}{0.9, 0.17, 0.31}
\definecolor{forestgreen(web)}{rgb}{0.13, 0.55, 0.13}
\definecolor{lavender(web)}{rgb}{0.9, 0.9, 0.98}
\definecolor{cosmiclatte}{rgb}{1.0, 0.97, 0.91}
\definecolor{jonquil}{rgb}{0.98, 0.85, 0.37}
\definecolor{khaki(x11)(lightkhaki)}{rgb}{0.94, 0.9, 0.55}
\definecolor{thistle}{rgb}{0.85, 0.75, 0.85}

\newcommand{\GCsp}{\text{GC}\ensuremath{_\mathrm{sp}}}
\newcommand{\GCph}{\text{GC}\ensuremath{_\mathrm{ph}}}
\newcommand{\de}{\mathrm{d}}
\newcommand{\bk}{\bm{k}}
\newcommand*{\veps} {\varepsilon}

\begin{document} 
%
%

\title{\Euclid\/: The search for primordial features\thanks{This paper is published on behalf of the \Euclid Consortium}}

\newcommand{\orcid}[1]{} 
\author{M.~Ballardini\orcid{0000-0003-4481-3559}$^{1,2,3}$\thanks{\email{mario.ballardini@unife.it}}, Y.~Akrami\orcid{0000-0002-2407-7956}$^{4,5}$, F.~Finelli\orcid{0000-0002-6694-3269}$^{3,6}$, D.~Karagiannis$^{7}$, B.~Li$^{8}$, Y.~Li\orcid{0000-0002-7091-9149}$^{9}$, Z.~Sakr\orcid{0000-0002-4823-3757}$^{10,11,12}$, D.~Sapone\orcid{0000-0001-7089-4503}$^{13}$, A.~Ach\'ucarro$^{14,15}$, M.~Baldi\orcid{0000-0003-4145-1943}$^{16,3,17}$
% ...
}

\institute{$^{1}$ Dipartimento di Fisica e Scienze della Terra, Universit\`a degli Studi di Ferrara, Via Giuseppe Saragat 1, 44122 Ferrara, Italy\\
$^{2}$ Istituto Nazionale di Fisica Nucleare, Sezione di Ferrara, Via Giuseppe Saragat 1, 44122 Ferrara, Italy\\
$^{3}$ INAF-Osservatorio di Astrofisica e Scienza dello Spazio di Bologna, Via Piero Gobetti 93/3, 40129 Bologna, Italy\\
$^{4}$ Instituto de F\'isica Te\'orica UAM-CSIC, Campus de Cantoblanco, 28049 Madrid, Spain\\
$^{5}$ CERCA/ISO, Department of Physics, Case Western Reserve University, 10900 Euclid Avenue, Cleveland, OH 44106, USA\\
$^{6}$ INFN-Bologna, Via Irnerio 46, 40126 Bologna, Italy\\
$^{7}$ Department of Physics and Astronomy, University of the Western Cape, Bellville, Cape Town, 7535, South Africa\\
$^{8}$ Department of Physics, Institute for Computational Cosmology, Durham University, South Road, DH1 3LE, UK\\
$^{9}$ Department of Physics \& Astronomy, University of Sussex, Brighton BN1 9QH, UK\\
$^{10}$ Universit\'e St Joseph; Faculty of Sciences, Beirut, Lebanon\\
$^{11}$ Institut de Recherche en Astrophysique et Plan\'etologie (IRAP), Universit\'e de Toulouse, CNRS, UPS, CNES, 14 Av. Edouard Belin, 31400 Toulouse, France\\
$^{12}$ Institut f\"ur Theoretische Physik, University of Heidelberg, Philosophenweg 16, 69120 Heidelberg, Germany\\
$^{13}$ Departamento de F\'isica, FCFM, Universidad de Chile, Blanco Encalada 2008, Santiago, Chile\\
$^{14}$ Institute Lorentz, Leiden University, PO Box 9506, Leiden 2300 RA, The Netherlands\\
$^{15}$ Departamento de F\'isica, Universidad del Pa\'is Vasco UPV-EHU, 48940 Leioa, Spain\\
$^{16}$ Dipartimento di Fisica e Astronomia, Universit\`a di Bologna, Via Gobetti 93/2, 40129 Bologna, Italy\\
$^{17}$ INFN-Sezione di Bologna, Viale Berti Pichat 6/2, 40127 Bologna, Italy\\
$^{18}$ Dipartimento di Fisica e Astronomia "G. Galilei", Universit\`a di Padova, Via Marzolo 8, 35131 Padova, Italy\\
$^{19}$ INFN-Padova, Via Marzolo 8, 35131 Padova, Italy\\
$^{20}$ INAF-Osservatorio Astronomico di Padova, Via dell'Osservatorio 5, 35122 Padova, Italy\\
$^{21}$ European Space Agency/ESTEC, Keplerlaan 1, 2201 AZ Noordwijk, The Netherlands\\
$^{22}$ Institute for Theoretical Particle Physics and Cosmology (TTK), RWTH Aachen University, 52056 Aachen, Germany\\
$^{23}$ Gran Sasso Science Institute (GSSI), Viale F. Crispi 7, L'Aquila (AQ), 67100, Italy\\
$^{24}$ INFN -- Laboratori Nazionali del Gran Sasso (LNGS), L'Aquila (AQ), 67100, Italy\\
$^{25}$ Institute of Theoretical Astrophysics, University of Oslo, P.O. Box 1029 Blindern, 0315 Oslo, Norway\\
$^{26}$ IFPU, Institute for Fundamental Physics of the Universe, via Beirut 2, 34151 Trieste, Italy\\
$^{27}$ INAF-Osservatorio Astronomico di Trieste, Via G. B. Tiepolo 11, 34143 Trieste, Italy\\
$^{28}$ SISSA, International School for Advanced Studies, Via Bonomea 265, 34136 Trieste TS, Italy\\
$^{29}$ INFN, Sezione di Trieste, Via Valerio 2, 34127 Trieste TS, Italy\\
$^{30}$ Department of Astronomy, Stockholm University, Albanova, 10691 Stockholm, Sweden\\
$^{31}$ Institut d'Astrophysique de Paris, 98bis Boulevard Arago, 75014, Paris, France\\
$^{32}$ Institut d'Astrophysique de Paris, UMR 7095, CNRS, and Sorbonne Universit\'e, 98 bis boulevard Arago, 75014 Paris, France\\
$^{33}$ The Institute of Mathematical Sciences, HBNI, CIT Campus, Chennai, 600113, India\\
$^{34}$ Homi Bhabha National Institute, Training School Complex, Anushakti Nagar, Mumbai 400085, India\\
$^{35}$ Department of Physics, P.O. Box 64, 00014 University of Helsinki, Finland\\
$^{36}$ Helsinki Institute of Physics, Gustaf H{\"a}llstr{\"o}min katu 2, University of Helsinki, Helsinki, Finland\\
$^{37}$ Institute of Cosmology and Gravitation, University of Portsmouth, Portsmouth PO1 3FX, UK\\
$^{38}$ INAF-Osservatorio Astronomico di Brera, Via Brera 28, 20122 Milano, Italy\\
$^{39}$ INAF-Osservatorio Astrofisico di Torino, Via Osservatorio 20, 10025 Pino Torinese (TO), Italy\\
$^{40}$ Dipartimento di Fisica, Universit\`a di Genova, Via Dodecaneso 33, 16146, Genova, Italy\\
$^{41}$ INFN-Sezione di Genova, Via Dodecaneso 33, 16146, Genova, Italy\\
$^{42}$ Department of Physics "E. Pancini", University Federico II, Via Cinthia 6, 80126, Napoli, Italy\\
$^{43}$ INAF-Osservatorio Astronomico di Capodimonte, Via Moiariello 16, 80131 Napoli, Italy\\
$^{44}$ INFN section of Naples, Via Cinthia 6, 80126, Napoli, Italy\\
$^{45}$ Instituto de Astrof\'isica e Ci\^encias do Espa\c{c}o, Universidade do Porto, CAUP, Rua das Estrelas, PT4150-762 Porto, Portugal\\
$^{46}$ Dipartimento di Fisica, Universit\`a degli Studi di Torino, Via P. Giuria 1, 10125 Torino, Italy\\
$^{47}$ INFN-Sezione di Torino, Via P. Giuria 1, 10125 Torino, Italy\\
$^{48}$ INAF-IASF Milano, Via Alfonso Corti 12, 20133 Milano, Italy\\
$^{49}$ Institut de F\'{i}sica d'Altes Energies (IFAE), The Barcelona Institute of Science and Technology, Campus UAB, 08193 Bellaterra (Barcelona), Spain\\
$^{50}$ Port d'Informaci\'{o} Cient\'{i}fica, Campus UAB, C. Albareda s/n, 08193 Bellaterra (Barcelona), Spain\\
$^{51}$ INAF-Osservatorio Astronomico di Roma, Via Frascati 33, 00078 Monteporzio Catone, Italy\\
$^{52}$ Dipartimento di Fisica e Astronomia "Augusto Righi" - Alma Mater Studiorum Universit\`a di Bologna, Viale Berti Pichat 6/2, 40127 Bologna, Italy\\
$^{53}$ Institute for Astronomy, University of Edinburgh, Royal Observatory, Blackford Hill, Edinburgh EH9 3HJ, UK\\
$^{54}$ European Space Agency/ESRIN, Largo Galileo Galilei 1, 00044 Frascati, Roma, Italy\\
$^{55}$ ESAC/ESA, Camino Bajo del Castillo, s/n., Urb. Villafranca del Castillo, 28692 Villanueva de la Ca\~nada, Madrid, Spain\\
$^{56}$ University of Lyon, Univ Claude Bernard Lyon 1, CNRS/IN2P3, IP2I Lyon, UMR 5822, 69622 Villeurbanne, France\\
$^{57}$ Institute of Physics, Laboratory of Astrophysics, Ecole Polytechnique F\'ed\'erale de Lausanne (EPFL), Observatoire de Sauverny, 1290 Versoix, Switzerland\\
$^{58}$ UCB Lyon 1, CNRS/IN2P3, IUF, IP2I Lyon, 4 rue Enrico Fermi, 69622 Villeurbanne, France\\
$^{59}$ Departamento de F\'isica, Faculdade de Ci\^encias, Universidade de Lisboa, Edif\'icio C8, Campo Grande, PT1749-016 Lisboa, Portugal\\
$^{60}$ Instituto de Astrof\'isica e Ci\^encias do Espa\c{c}o, Faculdade de Ci\^encias, Universidade de Lisboa, Campo Grande, 1749-016 Lisboa, Portugal\\
$^{61}$ Department of Astronomy, University of Geneva, ch. d'Ecogia 16, 1290 Versoix, Switzerland\\
$^{62}$ INAF-Istituto di Astrofisica e Planetologia Spaziali, via del Fosso del Cavaliere, 100, 00100 Roma, Italy\\
$^{63}$ Universit\'e Paris-Saclay, Universit\'e Paris Cit\'e, CEA, CNRS, AIM, 91191, Gif-sur-Yvette, France\\
$^{64}$ Max Planck Institute for Extraterrestrial Physics, Giessenbachstr. 1, 85748 Garching, Germany\\
$^{65}$ University Observatory, Faculty of Physics, Ludwig-Maximilians-Universit{\"a}t, Scheinerstr. 1, 81679 Munich, Germany\\
$^{66}$ Jet Propulsion Laboratory, California Institute of Technology, 4800 Oak Grove Drive, Pasadena, CA, 91109, USA\\
$^{67}$ von Hoerner \& Sulger GmbH, Schlo{\ss}Platz 8, 68723 Schwetzingen, Germany\\
$^{68}$ Technical University of Denmark, Elektrovej 327, 2800 Kgs. Lyngby, Denmark\\
$^{69}$ Cosmic Dawn Center (DAWN), Denmark\\
$^{70}$ Max-Planck-Institut f\"ur Astronomie, K\"onigstuhl 17, 69117 Heidelberg, Germany\\
$^{71}$ Aix-Marseille Universit\'e, CNRS/IN2P3, CPPM, Marseille, France\\
$^{72}$ Universit\'e de Gen\`eve, D\'epartement de Physique Th\'eorique and Centre for Astroparticle Physics, 24 quai Ernest-Ansermet, CH-1211 Gen\`eve 4, Switzerland\\
$^{73}$ NOVA optical infrared instrumentation group at ASTRON, Oude Hoogeveensedijk 4, 7991PD, Dwingeloo, The Netherlands\\
$^{74}$ Universit\"at Bonn, Argelander-Institut f\"ur Astronomie, Auf dem H\"ugel 71, 53121 Bonn, Germany\\
$^{75}$ Aix-Marseille Universit\'e, CNRS, CNES, LAM, Marseille, France\\
$^{76}$ Dipartimento di Fisica e Astronomia "Augusto Righi" - Alma Mater Studiorum Universit\`a di Bologna, via Piero Gobetti 93/2, 40129 Bologna, Italy\\
$^{77}$ Universit\'e Paris Cit\'e, CNRS, Astroparticule et Cosmologie, 75013 Paris, France\\
$^{78}$ Department of Physics and Astronomy, University of Aarhus, Ny Munkegade 120, DK-8000 Aarhus C, Denmark\\
$^{79}$ Centre for Astrophysics, University of Waterloo, Waterloo, Ontario N2L 3G1, Canada\\
$^{80}$ Department of Physics and Astronomy, University of Waterloo, Waterloo, Ontario N2L 3G1, Canada\\
$^{81}$ Perimeter Institute for Theoretical Physics, Waterloo, Ontario N2L 2Y5, Canada\\
$^{82}$ Universit\'e Paris-Saclay, Universit\'e Paris Cit\'e, CEA, CNRS, Astrophysique, Instrumentation et Mod\'elisation Paris-Saclay, 91191 Gif-sur-Yvette, France\\
$^{83}$ Space Science Data Center, Italian Space Agency, via del Politecnico snc, 00133 Roma, Italy\\
$^{84}$ Centre National d'Etudes Spatiales -- Centre spatial de Toulouse, 18 avenue Edouard Belin, 31401 Toulouse Cedex 9, France\\
$^{85}$ Institute of Space Science, Str. Atomistilor, nr. 409 M\u{a}gurele, Ilfov, 077125, Romania\\
$^{86}$ Universit\"ats-Sternwarte M\"unchen, Fakult\"at f\"ur Physik, Ludwig-Maximilians-Universit\"at M\"unchen, Scheinerstrasse 1, 81679 M\"unchen, Germany\\
$^{87}$ Universit\"at Innsbruck, Institut f\"ur Astro- und Teilchenphysik, Technikerstr. 25/8, 6020 Innsbruck, Austria\\
$^{88}$ Institut d'Estudis Espacials de Catalunya (IEEC), Carrer Gran Capit\'a 2-4, 08034 Barcelona, Spain\\
$^{89}$ Institute of Space Sciences (ICE, CSIC), Campus UAB, Carrer de Can Magrans, s/n, 08193 Barcelona, Spain\\
$^{90}$ Satlantis, University Science Park, Sede Bld 48940, Leioa-Bilbao, Spain\\
$^{91}$ AIM, CEA, CNRS, Universit\'{e} Paris-Saclay, Universit\'{e} de Paris, 91191 Gif-sur-Yvette, France\\
$^{92}$ Centro de Investigaciones Energ\'eticas, Medioambientales y Tecnol\'ogicas (CIEMAT), Avenida Complutense 40, 28040 Madrid, Spain\\
$^{93}$ Instituto de Astrof\'isica e Ci\^encias do Espa\c{c}o, Faculdade de Ci\^encias, Universidade de Lisboa, Tapada da Ajuda, 1349-018 Lisboa, Portugal\\
$^{94}$ Universidad Polit\'ecnica de Cartagena, Departamento de Electr\'onica y Tecnolog\'ia de Computadoras,  Plaza del Hospital 1, 30202 Cartagena, Spain\\
$^{95}$ Kapteyn Astronomical Institute, University of Groningen, PO Box 800, 9700 AV Groningen, The Netherlands\\
$^{96}$ Infrared Processing and Analysis Center, California Institute of Technology, Pasadena, CA 91125, USA\\
$^{97}$ Junia, EPA department, 41 Bd Vauban, 59800 Lille, France}

\abstract{
Primordial features, in particular oscillatory signals, imprinted in the primordial power spectrum of density 
perturbations represent a clear window of opportunity for detecting new physics at high-energy scales. Future 
spectroscopic and photometric measurements from the \Euclid space mission will provide unique constraints on 
the primordial power spectrum, thanks to the redshift coverage and high-accuracy measurement of nonlinear scales, 
thus allowing us to investigate deviations from the standard power-law primordial power spectrum.
We consider two models with primordial undamped oscillations superimposed on the matter power spectrum described 
by $1 + {\cal A}_{\rm X} \sin\left(\omega_{\rm X}\Xi_{\rm X} + 2\pi\phi_{\rm X}\right)$, one linearly spaced in 
$k$ space with $\Xi_{\rm lin} \equiv k/k_*$ where $k_* = 0.05\,{\rm Mpc}^{-1}$ and the other logarithmically spaced 
in $k$ space with $\Xi_{\rm log} \equiv \ln(k/k_*)$. We note that ${\cal A}_{\rm X}$ is the amplitude of the primordial feature, 
$\omega_{\rm X}$ is the dimensionless frequency, and $\phi_{\rm X}$ is the normalised phase, where 
$X=\{{\rm lin},{\rm log}\}$. We provide forecasts from spectroscopic and photometric primary \Euclid probes on 
the standard cosmological parameters $\Omega_{\rm m,0}$, $\Omega_{\rm b,0}$, $h$, $n_{\rm s}$, and $\sigma_8$, and 
the primordial feature parameters ${\cal A}_X$, $\omega_X$, and $\phi_X$.
We focus on the uncertainties of the primordial feature amplitude ${\cal A}_X$ and on the capability of \Euclid 
to detect primordial features at a given frequency. We also study a nonlinear density reconstruction method in 
order to retrieve the oscillatory signals in the primordial power spectrum, which are damped on small scales in 
the late-time Universe due to cosmic structure formation. Finally, we also include the expected measurements from 
\Euclid's galaxy-clustering bispectrum and from observations of the cosmic microwave background (CMB).
We forecast uncertainties in estimated values of the cosmological parameters with a Fisher matrix method applied 
to spectroscopic galaxy clustering (\GCsp), weak lensing (WL), photometric galaxy clustering (\GCph), the cross 
correlation (XC) between \GCph\ and WL, the spectroscopic galaxy clustering bispectrum, the CMB temperature and $E$-mode 
polarisation, the temperature-polarisation cross correlation, and CMB weak lensing. 
We consider two sets of specifications for the \Euclid probes (pessimistic and optimistic) and three different CMB 
experiment configurations, that is, {\em Planck}, Simons Observatory (SO), and CMB Stage-4 (CMB-S4).
We find the following percentage relative errors in the feature amplitude with \Euclid primary probes: for the 
linear (logarithmic) feature model, with a fiducial value of ${\cal A}_X = 0.01$, $\omega_X = 10$, and $\phi_X = 0$: 
21\% (22\%) in the pessimistic settings and 18\% (18\%) in the optimistic settings at a 68.3\% confidence level (CL) 
using \GCsp+WL+\GCph+XC. While the uncertainties on the feature amplitude are strongly dependent on the frequency 
value when single \Euclid probes are considered, we find robust constraints on ${\cal A}_{\rm X}$ from the combination 
of spectroscopic and photometric measurements over the frequency range of $(1,\,10^{2.1})$. Due to the inclusion of numerical 
reconstruction, the \GCsp\ bispectrum, SO-like CMB reduces the uncertainty on the primordial feature amplitude by 
32\%--48\%, 50\%--65\%, and 15\%--50\%, respectively.
Combining all the sources of information explored expected from \Euclid in combination with the future SO-like CMB 
experiment, we forecast ${\cal A}_{\rm lin} \simeq 0.010 \pm 0.001$ at a 68.3\% CL and 
${\cal A}_{\rm log} \simeq 0.010 \pm 0.001$ for \GCsp(PS rec + BS)+WL+\GCph+XC+SO-like for both the optimistic and 
pessimistic settings over the frequency range $(1,\,10^{2.1})$.}

%
%

\titlerunning{\Euclid\/: The search for primordial features}
\authorrunning{M. Ballardini et al.}

\maketitle
%
%
%
%

\section{Introduction} \label{sec:intro}
Future galaxy surveys are the new frontier in the study of initial conditions of the Universe. 
They are expected to improve significantly the constraints on many of the parameters characterising the physics 
of the early Universe. The \Euclid satellite, simultaneously performing a spectroscopic survey of galaxies and 
an imaging survey (targeting weak lensing and galaxy clustering using photometric redshifts), will have the unique 
opportunity of measuring ultra-large scales thanks to its large observed volume, as well as small scales of matter 
distribution in the full nonlinear regime. This opportunity will drastically improve our understanding of the 
early-Universe physics and of cosmic inflation 
\citep{Starobinsky:1980te,Guth:1980zm,Sato:1980yn,Linde:1981mu,Albrecht:1982wi,Hawking:1982ga,Linde:1983gd} through 
the study of the statistics hidden in the scalar density perturbations \citep{Mukhanov:1981xt}. \Euclid 
measurements are indeed expected to reduce uncertainties on the amplitude of primordial scalar perturbations $A_{\rm s}$, 
the scalar spectral index $n_{\rm s}$ and its derivatives (or scalar runnings), the amplitude of primordial 
non-Gaussianity $f_{\rm NL}$ (in particular of a local type), and the spatial curvature parameter $\Omega_K$; and 
to enable us to perform significantly more stringent tests of extended models beyond single-field slow-roll inflation 
\citep{EUCLID:2011zbd,Amendola:2016saw,Euclid:2019clj}.

The goal of this paper is to summarise the status of, motivations for, and challenges in searching for oscillatory 
features in the primordial fluctuations \citep[see][for reviews]{Chluba:2015bqa,2022arXiv220308128A} with 
large-scale structure (LSS) observations expected from \Euclid.  
Searches for primordial features based on the cosmic microwave background (CMB) angular power spectra 
\citep{Wang:2000js,Adams:2001vc,Peiris:2003ff,Mukherjee:2003ag,Covi:2006ci,Hamann:2007pa,Meerburg:2011gd,Planck:2013jfk,Meerburg:2013dla,Benetti:2013cja,Miranda:2013wxa,Easther:2013kla,Chen:2014joa,Achucarro:2014msa,Hazra:2014goa,Hazra:2014jwa,Hu:2014hra,Ade:2015lrj,Gruppuso:2015zia,Gruppuso:2015xqa,Hazra:2016fkm,Torrado:2016sls,Akrami:2018odb,Zeng:2018ufm,Canas-Herrera:2020mme,Braglia:2021ckn,Braglia:2021sun,Braglia:2021rej,Naik:2022mxn,Hamann:2021eyw} and bispectra 
\citep{Planck:2013wtn,Fergusson:2014tza,Planck:2015zfm,Meerburg:2015owa,Akrami:2018odb,Planck:2019izv} have so far 
not detected any statistically significant signal, setting constraints on deviations from a pure power-law primordial 
power spectrum (PPS) at the few percent level (depending on the methodology applied) for wavenumbers in the range 
$0.005 < k/(h\,{\rm Mpc}^{-1}) < 0.2$. However, it is important to stress that there are interesting models of primordial 
features that are able to reproduce some of the anomalous features observed in the CMB temperature angular power spectrum 
at a marginal statistical significance of 99.7\%. These include the dip in power in the multipole range $\ell \sim 20$--40 
and an oscillatory pattern at the intermediate scales $\ell \sim 700$--800. Models with primordial features have also 
been studied in the context of the anomalous CMB lensing smoothing excess, as well as the $H_0$ and $S_8$ tensions 
\citep{Akrami:2018odb,Domenech:2019cyh,Liu:2019dxr,Domenech:2020qay,Keeley:2020rmo,Hazra:2022rdl,Antony:2022ert,Ballardini:2022vzh}.

In the context of primordial features, large-volume photometric and spectroscopic surveys not only provide a complementary 
look at the structure on very large scales, but they can also provide a precise measurement of the power spectrum on small 
scales, complementing the CMB measurements and improving the sensitivity for high-frequency signals. 
This has been studied extensively with forecast analyses in 
\citet{Wang:1998gb}, \citet{Zhan:2005rz}, \citet{Huang:2012mr}, \citet{Chen:2016vvw}, \citet{Chen:2016zuu}, \citet{Ballardini:2016hpi}, \citet{Xu:2016kwz}, \citet{Fard:2017oex}, \citet{Palma:2017wxu}, \citet{Ballardini:2017qwq}, \citet{Ballardini:2018noo}, \citet{Beutler:2019ojk}, \citet{Ballardini:2019tuc}, \citet{Debono:2020emh}, and \citet{Li:2021jvz} and demonstrated on real 
data in \citet{Beutler:2019ojk}, \citet{Ballardini:2022wzu}, and \citet{Mergulhao:2023ukp} showing that current galaxy clustering data of the 
Baryon Oscillation Spectroscopic Survey (BOSS) alone can already provide constraints that are competitive with those 
derived from the {\em Planck} CMB data. Given the potential of the \Euclid mission, it is imperative to have a good 
description of the LSS observables on all observed scales, and to do so, nonlinear corrections need to be correctly 
accounted for 
\citep{Vlah:2015zda,Vasudevan:2019ewf,Beutler:2019ojk,Ballardini:2019tuc,Chen:2020ckc,Li:2021jvz,Ballardini:2022vzh}.

In this paper, we forecast how well the surveys from the \Euclid mission can constrain two templates of primordial 
undamped linear and logarithmic oscillations. In addition to the \Euclid's primary probes, i.e. the spectroscopic galaxy 
clustering and the combination of photometric surveys, we study the further constraints that will be added by \Euclid's 
measurements of the galaxy bispectrum and the information provided by future CMB experiments. The bispectrum is shown 
to be able to provide essential information on the parameters of the feature models, highlighting the importance of a 
high-order statistics analysis on the spectroscopic sample.

We structure the paper as follows. 
In Sect.~\ref{sec:theory}, we introduce the physics of the adiabatic mode and we briefly review the diverse theoretical 
mechanisms that naturally generate features in the PPS, in particular realisations of inflation. In Sect.~\ref{sec:probes}, 
we review \Euclid specifications and calculation of the theoretical observables used in the Fisher forecast analysis. 
In Sect.~\ref{sec:modelling} we start introducing the parametrised templates for primordial features with linear and 
logarithmic undamped oscillations. We then study the behaviour of these models on nonlinear scales of the matter power 
spectrum comparing the results from perturbation theory in Sect.~\ref{sec:PT} to N-body simulations in 
Sect.~\ref{sec:PT_vs_sim}, and we study the numerical reconstruction of primordial features in 
Sect.~\ref{sec:reconstruction}. We present our results from primary \Euclid observables in Sect.~\ref{sec:results}. We 
complement those by adding the information from the galaxy bispectrum in Sect.~\ref{sec:bispectrum} and combining \Euclid 
results with future CMB measurements in Sect.~\ref{sec:CMB}. We conclude in Sect.~\ref{sec:conclusions}.

\section{\label{sec:theory}Primordial features and the early Universe}
As mentioned in Sect.~\ref{sec:intro}, the search for features in the primordial power spectrum has so far been 
statistically unsuccessful. On the largest scales, CMB observations agree with a red-tilted power-law PPS, 
${\cal P}_{{\cal R},\,0}(k)$, with the following parametrisation 
\begin{equation} \label{eq:PPS}
    {\cal P}_{{\cal R},\,0}(k) = A_{\rm s} \left(\frac{k}{k_*}\right)^{n_{\rm s}-1} \,,
\end{equation}
where $A_{\rm s}$ and $n_{\rm s}$ are the amplitude and the spectral index of the comoving curvature perturbations 
${\cal R}$ on superhorizon scales at the pivot scale $k_* = 0.05\,{\rm Mpc}^{-1}$. 

The simplest class of models that can produce such a spectrum is known as canonical single-field slow-roll inflation. 
In these models a single scalar field — the inflaton — is responsible for both the primordial perturbations and the 
background evolution during inflation.\footnote{These models are based additionally on other simplifying assumptions 
about the inflaton field, such as canonical kinetic terms, minimal coupling to gravity and the \emph{Bunch-Davies} 
quantum vacuum for the perturbations.}

Unless the inflaton is the Higgs field \citep{Bezrukov:2007ep}, inflation always takes us beyond the Standard Model 
(SM) of particle physics. While these scenarios can support cosmic inflation, it is rarely of the simplest kind. 
Beyond-SM models often involve multiple fields that interact with each other, and with the inflaton, and they can 
leave detectable imprints on the primordial perturbations including, for example, isocurvature modes and localised features. 
It is essential to have a bottom-up theoretical framework enabling us to analyse and classify all possible departures 
from the simplest scenario of Eq.~\eqref{eq:PPS} that are compatible with observations. Here we focus on the fluctuations 
as those are what we observe in the CMB and LSS. An important distinction is whether the curvature perturbations on 
superhorizon scales are generated by a single, possibly effective, degree of freedom. These are called effectively 
single field or `single clock' models. 

It can be shown that in the effective single-field slow-roll models the effective action for the comoving curvature 
perturbations is, to quadratic order, 
\begin{equation}\label{eq:S2_EFT}
S = \int{{\rm d}^4x\, a^3M_\mathrm{Pl}^2\epsilon_1 \left[\frac{\dot{\mathcal{R}}^2}{c_{\rm s}^2}-\frac{(\partial_i \mathcal{R})^2}{a^2}\right]} + \dots \,,
\end{equation}
where $a(t)$ is the scale factor, $c_{\rm s}(t)$ is the speed of sound for curvature perturbations, and 
$\epsilon_1(t) \equiv -\dot{H} / H^2$ is the first slow-roll parameter; $H \equiv \dot{a}/a$ is the Hubble expansion 
rate, where the overdot denotes derivative with respect to the cosmic time $t$.

Eq.~\eqref{eq:S2_EFT} is well known in the context of single-field slow-roll inflation, but it is much more general 
than that. It is the first term in a perturbative expansion — the effective field theory (EFT) of inflationary 
perturbations \citep{2008JHEP...03..014C} — where all the information about the background is systematically encoded 
in a set of functions ($\epsilon_1$ and $c_{\rm s}$ being the first two) that describe what is seen by the perturbations.  

This action describes almost all models of slow-roll inflation in which the primordial perturbations are generated by 
a single quantum field. It includes canonical single-field slow-roll models as a particular case ($c_{\rm s} = 1$) but 
also any multi-field model in which the primordial perturbations are generated by a single quantum field (typically an 
effective low energy degree of freedom involving multiple high energy fields). The action in Eq.~\eqref{eq:S2_EFT} is 
then obtained by integrating out the fast, or heavy, high energy degrees of freedom; see \citet{Achucarro:2012sm}.

The idea behind the EFT approach is that what the perturbations see is an almost time-independent background, and this 
symmetry under time translations completely dictates the form of the action in Eq.~\eqref{eq:S2_EFT}, to all orders. 
The red tilt is a small deviation from perfect scale invariance, measured by the smallness of the slow-roll parameters. 
Any deviations from time-independence in the background functions result in features in the spectrum that can be calculated 
and cross-correlated among different observables; see \citet{Bartolo:2013exa,Cannone:2014qna} for applications of the 
EFT in the context of primordial oscillatory features. 

In general, and depending on their origin, oscillatory features fall into two main classes which are the focus of this 
paper \cite[see][for details and references]{2022arXiv220308128A}. A small, abrupt change in the background functions 
$\epsilon_1(t)$ and $c_{\rm s}(t)$ at a particular moment during inflation results in a transient oscillatory feature, 
linearly spaced in $k$, superimposed on the power spectrum \eqref{eq:PPS}. Two well studied examples are  a step in the 
inflationary potential \citep{Starobinsky:1992ts,Adams:2001vc} and a localised turn in the inflationary trajectory 
\citep{Achucarro:2010da,Chen:2011zf}. The range of $k$ where the oscillation persists increases with the sharpness of 
the change. But the sharpness cannot be arbitrarily large, since it should not excite the high energy modes that have 
been integrated out, nor enhance higher order terms that have been neglected in the perturbative expansion. On the other 
hand, oscillations logarithmically spaced in $k$ are usually associated with periodicity in the background functions, 
for instance if there is a periodic modulation in the inflationary potential \citep{Freese:1990rb,Chen:2008wn}. More 
complicated combinations are also possible, that require a more tailored analysis. The upshot is that, if a feature is 
detected in the power spectrum, establishing its high energy origin will require stringent self-consistency checks, as 
well as correlated detections in other observables. 

It is worth emphasising that the power of the EFT approach is that, since it is completely agnostic about the origin of 
the background expansion, it tests entire classes of models as opposed to individual ones. For example in multi-field 
models, $\epsilon_1$ is related to the flatness of the potential along the inflationary trajectory, and $c_{\rm s}$ is 
related to the rate of turning of this trajectory in multi-field space. But the high energy origin of $\epsilon_1$ and 
$c_{\rm s}$ may be completely different in other models.

\section{\label{sec:probes}\textbf{\textit{Euclid}} probes}
\Euclid is one of the next generation deep- and large-field galaxy surveys. It will be able to measure up to $30$ 
million spectroscopic redshifts, which can be used for galaxy clustering measurements, and $2$ billion photometric 
galaxy images, which can be used for weak lensing observations.

\Euclid will use near-IR (NIR) slitless spectroscopy to collect large samples of emission-line galaxies and to 
perform spectroscopic galaxy clustering measurements. The great advantage of this spectroscopic method is the high 
precision on measuring the redshift of the sources. However, one of the difficulties in inferring the cosmological 
parameters is given to the knowledge of the number density of the $H\alpha$ targets. The total number of galaxies that 
can be used in the analysis mostly depends on two factors: completeness and purity. Completeness is the fraction of 
objects correctly identified relative to the true number, whereas purity is defined as the fraction of objects correctly 
identified relative to the total number of detected objects. In practice, completeness represents the size of a sample, 
whereas purity is its quality. A lower value of the number density $n(z)$ of observed galaxies leads to an increase 
of the shot noise, resulting as a degradation on the constraints of the cosmological parameters.

One of the expected lensing observations is the cosmic shear, which is the distortion in the observed shapes of 
distant galaxies due to weak gravitational lensing by the LSS while the other probe comes from galaxy clustering 
using the positions of objects detected by the photometric measurements of redshifts. Given that both depend on 
the LSS density fluctuations as well as the geometry of the Universe, this will allow us to constrain the cosmological 
parameters. One of the main difficulties in using the weak lensing observable is the ability to accurately model 
the intrinsic alignments (IA) of galaxies that mimic the cosmological lensing signal.
While for the galaxy clustering from photometric measurements, the main effects that need to be taken into account 
are the galaxy bias, accounting for the relation between the galaxy distribution and the underlying total matter 
distribution, and the photometric-redshift uncertainties, since redshift in this case is estimated from observing 
through multi-band filters instead of the full spectral energy distribution. Finally, an important issue for both 
the weak lensing and galaxy clustering probes is the modelling of the small-scale nonlinear clustering on the weak 
lensing two-point statistics that the high density of detected galaxies will allow us to reach.

In the following we detail on the prescription used in this work for our model for these two probes.

\subsection{\label{sec:GCsp}Spectroscopic galaxy clustering}
Following \citet{Euclid:2019clj}, we define the observed galaxy power spectrum as
\begin{align} \label{eq:GC:pkobs}
    P_\text{obs}(k_{\rm ref},\mu_{\rm ref};z) =
    & \frac{1}{q_\perp^2 q_\parallel} \frac{1}{1+k^2\mu^2f^2(z)\sigma_{\rm p}^2(z)} \notag\\ 
    & \times \left[b(z)\sigma_8(z)+f(z)\sigma_8(z)\mu^2\right]^2 \notag\\
    & \times \frac{P^{\rm IR\,res,\,LO}(k,\mu;z)}{\sigma^2_8(z)} F_z(k,\mu;z) \notag\\
    & + P_{\rm s}(z) \,,
\end{align}
where $\mu$ is the cosine of the angle of the wave mode with respect to the line of sight pointing into the direction 
$\hat{r}$. Here $b(z)$ is the linear clustering bias, $f(z)$ is the growth rate, and $\sigma_8(z)$ is the root mean 
square linearly evolved density fluctuations in spheres of $8\,h^{-1}\,{\rm Mpc}$.\footnote{$h$ is the dimensionless 
Hubble parameter defined as $H_0 = 100 h\,{\rm km}\,{\rm s}^{-1}\,{\rm Mpc}^{-1}$.}

The Alcock-Paczynski (AP) effect \citep{Alcock:1979mp}, used to account for deviations of the cosmological models from 
the fiducial one, is parametrized through rescaling of the angular diameter distance $D_{\rm A}(z)$ and the Hubble 
parameter $H(z)$, and enters as multiplicative factor through
\begin{equation}\label{eq:AP1}
    q_\perp(z) = \frac{D_{\rm A}(z)}{D_{\rm A,\, ref}(z)} 
    \qquad \text{and} \qquad q_\parallel(z) = \frac{H_{\rm ref}(z)}{H(z)} \,,
\end{equation}
where the subscript $\rm ref$ refers to the fiducial cosmology. This leads also to a rescaling of the wavevector 
components as
\begin{equation}\label{eq:AP2}
    k_\perp = \frac{k_{\perp,{\rm ref}}}{q_\perp} \qquad \text{and} \qquad k_\parallel = \frac{k_{\parallel,{\rm ref}}}{q_\parallel} \,.
\end{equation}
Using Eqs.~\eqref{eq:AP1} and \eqref{eq:AP2} we can convert the known reference cosmology ($k_{\rm ref}$ and $\mu_{\rm ref}$) to the true unknown cosmology ($k$ and $\mu$),
\begin{align}
    &k(k_{\rm ref},\mu_{\rm ref}) = k_{\rm ref}\alpha(\mu_{\rm ref}) \,,\\
    &\mu(\mu_{\rm ref}) = \frac{\mu_{\rm ref}}{\alpha(\mu_{\rm ref})q_\parallel} \,,
 \end{align}
 where
 \begin{equation}
         \alpha(\mu) = \left[\frac{\mu^2}{q_\parallel^2}+\frac{\left(1-\mu^2\right)}{q_\perp^2}\right]^{1/2} \,.  
 \end{equation}
 The relations above can be used to define the effect of the choice of reference cosmology on the observed power spectrum via
 \begin{equation}
     P_{\rm obs}(k_{\rm ref},\mu_{\rm ref};z)=\frac{1}{q_\perp^2q_\parallel}P_{\rm g}\left(k(k_{\rm ref}),\mu(\mu_{\rm ref});z\right) \,.
 \end{equation}

Galaxy bias, connecting the underlying dark matter (DM) power spectrum to the $H\alpha$-line emitter galaxies detected 
by \Euclid, is modelled by a simple linear clustering bias redshift-dependent coefficient $b(z)$. The anisotropic 
distortion to the density field due to the line-of-sight effects of the peculiar velocity of the observed galaxy 
redshifts, known as redshift-space distortions (RSD), is modelled in the linear regime by the contribution in the 
square brackets introduced by \citet{Kaiser:1987qv}. Linear RSD are corrected for the nonlinear finger-of-God (FoG) 
effect under the assumption of an exponential galaxy velocity distribution function as a Lorentzian \citep{Hamilton:1997zq}. 
The nonlinearities due to gravitational instabilities have the effect of smearing the baryon acoustic oscillations 
(BAO) signal, this is modelled through time-sliced perturbation theory (TSPT); we discuss the modelling of 
$P^{\rm IR\,res,\,LO}$ in Sect.~\ref{sec:PT}. The pairwise velocity dispersion, $\sigma_{\rm p}(z)$, is evaluated from 
the linear matter power spectrum as 
\begin{equation} \label{eq:sigmas}
    \sigma^2_{\rm p}(z)  
    = \frac{1}{6\pi^2}\int_0^{k_{\rm max}}{\rm d} k\, P_{\rm lin}(k,z) \,.
\end{equation}
The total galaxy power spectrum in Eq.~\eqref{eq:GC:pkobs} includes errors on the measurement of the redshift through 
the exponential factor 
\begin{equation}\label{eq:zerror}
    F_z(k, \mu;z) = \text{e}^{-k^2\mu^2\sigma_{r}^2(z)} \,,
\end{equation}
where $\sigma_{r}^2(z) = c(1+z)\sigma_{0,z}/H(z)$ with $\sigma_{0,z}$ the error on the measured redshifts. Finally, 
we introduce a shot-noise term $P_{\rm s}(z)$ due to imperfect removal of the Poisson sampling and the imperfect 
modelling of small scales.

The final Fisher matrix for the spectroscopic galaxy clustering (\GCsp) observable for one redshift bin $z_i$ is 
\begin{eqnarray} \label{eq:fisher-gc}
    F_{\alpha\beta}(z_i) &=& \frac{1}{8\pi^2}\int_{-1}^{1}{\rm d}\mu\int_{k_{\rm min}}^{k_{\rm max}} k^2{\rm d}k V_{\rm eff}(z_i,k) \nonumber \\
    &&\times \frac{\partial \ln P_{\rm obs}(k,\mu;\,z_i)}{\partial \theta_\alpha}
    \frac{\partial \ln P_{\rm obs}(k,\mu;\,z_i)}{\partial \theta_\beta} \,,
\end{eqnarray}
where the derivatives are evaluated at the parameter values of the fiducial model and $V_{\rm eff}$ is the effective 
volume of the survey, given by
\begin{equation}
    V_{\rm eff}(k,\mu;\,z) = V_{\rm s}\left[\frac{n(z)P_{\rm obs}(k,\mu;\,z)}{1+n(z)P_{\rm obs}(k,\mu;\,z)}\right]^2 \,,
\end{equation}
where $V_{\rm s}$ is the volume of the survey and $n(z)$ is the number of galaxies in a redshift bin. Assuming that 
the observed power spectrum follows a Gaussian distribution, we write in Eq.~\eqref{eq:fisher-gc} its covariance matrix 
as
\begin{equation}
    {\rm Cov}(k,k') \approx \frac{2(2\pi)^3}{V_{\rm eff}(k,\mu;\,z)} P_{\rm obs}^2(k,\mu;\,z) \delta^{\rm D}(k-k') \,,
\end{equation}
where $\delta^{\rm D}(k-k')$ is the Dirac delta function.

We model the observed matter power spectrum of Eq.~\eqref{eq:GC:pkobs} in bandpowers averaged over a bandwidth of 
$\Delta k$ with a top-hat window function as in \citet{Huang:2012mr,Ballardini:2016hpi},
\begin{equation}
    \hat{P}_{\rm obs}(k,\mu;\,z) = \frac{1}{\Delta k}\int^{k+\Delta k/2}_{k-\Delta k/2} {\rm d}k' 
    P_{\rm obs}(k',\mu;\,z) \,.
\end{equation}
The size of the $k$-bin $\Delta k$ should be equal or larger than the effective fundamental frequency defined as 
$k_{\rm eff} \equiv 2\pi/V^{1/3}$ for an ideal cubic volume in order to guarantee uncorrelated measurements of the 
observed power spectrum. The values of $k_{\rm eff}$ span in the range 0.0025--$0.0031 \,h\,{\rm Mpc}^{-1}$ for a 
cubic volume for the four redshift bins. We assume a bin width of $\Delta k = 0.004\,h\,{\rm Mpc}^{-1}$ with 
$k_{\rm min} = 0.002 \,h\,{\rm Mpc}^{-1}$. Finally, we can rewrite the Fisher matrix of Eq.~\ref{eq:fisher-gc} for 
the discrete number of averaged bandpower for the redshift bin $z_i$ as
\begin{align}
    F_{\alpha\beta} (z_i) = &\frac{1}{8\pi^2}\int_{-1}^{1} {\rm d}\mu\sum_{k = k_{\rm min}}^{k_{\rm max}} k^2\Delta k V_{\rm eff}(z_i,k) \notag\\
    &\times\frac{\partial \ln \hat{P}_{\rm obs}(k,\mu;\,z_i)}{\partial \theta_\alpha}
    \frac{\partial \ln \hat{P}_{\rm obs}(k,\mu;\,z_i)}{\partial \theta_\beta} \,.
\end{align}
The total spectroscopic Fisher matrix is then calculated by summing over the redshift bins as 
$F_{\alpha\beta}^{\rm sp} = \sum_i F_{\alpha\beta} (z_i)$, assuming that the redshift bins are independent. Since we 
are interested in the standard cosmological parameters and the feature parameters, we marginalise the \GCsp\ Fisher 
matrix over two redshift-dependent parameters $b\sigma_8(z_i)$ and $P_{\rm s}(z_i)$ for each of the four redshift bins.

\subsection{Photometric galaxy clustering and weak lensing}
Both the forecasting method and the tools used for the photometric analyses are the same as the ones in 
\citet{Euclid:2019clj} apart from the changes in the power spectrum due to the primordial features in the predicted 
\Euclid observables. Here we only remind the reader of the main steps.

The observables we consider are the angular power spectra \smash{$C_{ij,\ell}^{XY}$} between probe $X$ in the $i$-th 
redshift bin and probe $Y$ in the $j$-th redshift bin, where the probes $X$ and $Y$ are L for weak lensing or G for 
photometric galaxy clustering; \smash{$C_{ij,\ell}^{XY}$} therefore refers to both auto- and cross-correlations of these 
probes. Relying on the Limber approximation and within the flat sky limit, the spectra are given by
\begin{equation}\label{eq:ISTrecipe}
    C^{XY}_{ij,\ell} = c\int_{z_{\rm min}}^{z_{\rm max}}\de z\,{\frac{W_i^X(z)W_j^Y(z)}{H(z)r^2(z)}P_{\rm NL}(k_\ell,z)} \,,
\end{equation}
with $k_\ell=(\ell+1/2)/r(z)$, $r(z)$ the comoving distance to redshift $z$, and $P_{\rm NL}(k_\ell,z)$ the nonlinear 
power spectrum of matter density fluctuations at wave number $k_{\ell}$ and redshift $z$. The \GCph\, and WL window 
functions read 
\begin{align}\label{eq:wg_mg}
    W_i^{\rm G}(z) =&\; b_i(z)\,n_i(z)\frac{H(z)}{c} \,, \\  
    W_i^{\rm L}(z) =&\; \frac{3}{2}\Omega_{\mathrm{m},0} \left(\frac{H_0}{c}\right)^2(1+z)\,r(z)\,
\int_z^{z_{\rm max}}{\de z'n_i(z')\left[1-\frac{r(z)}{r(z')}\right]}\nonumber\\
    &+W^{\rm IA}_i(z) \,, \label{eq:wl_mg}
\end{align}
where $b_i(z)$ is the galaxy bias in the $i$-th redshift bin, and $W^{\rm IA}_i(z)$ encodes the contribution of IA to 
the WL power spectrum. The normalised number density distribution $n_i(z)$ of observed galaxies in the $i$-th redshift 
bin is given by
\begin{equation}
    n_i(z) = \frac{\int_{z_i^-}^{z_i^+}\de z'n(z)p_{\rm ph}(z'|z)}{\int_{z_{\rm min}}^{z_{\rm max}}\de z\int_{z_i^-}^{z_i^+}\de z'n(z)p_{\rm ph}(z'|z)} \,,
\end{equation}
where $(z_i^-,z_i^+)$ are the edges of the $i$-th redshift bin and $n(z)$ is the underlying true redshift distribution. 
The true number density of galaxies is convolved with the probability distribution function $p_{\rm ph}(z'|z)$ is given 
following the parameterization in \citet{Euclid:2019clj}.

The IA contribution is computed following the extended nonlinear alignment (eNLA) model adopted in \citet{Euclid:2019clj} 
so that the corresponding window function is
\begin{equation}\label{eq:IA}
    W^{\rm IA}_i(z)=-\frac{\mathcal{A}_{\rm IA}\,\mathcal{C}_{\rm IA}\,\Omega_{\rm m,0}\,\mathcal{F}_{\rm IA}(z)}{D(z)}n_i(z)\frac{H(z)}{c} \,,
\end{equation}
where 
\begin{equation}
    \mathcal{F}_{\rm IA}(z) = (1+z)^{\eta_{\rm IA}}\left[\frac{\langle L\rangle(z)}{L_\star(z)}\right]^{\beta_{\rm IA}} \,,
\end{equation}
with $\langle L\rangle(z)$ the redshift-dependent mean, and $L_\star(z)$ the characteristic luminosity of source galaxies 
as computed from the luminosity function. $\mathcal{A}_{\rm IA}$, $\beta_{\rm IA}$ and $\eta_{\rm IA}$ are the nuisance 
parameters of the model, and $\mathcal{C}_{\rm IA}$ is a constant accounting for dimensional units.

We consider a Gaussian-only covariance whose elements are given by
\begin{align}
    \text{Cov}\left[C_{ij,\ell}^{\rm AB},C_{mn,\ell'}^{\rm CD}\right]=&\frac{\delta_{\ell\ell'}^{\rm K}}{(2\ell+1)f_{\rm sky}\Delta \ell} \notag\\
    &\times\left\{\left[C_{im,\ell}^{\rm AC}+{\cal N}_{im,\ell}^{\rm AC}\right]\left[C_{jn,\ell'}^{\rm BD}+{\cal N}_{jn,\ell'}^{\rm BD}\right]\right. \notag\\
    &+\left.\left[C_{in,\ell}^{\rm AD}+{\cal N}_{in,\ell}^{\rm AD}\right]\left[C_{jm,\ell'}^{\rm BC}+{\cal N}_{jm,\ell'}^{\rm BC}\right]\right\} \,,
\end{align}
where the upper- and lower-case Latin indices run over L and G (all tomographic bins), $\delta_{\ell\ell'}^{\rm K}$ 
is the Kronecker delta coming from the lack of correlation between different multipoles $(\ell, \ell')$, $f_{\rm sky}$ 
is the survey's sky fraction, and $\Delta \ell$ denotes the width of the logarithmic equi-spaced multipole bins. We 
consider a white noise,
\begin{align}
    {\cal N}_{ij,\ell}^{\rm LL} &= \frac{\delta_{ij}^{\rm K}}{\bar{n}_i}\sigma_\epsilon^2 \,, \quad
    {\cal N}_{ij,\ell}^{\rm GG} = \frac{\delta_{ij}^{\rm K}}{\bar{n}_i} \,, \quad
    {\cal N}_{ij,\ell}^{\rm GL} = 0 \,,
\end{align}
where $\sigma_\epsilon$ is the variance of observed ellipticities.

For evaluating the Fisher matrix $F_{\alpha\beta}^{\rm ph}$ for the observed galaxy power spectrum, we use
\begin{equation}
    F_{\alpha\beta}^{\rm ph} = \sum_{\ell=\ell_{\rm min}}^{\ell_{\rm max}} \sum_{ij,mn}
    \frac{\partial C_{ij,\ell}^{\rm AB}}{\partial \theta_\alpha}
    \text{Cov}^{-1}\left[C_{ij,\ell}^{\rm AB},C_{mn,\ell}^{\rm CD}\right]
    \frac{\partial C_{mn,\ell}^{\rm CD}}{\partial \theta_\beta} \,.
\end{equation}

\subsection{Survey specifications}
Following \citet{Euclid:2019clj}, we consider for the spectroscopic sample four redshift bins centred at 
$\{1.0,1.2,1.4,1.65\}$, whose widths are $\Delta z = 0.2$ for the first three bins and $\Delta z = 0.3$ 
for the last bin. The H$\alpha$ bias, evaluated at the central redshift of the bins, corresponds to 
$\{1.46,1.61,1.75,1.90\}$. The number density of galaxies $n(z)$ corresponds to $\{6.86,5.58,4.21,2.61\}\times 10^{-4}$.

For the photometric probes, the sources are split into 10 equi-populated redshift bins whose limits are 
obtained from the redshift distribution 
\begin{equation}
    n(z)\propto\left(\frac{z}{z_0}\right)^2\,\text{exp}\left[-\left(\frac{z}{z_0}\right)^{3/2}\right] \,,
\end{equation}
with $z_0=0.9/\sqrt{2}$ and the normalisation set by the requirement that the surface density of galaxies 
is $\bar{n}_g=30\,\mathrm{arcmin}^{-2}$. This is then convolved with the sum of two Gaussians to account 
for the effect of photometric redshift \cite[see][for details]{Euclid:2019clj}. The galaxy bias is assumed 
to be constant within each redshift bin, with fiducial values $b_i = \sqrt{1 + \bar{z}_i}$, where $\bar{z}_i$ 
is the bin centre. We consider $100$ logarithmic equi-spaced multipole bins with $\sigma_\epsilon=0.3$ the 
variance of observed ellipticities. We set $(z_{\rm min}, z_{\rm max}) = (0.001, 4)$, which spans the full 
range where the source redshift distributions $n_i(z)$ are non-vanishing.

For all the \Euclid probes the survey's sky fraction is $f_{\rm sky}\simeq 0.36$ \citep{Euclid:2021icp}.

\section{\label{sec:modelling}Theoretical modelling of primordial features in galaxy surveys}
As representative models for features in primordial fluctuations in this paper we consider oscillations linearly 
and logarithmically spaced in Fourier space with a constant amplitude superimposed to a power spectrum described 
by a power-law.

\subsection{\label{sec:features}Models}
We consider a superimposed pattern of oscillations as
\begin{equation}\label{eq:feature_PS}
    {\cal P}_{\cal R}(k) = {\cal P}_{{\cal R},\,0}(k) \left[1 + \delta P^{\rm X}(k)\right] \,,
\end{equation}
where ${\cal P}_{{\cal R},\,0}(k)$ is the standard power-law PPS of the comoving curvature perturbations $\cal R$ 
on superhorizon scales, given in Eq.~\eqref{eq:PPS}. We study the following templates with superimposed oscillations 
on the PPS
\begin{equation} \label{eqn:pk_wigg}
    \delta P^{\rm X}(k) = {\cal A}_{\rm X} \sin\left(\omega_{\rm X}\Xi_{\rm X} + 2\pi\phi_{\rm X}\right) \,,
\end{equation}
where ${\rm X} = \{{\rm lin,\,log}\}$ and $\Xi_{\rm X} = \{k/k_*,\,\ln(k/k_*)\}$.

We choose the fiducial cosmological parameters according to {\em Planck} DR3 mean values of the marginalised 
posterior distributions \citep{Planck:2018vyg}: $\omega_{\rm b} = 0.02237$, $\omega_{\rm c} = 0.1200$, 
$H_0 = 67.36\, \kmsMpc$, $\ln (10^{10}A_{\rm s}) = 3.044$, $n_{\rm s} = 0.9649$, $\sigma_8 = 0.8107$, and 
$m_\nu = 0.06$\,\si{eV} (with one massive neutrino and two massless ones). In what follows we specify the fiducial 
values of the model parameters.
\begin{enumerate}
 \item Linear oscillations:
  \begin{equation}
   \Theta_{\rm lin} = \Bigl\{ {\cal A}_{\rm lin} = 0.01,\, \omega_{\rm lin} = 10,\, \phi_{\rm lin} = 0 \Bigr\} \,.
   \label{eqn:fiducial_lin}
  \end{equation}
 \item Logarithmic oscillations:
  \begin{equation}
   \Theta_{\rm log} = \Bigl\{ {\cal A}_{\rm log} = 0.01,\, \omega_{\rm log} = 10,\, \phi_{\rm log} = 0 \Bigr\} \,.
   \label{eqn:fiducial_log}
  \end{equation}
\end{enumerate}

\begin{figure}
\centering
\includegraphics[width=0.48\textwidth]{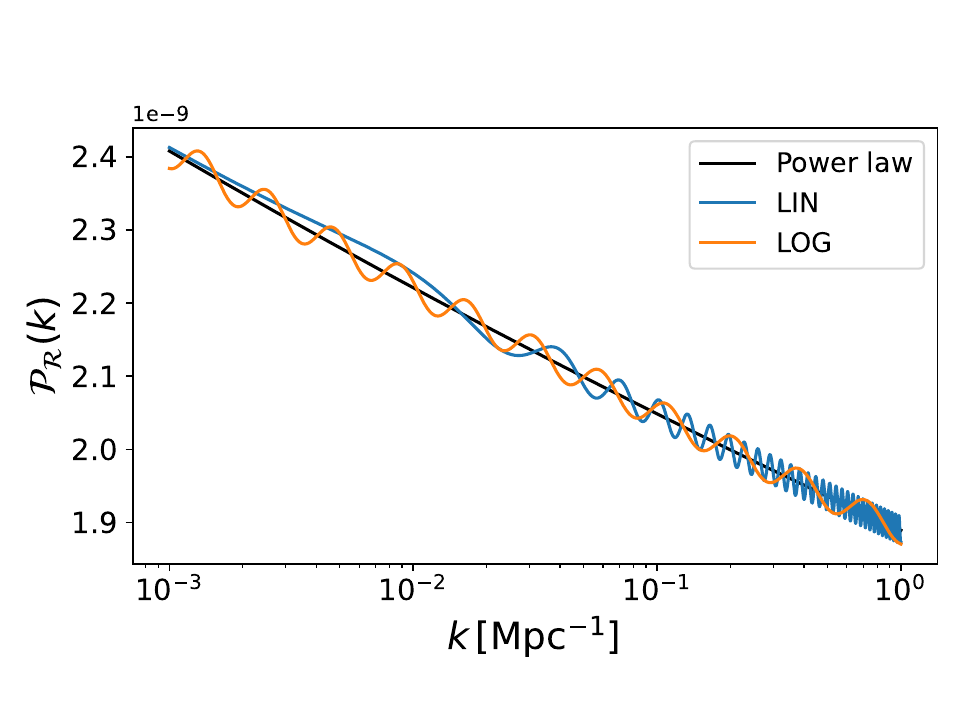}
\caption{Primordial power spectrum of curvature perturbations for linear (LIN) and logarithmic (LOG) feature 
models with fiducial parameter values given in \eqref{eqn:fiducial_lin} and \eqref{eqn:fiducial_log}.}
\label{fig:pps}
\end{figure}

We show the effect of the feature parameters on the PPS in Fig.~\ref{fig:pps}.

\subsection{\label{sec:PT}Modelling nonlinear scales in perturbation theory in the presence of primordial oscillatory features}
The linearly propagated matter power spectrum is given by
\begin{equation}\label{eq:Plin}
    P_{\rm lin}(k,z)=M^2(k,z)P_\Phi(k) \,,
\end{equation}
where $M(k,z)=2k^2c^2T(k)D(z)/(3\Omega_{\rm m}H_{0}^2)$, with $T(k)$ the matter transfer function normalised to 
unity at large scales (i.e. $k \rightarrow 0$) and $c$ the speed of light. The gravitational potential power spectrum 
will then be derived from the two oscillatory models considered: $P_\Phi=9/25(2\pi^2){\cal P}_{\cal R}(k)/k^3$.

Our analytic model for the matter power spectrum is based on TSPT and it takes into account the damping of oscillations 
by infrared (IR) resummation of the large-scale bulk flows 
\citep{Crocce:2007dt,Creminelli:2013poa,Baldauf:2015xfa,Blas:2016sfa,Senatore:2017pbn}. The implementation of the 
IR resummation can be done following the approach to BAO in the context of TSPT \citep{Blas:2015qsi,Blas:2016sfa}. 
We start by decomposing the linear matter power spectrum into a smooth (or no-wiggle; nw) and an oscillating (w) 
contribution,
\begin{equation} \label{eqn:split1}
    P_{\rm lin}(k,z) = D^2(z) \left[P_{\rm nw}(k) + P_{\rm w}(k)\right] \,,
\end{equation}
where $P_{\rm nw}$ is the no-wiggle power spectrum, and the oscillatory part $P_{\rm w}$ describes both the BAO feature 
and the primordial oscillating feature as
\begin{equation} \label{eqn:split2}
    P_{\rm w}(k) \equiv P_{\rm nw}\left[\delta P_{\rm w}^{\rm BAO}(k) 
    + \delta P_{\rm w}^{\rm X}(k) + \delta P_{\rm w}^{\rm BAO}(k)\delta P_{\rm w}^{\rm X}(k)\right] \,.
\end{equation}
Here we have factored out the time-dependence given by the growth factor $D(z)$. We have neglected the cross term in 
Eq.~\eqref{eqn:split2} as it is proportional to $A_{\rm BAO} \times {\cal A}_{\rm lin}$ and therefore subdominant. 
We filter the BAO feature from the linear matter power spectrum using a Savitzky-Golay filter; see \citet{Boyle:2017lzt}. 

At next-to-leading order (NLO), the IR resummed power spectrum for linear oscillations can be written as 
\citep{Blas:2016sfa,Beutler:2019ojk}
\begin{align} \label{eqn:lin_NLO}
    &P^{\rm IR\, res,\, LO+NLO}(k,z) = D^2(z) P_{\rm nw}(k) \notag\\
    &\qquad\times\left\{1 + \left[1+k^2 D^2(z) \Sigma^2_{\rm BAO}\right]\mathrm{e}^{-k^2 D^2(z) \Sigma^2_{\rm BAO}} \delta P_{\rm w}^{\rm BAO}(k)\right. \notag\\
    &\qquad \left.+\left[1+k^2 D^2(z) \Sigma^2_{\rm lin}\right]\mathrm{e}^{-k^2 D^2(z) \Sigma^2_{\rm lin}}\delta P_{\rm w}^{\rm lin}(k)\right\} \notag\\
    &\qquad+ D^4(z) P^{\rm 1-loop}\left[ P^{\rm IR\, res,\, LO}(k) \right] \,,
\end{align}
and for logarithmic oscillations as \citep{Vasudevan:2019ewf,Beutler:2019ojk}
\begin{align} \label{eqn:log_NLO}
    &P^{\rm IR\, res,\, LO+NLO}(k,z) = D^2(z) P_{\rm nw}(k) \notag\\
    &\qquad\times\Biggl\{1 + \left[1+k^2 D^2(z) \Sigma^2_{\rm BAO}\right]\mathrm{e}^{-k^2 D^2(z) \Sigma^2_{\rm BAO}} \delta P_{\rm w}^{\rm BAO}(k) \notag\\
     &\qquad+\left[1+k^2 D^2(z) \Sigma^2_{\rm log}\right]\mathrm{e}^{-k^2 D^2(z) \Sigma^2_{\rm log}} \biggl[\cos\left(k^2 D^2(z)\hat{\Sigma}_{\rm log}^2\right)\delta P_{\rm w}^{\rm log}(k) \notag\\
    &\qquad-\sin\left(k^2 D^2(z)\hat{\Sigma}^2_{\rm log}\right)
    \frac{{\rm d}\, \delta P_{\rm w}^{\rm log}(k)}{\omega_{\rm log}\,{\rm d}\, \ln\, k} \biggr]\Biggr\} \notag\\
    &\qquad+ D^4(z) P^{\rm 1-loop}\left[ P^{\rm IR\, res,\, LO}(k) \right] \,.
\end{align}
$P^{\rm 1-loop}$ is the standard one-loop result, but computed with the leading-order (LO) IR resummed power spectrum. 
We take the usual expression $P^{\rm 1-loop} = P_{22} + 2 P_{13}$ with
\begin{align} \label{eqn:P22}
    &P_{22}(k) = \int \frac{{\rm d}^3 q}{4\pi^3} F_2^2({\bf q},{\bf k}-{\bf q})P_{\rm lin}(q)P_{\rm lin}(|{\bf k}-{\bf q}|) \,, \\ \label{eqn:P13}
    &P_{13}(k) =  3P_{\rm lin}(k) \int \frac{{\rm d}^3 q}{4\pi^3} F_3({\bf q},-{\bf q},{\bf k})P_{\rm lin}(q) \,,
\end{align}
while evaluating the loop integrals $P_{22}$ and $P_{13}$ with the input spectrum $P^{\rm IR\, res,\, LO}$ instead 
of the linear spectrum \citep{Blas:2015qsi}. $F_n$ are the usual perturbation theory (PT) kernels 
\citep{Bernardeau:2001qr}. We calculate these quantities, i.e. Eqs.~\eqref{eqn:P22} and \eqref{eqn:P13}, with the 
publicly available code {\tt FAST-PT} \citep{McEwen:2016fjn,Fang:2016wcf}.\footnote{\url{https://github.com/JoeMcEwen/FAST-PT}} 
The IR resummed power spectrum at leading order is given by
\begin{align} \label{eqn:IRlin_LO_2}
    &P^{\rm IR\, res,\, LO}(k,z) = D^2(z) P_{\rm nw}(k) \notag\\
    &\qquad\times\left[1 + \mathrm{e}^{-k^2 D^2(z) \Sigma^2_{\rm BAO}} \delta P_{\rm w}^{\rm BAO}(k)
    +\mathrm{e}^{-k^2 D^2(z) \Sigma^2_{\rm lin}}\delta P_{\rm w}^{\rm lin}(k)\right]
\end{align}
for the linear oscillations \citep{Blas:2016sfa,Beutler:2019ojk} and by
\begin{align} \label{eqn:IRlog_LO_2}
    &P^{\rm IR\, res,\, LO}(k,z) = D^2(z) P_{\rm nw}(k) \notag\\
    &\qquad\times\left[1 + \mathrm{e}^{-k^2 D^2(z) \Sigma^2_{\rm BAO}} \delta P_{\rm w}^{\rm BAO}(k)\right. \notag\\
    &\qquad\left.+\mathrm{e}^{-k^2 D^2(z) \Sigma^2_{\rm log}}\cos\left(k^2 D^2(z)\hat{\Sigma}_{\rm log}^2\right)\delta P_{\rm w}^{\rm log}(k)\right. \notag\\
    &\qquad\left.-\mathrm{e}^{-k^2 D^2(z)\Sigma_{\rm log}^2}\sin\left(k^2 D^2(z)\hat{\Sigma}^2_{\rm log}\right)
    \frac{{\rm d}\, \delta P_{\rm w}^{\rm log}(k)}{\omega_{\rm log}\,{\rm d}\, \ln\, k} \right]
\end{align}
for the logarithmic oscillations \citep{Vasudevan:2019ewf,Beutler:2019ojk}. The result of the IR resummation at LO 
is given by a first contribution corresponding to the smooth part of the linear power spectrum and a second contribution 
corrected by the exponential damping of the oscillatory part due to the effect of IR enhanced loop contributions. 
The result for the logarithmic oscillations has additional contributions, on top of scale-dependent damping factors, 
due to the non-trivial oscillatory behaviour in real space. The damping factors above correspond to
\begin{align}
    &\Sigma_{\rm BAO}^2(k_{\rm S}) \equiv \int_0^{k_{\rm S}}\frac{{\rm d}q}{6\pi^2} P_{\rm nw}(q) \left[1-j_0\left(q r_{\rm s}\right)+2j_2\left(q r_{\rm s}\right)\right] \label{eqn:sigma_bao}\,,\\
    &\Sigma_{\rm lin}^2(\omega_{\rm lin},k_{\rm S}) \equiv \notag\\
    &\qquad \int_0^{k_{\rm S}}\frac{{\rm d}q}{6\pi^2} P_{\rm nw}(q) \left[1-j_0\left(q \frac{\omega_{\rm lin}}{k_*}\right)+2j_2\left(q \frac{\omega_{\rm lin}}{k_*}\right)\right] \label{eqn:sigma_lin}\,,\\
    &\Sigma^2_{\rm log}(k,\omega_{\rm log},k_{\rm S}) \equiv \int_0^{k_S} \frac{{\rm d}q}{4\pi^2} P_{\rm nw}(q) \notag\\
    &\qquad \times\int_{-1}^1 {\rm d}\mu \mu^2 \left\{ 1-\cos\left[\omega_{\rm log}\ln\left(1-\frac{q\mu}{k}\right)\right] \right\} \label{eqn:sigma_log}\,,\\
    &\hat{\Sigma}^2_{\rm log}(k,\omega_{\rm log},k_{\rm S}) \equiv \notag\\ 
    &-\int_0^{k_S}\frac{{\rm d}q}{4\pi^2} P_{\rm nw}(q) 
    \int_{-1}^1 {\rm d}\mu \mu^2 \sin\left[\omega_{\rm log}\ln\left(1-\frac{q\mu}{k}\right)\right] \label{eqn:sigma_log2}\,,
\end{align}
where $j_n$ are spherical Bessel functions, $r_{\rm s} \simeq 147$\,\si{Mpc} is the scale setting the period of the 
BAO \citep{Planck:2018vyg}, and $k_{\rm S}$ is the separation scale controlling the modes which are to be resummed. 
The dependence on $k_{\rm S}$ can be connected with an estimate of the theoretical perturbative uncertainties. For 
this reason and since IR expansions are valid for $q \ll k$, we assume 
$k_{\rm S} = \epsilon k$ with $\epsilon \in [0.3,0.7]$ \citep{Baldauf:2015xfa,Vasudevan:2019ewf}.

In redshift space, the damping factor becomes also dependent on the cosine $\mu$ since peculiar velocities additionally 
wash out wiggle signals along the line of sight. In this case, we have at leading order \citep{Eisenstein:2006nk} in 
Eqs.~\eqref{eqn:IRlin_LO_2} and \eqref{eqn:IRlog_LO_2} $D(z)\Sigma_{\rm X} \to D(z)\Sigma_{\rm X}(\mu)$ with
\begin{equation} \label{eq:gmu}
    \Sigma_{\rm X}^2(k,\mu,z) = \Sigma_{\rm X}^2(z)\left\{1 - \mu^2 + \mu^2\left[1+f(k,z)\right]^2\right\} \,.
\end{equation}

We note that while in the next subsection, Sect.~\ref{sec:PT_vs_sim}, we present both PT results at LO and NLO, that is Eqs.~\eqref{eqn:IRlin_LO_2}-\eqref{eqn:IRlog_LO_2}-\eqref{eqn:lin_NLO}-\eqref{eqn:log_NLO}, and we compare them 
to N-body simulations. Furthermore, we only used the LO to derive the Fisher-matrix uncertainties in Sect.~\ref{sec:results}.

For the nonlinear modelling for photometric probes, entering Eq.~\eqref{eq:ISTrecipe}, we use a modified version of {\tt HMCODE} 
\citep{Mead:2016zqy} where the $\Lambda$CDM nonlinear matter power spectrum is dressed with the primordial 
oscillations according to
\begin{equation}
    P_{\rm NL}(k,z) = \frac{P_{\rm NL}^{\rm \Lambda CDM}(k,z)}{P_{\rm lin}^{\rm \Lambda CDM}(k,z)} P^{\rm IR\, res,\, LO}(k,z) \,.
\end{equation}

\subsection{\label{sec:PT_vs_sim}Comparing predictions of perturbation theory with N-body simulations}
\begin{figure}
\centering
\includegraphics[width=0.4\textwidth]{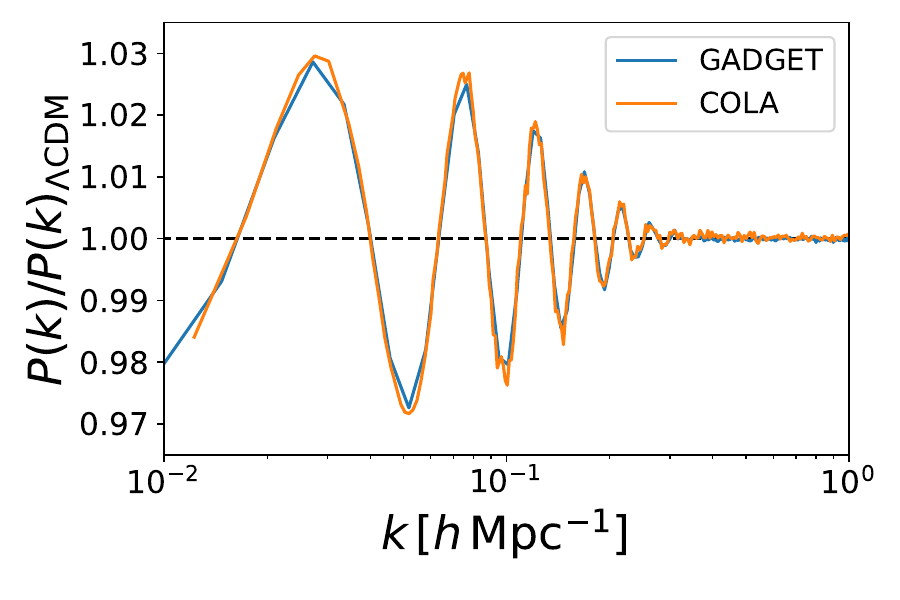}
\caption{Ratio with respect to the $\Lambda$CDM case of the nonlinear matter power spectra in Fourier space at 
redshift $z=0$ calculated with the approximate N-body method COLA and with the full N-body code {\tt GADGET-3} 
for the linear oscillations with ${\cal A}_{\rm lin} = 0.03$ and $\omega_{\rm lin} = 10$ \citep{Ballardini:2019tuc}.}
\label{fig:Comparison_Nbody}
\end{figure}

We produce cosmological simulations based on the COmoving Lagrangian Approximation (COLA) method 
\citep{Tassev:2013pn,Tassev:2015mia,Winther:2017jof,Wright:2017dkw} in order to assess the accuracy of the 
predictions of PT at LO and NLO for the template \eqref{eqn:pk_wigg} using a modified version of the publicly 
available code {\tt L-PICOLA} \citep{Howlett:2015hfa}.\footnote{\url{https://github.com/CullanHowlett/l-picola}} 
We fix the standard cosmological parameters and the amplitude and phase of the feature, and we study the effect 
of varying the frequency of the primordial oscillations over the set of values 
$\omega_{\rm X} \in \{0.2,0.4,0.6,0.8,1,1.2,1.4,1.6,1.8,2\}$. 
Following the analysis done by \citet{Ballardini:2022vzh}, we run each simulation with $1024^3$ dark matter 
particles in a comoving box with side length of $1024$\,\si{\hMpc} evolved with 30 time steps. We set at redshift 
$z = 9$ the initial conditions that we generate using second-order Lagrangian perturbation theory, with the 
{\tt 2LPTic} code~\citep{Crocce:2006ve}. Finally, we use spectra averaged over pairs of simulations with the same 
initial seeds and inverted initial conditions, and with amplitude fixing in order to minimise the cosmic variance 
\citep{Viel:2010bn,Villaescusa-Navarro:2018bpd}. 
In Fig.~\ref{fig:Comparison_Nbody}, we show a comparison between the nonlinear matter power spectrum at $z = 0$ 
simulated with COLA and the one obtained by \citet{Ballardini:2019tuc} at the same resolution with the N-body code 
{\tt GADGET-3}, a modified version of the publicly available code {\tt GADGET-2} 
\citep{Springel:2000qu,Springel:2005mi},\footnote{\url{https://wwwmpa.mpa-garching.mpg.de/gadget}} for 
${\cal A}_{\rm lin} = 0.03$ and $\omega_{\rm lin} = 10$. 

\begin{figure}
\centering
\includegraphics[width=0.4\textwidth]{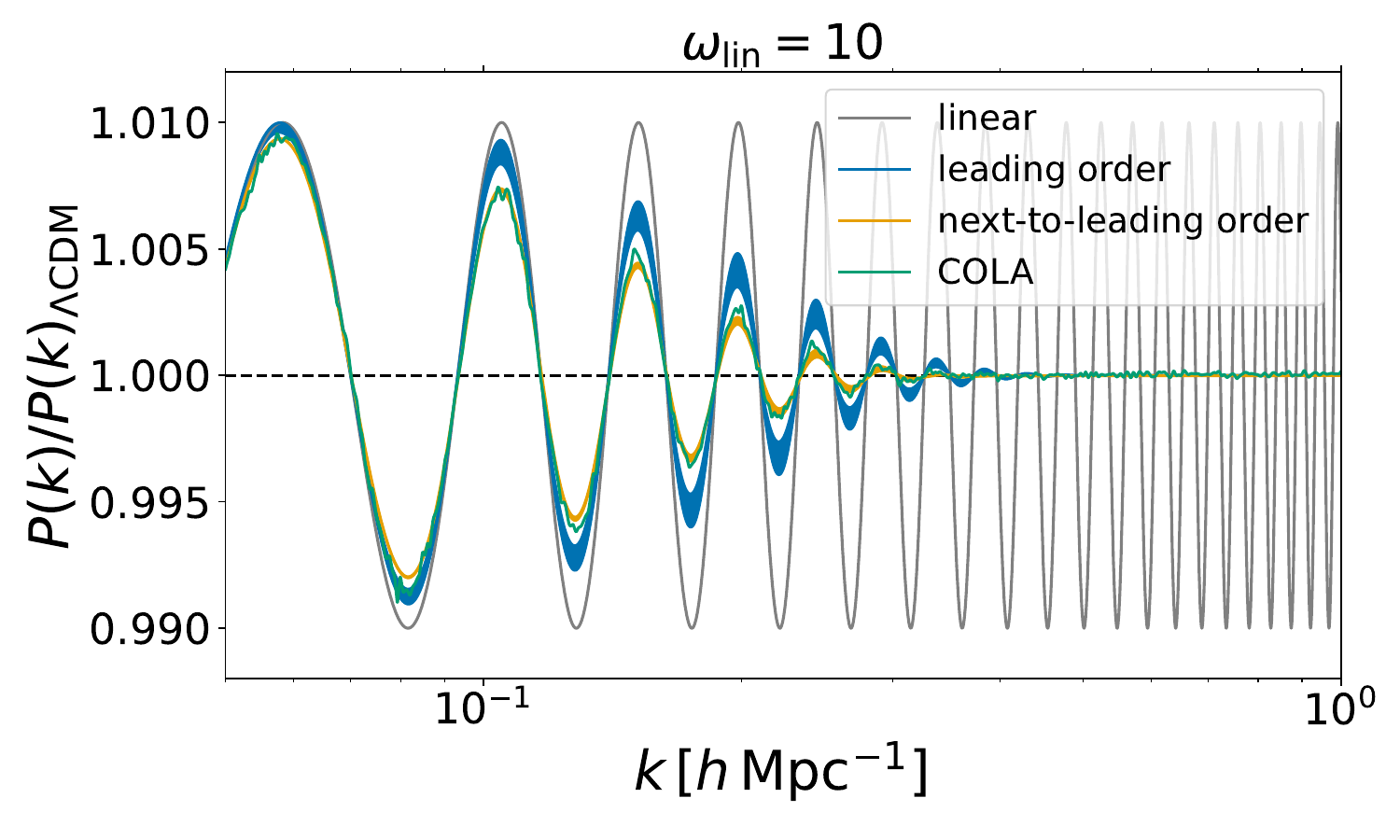}
\includegraphics[width=0.4\textwidth]{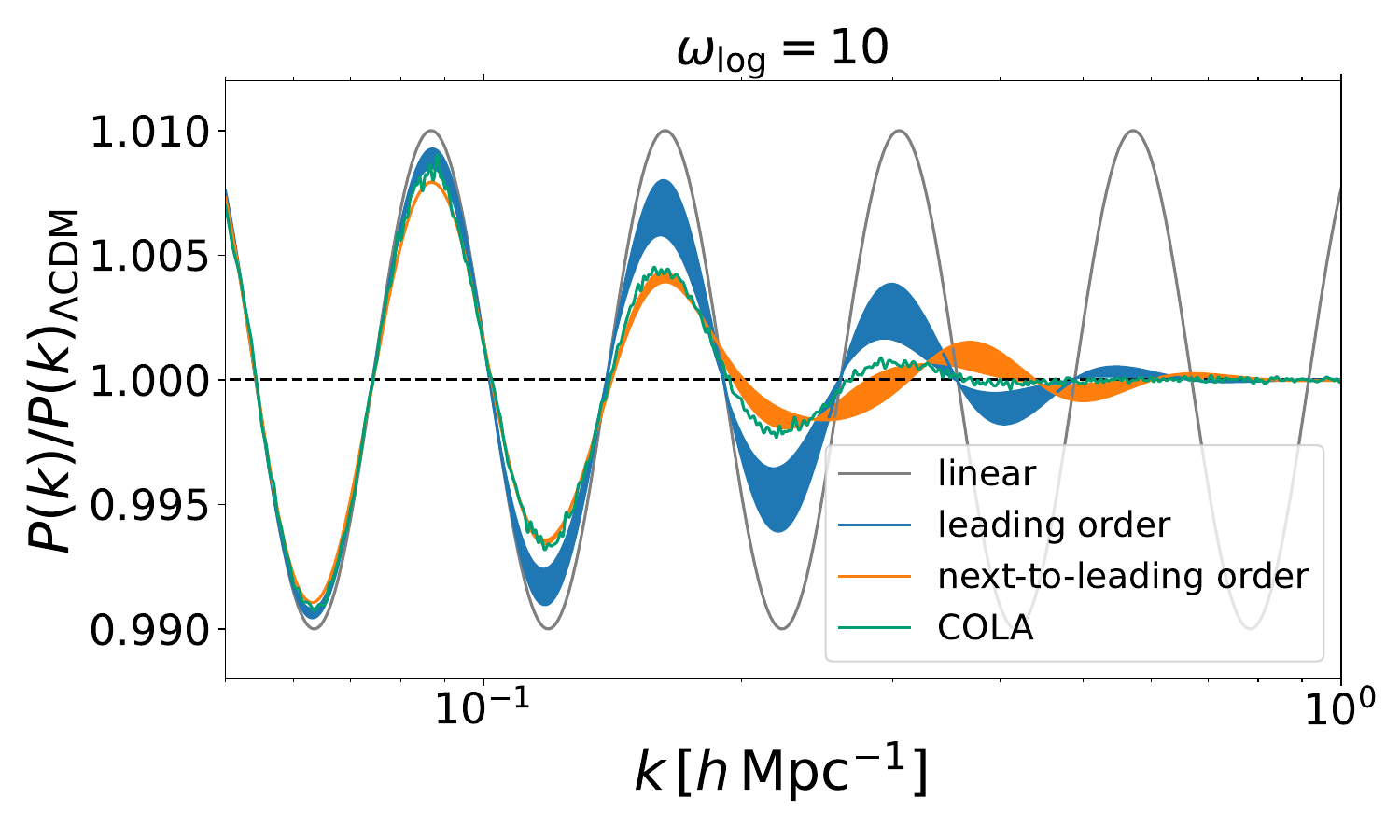}
\caption{Ratio of IR resummed matter power spectrum at LO (blue) and NLO (orange) obtained for linear (top panel) 
and logarithmic (bottom panel) oscillations to the one obtained with a power-law PPS at redshift $z = 0$ when the 
frequency $\omega_{\rm X} = 10$ and the IR separation scale $k_{\rm S} = \epsilon k$ is varied (with $\epsilon \in [0.3,0.7]$). 
We also show the results of the linear theory (grey) and the ones obtained from N-body simulations (green).}
\label{fig:Comparison}
\end{figure}

We show in Fig.~\ref{fig:Comparison} the ratio between the matter power spectrum with primordial oscillations 
and the one with featureless PPS calculated at redshift $z = 0$ for the results both at LO and NLO for the two 
templates with linear and logarithmic oscillations. We collect in Appendix~\ref{sec:appendix1} the results for 
different frequencies; see Figs.~\ref{fig:Comparison_lin} and \ref{fig:Comparison_log}.

For the template with undamped linear oscillations, the agreement between N-body simulations and PT results 
is remarkable for the entire range of frequencies considered, i.e. $\log_{10}\omega_{\rm lin} \in (0.2,\,2)$.
At LO, we find differences for the matter power spectrum in Fourier space at redshift $z = 0$ less than $1\%$ 
for $\log_{10}\omega_{\rm lin} > 0.6$ and between $1\%$ and $2\%$ for $\log_{10}\omega_{\rm lin} \leq 0.6$.
At NLO, we find an agreement better than $1\%$ for all the frequencies and differences $< 0.5\%$ for 
$\log_{10}\omega_{\rm lin} > 0.4$. 
For the template with undamped logarithmic oscillations, we find a similar agreement for the frequencies 
$\log_{10}\omega_{\rm log} \geq 1$, less than $1\%$ for the LO and less than $0.5\%$ for the NLO. 
In order to improve the accuracy of the results for low frequencies \cite[see][for a discussion on the validity 
of perturbation approach for small frequencies]{Beutler:2019ojk} we can fit a Gaussian damping to the N-body 
simulation as done by \citet{Ballardini:2019tuc}.

While we restrict the validation of results to the matter power spectrum in real space, a validation for the 
halo power spectrum in redshift space has been performed by \citet{Chen:2020ckc}, showing that current results 
from PT are able to describe redshift-space clustering observables also in the presence of primordial features.

\subsection{\label{sec:reconstruction}Nonlinear reconstruction of primordial features}
An alternative to the modelling approach described above is through reconstruction. This takes the opposite logic, 
by starting from an evolved, nonlinear, matter field where the primordial oscillations have been damped, it tries 
to recover their undamped states. Reconstruction has long been adopted as a useful technique to undo the effect of 
structure evolution and retrieve sharper BAO features, hence improving their use as a standard ruler to constrain 
cosmological parameters \citep{Eisenstein:2006nk,Wang:2017apjl,Sarpa:2018ucb}, with an even longer history 
of applications in other subfields of cosmology and astrophysics \citep{Peebles:1989apj,Croft:1997mn,Monaco:1999mn,Nusser:1999hm,Brenier:2003mn,Mohayaee:2006aap,Schmittfull:2017prd,Zhu:2017prd,Shi:2017gqs,Birkin:2018nag,Mao:2020vdp}.

From the point of view of reconstruction, there is no fundamental difference between BAO features and features that 
exist in the primordial density fluctuations on similar length scales: both are present at the time of last scattering, 
the preservation of both is negatively impacted by late-time cosmic structure formation, and both can be at least partially 
recovered if such an impact is undone through processes such as reconstruction \citep{Beutler:2019ojk,Li:2021jvz}. Therefore, we expect that reconstruction can find uses in the study of primordial features, at least when the oscillations 
have similar frequencies as the BAO peaks, $0.05\lesssim{k}/(h\,{\rm Mpc}^{-1})\lesssim0.5$ --- at scales much larger 
than this range any significant damping of the features due to cosmological evolution is yet to happen, because structure 
formation is hierarchical and affects smaller scales first, while on much smaller scales reconstruction becomes unreliable. 

Another useful property of reconstruction is that it is possible to include galaxy bias \citep{Birkin:2018nag} 
and RSD \citep{Monaco:1999mn,Burden:2014cwa,Hada:2018fde,Hada:2018ziy,Wang:2019zuq} in the pipeline, so that 
two of the most important theoretical systematic effects in analysing the galaxy redshift power spectrum are taken into 
account. Of course, caution is due here to ensure that the range of length scales where reconstruction from a catalogue 
of redshift-space biased tracers can be done reliably covers the range where we hope reconstruction to also benefit the 
recovery of the primordial features of interest. Checking this naturally requires a more detailed analysis, which may 
also depend on the reconstruction algorithm in actual use: while a quantitative analysis of this issue is not the topic 
here, our experience is that the scale range $k\lesssim0.2\,h\,{\rm Mpc}^{-1}$ or $s\gtrsim15$--$20\,\si{\hMpc}$ satisfies these requirements. 

In this work, we exemplify using the nonlinear reconstruction method described in \citet{Shi:2017gqs,Birkin:2018nag,Wang:2019zuq}. Other algorithms may give quantitatively different results, though we 
do not expect a huge variation among the ones that are at an advanced stage of development. The problem of reconstruction 
can be reduced to identifying a mapping between the initial, Lagrangian coordinate of some particle, $\boldsymbol{q}$, 
and its Eulerian coordinate after evolution, $\boldsymbol{x}(t)$, at a later time $t$. If trajectory crossings of particles 
have not happened, this mapping is unique and can be obtained by solving the mass conservation equation
\begin{equation}\label{eq:mass_conservation}
    \rho(\boldsymbol{x})\,{\rm d}^{3}\boldsymbol{x} = \rho(\boldsymbol{q})\,{\rm d}^{3}\boldsymbol{q} \approx \bar{\rho}\,{\rm d}^{3}\boldsymbol{q} \,,
\end{equation}
where $\rho(\boldsymbol{q})$ and $\rho(\boldsymbol{x})$ are, respectively, the initial density field in some infinitesimal 
volume element ${\rm d}\boldsymbol{q}$, and the density field in the same volume element at time $t$, which now has been 
moved, and possibly deformed, and described by ${\rm d}\boldsymbol{x}$. As the density field in the early Universe is 
homogeneous to a very good approximation, one can take $\rho({\bf q})$ to be a constant, i.e. 
$\rho(\boldsymbol{q})\simeq\bar{\rho}$. 

We define the displacement field $\boldsymbol{\Psi}$ between the final and initial positions of a particle as 
$\boldsymbol{\Psi}(\boldsymbol{x})=\boldsymbol{x}-\boldsymbol{q}$, which can be rewritten as
\begin{equation}\label{eq:Theta_defination}
    \boldsymbol{\nabla}_{\boldsymbol{x}}{\Theta}(\boldsymbol{x}) \equiv \boldsymbol{q} = \boldsymbol{x} - \boldsymbol{\Psi}(\boldsymbol{x}) \,,
\end{equation}
with $\Theta({\bf x})$ being the displacement potential. Here we have assumed the displacement field to be curl-free, 
i.e. $\boldsymbol{\nabla}\times\boldsymbol{\Psi}=\boldsymbol{0}$, which is valid only on large scales.\footnote{This 
is similar to the assumption of no particle trajectory crossing, which must break down on small scales. Together, 
these assumptions mean that the reconstruction method described here, similar to other reconstruction methods, should only 
be expected to work for relatively large scales, and its performance progressively degrades if one goes to smaller scales.} 
Combining Eq.~\eqref{eq:Theta_defination} and Eq.~\eqref{eq:mass_conservation} gives 
\begin{equation}\label{eq:jacobian}
    \det[\nabla^{i}\nabla_{j}\Theta({\bf x})] = \frac{\rho({\bf x})}{\bar{\rho}} \equiv 1 + \delta({\bf x}) \,,
\end{equation}
where $\delta(\boldsymbol{x})=\rho(\boldsymbol{x})/\bar{\rho}-1$ is the density contrast at time $t$, `${\det}$' is 
the determinant of a matrix, here the Hessian of $\Theta({\bf x})$, and $i,j=1,2,3$. \citet{Shi:2017gqs} proposed to 
solve this equation as a nonlinear partial differential equation with cubic power of second-order derivatives of $\Theta$. 
This was later extended by \citet{Birkin:2018nag} to cases where $\delta(\boldsymbol{x})$ in the above can be the density 
contrast of some biased tracers of the dark matter field, such as galaxies and dark matter haloes. 

Once $\Theta(\boldsymbol{x})$ is solved, one can obtain $\boldsymbol{\Psi}(\boldsymbol{x})$ and the reconstructed density 
field is given by
\begin{equation} \label{eq:delta_r}
    \delta_{\rm r} = -\nabla_{\boldsymbol{q}}\cdot\boldsymbol{\Psi}(\boldsymbol{q}) \,,
\end{equation}
where we express ${\bf\Psi}$ in terms of the Lagrangian coordinate $\boldsymbol{q}$, since the divergence 
$\nabla_{\boldsymbol{q}}$ therein is with respect to $\boldsymbol{q}$. This calculation means that we need to have 
$\boldsymbol{\Psi}(\boldsymbol{q})$ on a regular $\boldsymbol{q}$-grid, which can be done using, for example the Delaunay 
Tessellation Field Estimator code \citep[DTFE;][]{Schaap:2000aap,vandeWeygaert:2007ze,Cautun:2011dtfe}, by 
interpolating $\boldsymbol{\Psi}(\boldsymbol{x})$ to the target $\boldsymbol{q}$-grid. 

The quantity $\delta_{\rm r}$ obtained above is an approximation to the initial linear matter density contrast, 
linearly extrapolated to time $t$ (bearing in mind that all the discussion in this subsection is for $\Lambda$CDM, 
for which the linear growth factor is scale-independent). As a result, part of the structure-formation-induced damping 
of the BAO or primordial features imprinted in the density field can be undone this way.

The damping of the BAO or primordial wiggles can be described by a Gaussian fitting function \citep[cf.~Sect.~\ref{sec:PT_vs_sim};][]{Vasudevan:2019ewf,Beutler:2019ojk,Ballardini:2019tuc}, effectively 
multiplying the undamped power spectrum wiggles $O_{\rm w}^{\rm undamped}$ by a Gaussian function
\begin{equation} \label{eq:Gaussian_damping}
    {O}_{\rm w}^{\rm damped}(k,z) =  {O}_{\rm w}^{\rm undamped}(k,z) \exp \left[-\frac{k^{2}\zeta(z)^{2}}{2} \right] \,,
\end{equation}
where $\zeta(z)$ is the parameter quantifying the level of damping. In the linearly evolved density field with no 
damping, one has $\zeta=0$. For the evolved nonlinear matter or galaxy field, $\zeta>0$ with its values depending 
on redshift, tracer type, tracer number density etc., and the reconstructed density field generally features a smaller 
yet positive value of $\zeta$. 

We generate a suite of N-body simulations using the initial conditions described in Sect.~\ref{sec:PT_vs_sim}, which 
includes a no-wiggle model and 20 wiggled models. The simulations are performed using the parallel N-body code {\tt ramses} 
\citep{teyssier2002}. We output 4 snapshots of the DM respectively at $z = 1.5$, $1.0$, $0.5$ and $0$, and measure the 
matter power spectrum of each of the snapshots. We follow the same reconstruction pipeline employed in \citet{Li:2021jvz} 
to reconstruct the density fields from each of the snapshots for all the models (shown in Fig.~\ref{fig:reconstruction}). 
We have also performed reconstruction from halo catalogues, though the results are noisier and not shown here; the DM 
reconstruction result can be considered as an ideal scenario, which serves as a limit of what can be achieved from halo 
or galaxy reconstruction with increasing tracer number density. We present the reconstruction from mock galaxy 
catalogues in redshift space based on the same set of simulations in a separate work. Fig.~\ref{fig:reconstruction} shows 
that reconstruction can help recover the weakened wiggles down to $k\simeq1\,h\,{\rm Mpc}^{-1}$. Note that, in order to 
avoid artificial features in the $P(k)$ of the models with high-frequency wiggles, we have binned $P(k)$ in $k$ bins of 
width $0.001\,h\,{\rm Mpc}^{-1}$. 

To estimate the best-fit values of $\zeta(z)$, the Gaussian function is applied to fit the envelopes of the unreconstructed 
and reconstructed wiggled spectra, $O_{\rm w}^{\rm damped}(k,z)$. We apply the Hilbert transform to measure the wiggle 
envelopes for each wiggled model and each redshift. Nevertheless, we expect that the damping parameter $\zeta(z)$, for 
both the unreconstructed and the reconstructed cases, should mostly depend on the nonlinear structure formation and be 
insensitive to the wiggle model, given that the wiggles are weak. We have explicitly checked this by comparing the envelopes 
of $O^{\rm damped}_{\rm w}$ and $O^{\rm undamped}_{\rm w}$ of the different models, and finding them to agree very well. 
As a result, to get $\zeta(z)$, the $O_{\rm w}$ envelopes of all wiggle models at redshift $z$ could be combined to do a 
single least-squares fitting. In practice, because the measurement of the envelopes is not very reliable for the 
low-frequency featured models due to the very few wiggles in the $k$ range of fitting, $k = (0.05$--$1)\, h\, \rm Mpc^{-1}$, 
we select only the 10 high-frequency feature models to fit $\zeta(z)$, and the result is given in 
Table~\ref{tab:damping parameter}. We nevertheless have checked that the fitted envelope also well describes the wiggles 
of low-frequency models.

\begin{table}
\centering
\caption{Best-fit values of $\zeta(z)$ for unreconstructed (reconstructed) wiggle spectra.}
\label{tab:damping parameter}
\begin{tabular}{c|c|c|c}
\hline
$z$ & ${\rm unrec}$ & ${\rm rec}$ & ${\rm efficiency}$\\
\hline
$0.0$ & $8.16$ & $2.04$ & $4.00$ \\
$0.5$ & $6.49$ & $1.69$ & $3.84$ \\
$1.0$ & $5.23$ & $1.41$ & $3.71$ \\
$1.5$ & $4.32$ & $1.21$ & $3.57$ \\
\hline
\end{tabular}
\tablefoot{Columns respectively denote 
(1) redshift, (2) unreconstructed wiggles, (3) reconstructed wiggles, and (4) reconstruction efficiency (defined as 
$\zeta_{\rm unrec} / \zeta_{\rm rec}$).}
\end{table}

\section{\label{sec:results}Expected constraints from \Euclid primary probes}
In this section, we show the results of the Fisher analysis for the cosmological parameters of interest corresponding 
to a flat $\Lambda$CDM,
\begin{equation}
    \Theta_{\rm final} = \left\{\Omega_{\rm m,0}, \Omega_{\rm b,0}, h, n_{\rm s}, \sigma_8, {\cal A}_X, \omega_X, \phi_X\right\} \,,
\end{equation}
after marginalisation over spectroscopic and photometric nuisance parameters. We compute the spectroscopic galaxy 
clustering (\GCsp), the photometric galaxy clustering (\GCph), the weak lensing (WL), and the cross-correlation 
between the two photometric probes (XC) in two configurations: a pessimistic setting and an optimistic one. For 
the pessimistic setting, for \GCsp\ we have $k_{\rm max} = 0.25\,h\,{\rm Mpc}^{-1}$, for \GCph\ and XC we have 
$\ell_{\rm max} = 750$, and for WL we have $\ell_{\rm max} = 1500$. In addition, we impose a cut in redshift of 
$z<0.9$ to \GCph\ to limit any possible cross-correlation between spectroscopic and photometric data. For the 
optimistic setting, we extend \GCsp\ to $k_{\rm max} = 0.30\,h\,{\rm Mpc}^{-1}$, \GCph\ and XC to 
$\ell_{\rm max} = 3000$, and WL to $\ell_{\rm max} = 5000$, without imposing any redshift cut to \GCph.

\begin{table*}[htbp]
\caption{Fisher-forecast 68.3\% CL marginalised uncertainties on cosmological and primordial feature parameters, 
relative to their corresponding fiducial values.}
\begin{tabularx}{\textwidth}{Xcccccccc}
\hline
\hline
\rowcolor{jonquil} \multicolumn{9} {c}{{{\mbox{\textbf{LIN}}\,\,\, \boldsymbol{$\omega_{\rm lin}=10$}}}} \\
& \multicolumn{1}{c}{$\Omega_{\rm m,0}$} & \multicolumn{1}{c}{$\Omega_{\rm b,0}$} & \multicolumn{1}{c}{$h$} & \multicolumn{1}{c}{$n_{\rm s}$} & \multicolumn{1}{c}{$\sigma_{8}$} & \multicolumn{1}{c}{${\cal A}_{\rm lin}$} & \multicolumn{1}{c}{$\omega_{\rm lin}$} & \multicolumn{1}{c}{$\phi_{\rm lin}$}\\
\hline
\rowcolor{lavender(web)}\multicolumn{9}{l}{{{Pessimistic setting}}} \\ 
\GCsp\,$(k_{\rm max} = 0.25\,h\,{\rm Mpc}^{-1})$                &   1.2\%   &   1.7\%  &   1.4\%   &   1.5\%   &   0.83\%  &   23\%  &   4.1\%  &   0.11   \\
WL+\GCph+XC                                                     &   0.83\%  &   5.7\%  &   4.0\%   &   1.6\%   &   0.39\%  &   32\%  &   5.7\%  &   0.12  \\
\GCsp+WL+\GCph+XC($z<0.9$)                                      &   0.61\%  &   1.3\%  &   0.59\%  &   0.70\%  &   0.28\%  &   21\%  &   2.9\%  &   0.083   \\
\hline
\rowcolor{lavender(web)}\multicolumn{9}{l}{{{Optimistic setting}}}  \\ 
\GCsp\, $(k_{\rm max} = 0.3\,h\,{\rm Mpc}^{-1})$    &   1.1\%   &   1.6\%  &   1.1\%   &   1.3\%   &   0.76\%  &   23\%  &   3.6\%  &   0.10   \\
WL+\GCph+XC                                         &   0.27\%  &   4.4\%  &   2.5\%   &   0.66\%  &   0.13\%  &   31\%  &   2.8\%  &   0.072   \\
\GCsp+WL+\GCph+XC                                   &   0.23\%  &   1.1\%  &   0.45\%  &   0.28\%  &   0.11\%  &   18\%  &   1.2\%  &   0.047   \\
\hline
\hline
\rowcolor{jonquil} \multicolumn{9} {c} {{{\mbox{\textbf{LOG}}\,\,\, \boldsymbol{$\omega_{\rm log}=10$}}}} \\ 
& \multicolumn{1}{c}{$\Omega_{\rm m,0}$} & \multicolumn{1}{c}{$\Omega_{\rm b,0}$} & \multicolumn{1}{c}{$h$} & \multicolumn{1}{c}{$n_{\rm s}$} & \multicolumn{1}{c}{$\sigma_{8}$} & \multicolumn{1}{c}{${\cal A}_{\rm log}$} & \multicolumn{1}{c}{$\omega_{\rm log}$} & \multicolumn{1}{c}{$\phi_{\rm log}$}\\
\hline
\rowcolor{lavender(web)}\multicolumn{9}{l}{{{Pessimistic setting}}} \\
\GCsp\,$(k_{\rm max} = 0.25\,h\,{\rm Mpc}^{-1})$                &   1.1\%   &   1.8\%  &   1.4\%   &   1.5\%   &   0.82\%  &   25\%  &   4.7\%  &   0.053   \\
WL+\GCph+XC                                                     &   0.83\%  &   5.7\%  &   4.0\%   &   1.6\%   &   0.38\%  &   32\%  &   5.7\%  &   0.12  \\
\GCsp+WL+\GCph+XC($z<0.9$)                                      &   0.61\%  &   1.4\%  &   0.59\%  &   0.71\%  &   0.28\%  &   22\%  &   2.4\%  &   0.044   \\
\hline
\rowcolor{lavender(web)}\multicolumn{9}{l}{{{Optimistic setting}}}  \\ 
\GCsp\, $(k_{\rm max} = 0.3\,h\,{\rm Mpc}^{-1})$    &   1.1\%   &   1.8\%  &   1.1\%   &   1.3\%   &   0.76\%  &   25\%  &   4.7\%  &  0.052   \\
WL+\GCph+XC                                         &   0.27\%  &   4.4\%  &   2.5\%   &   0.66\%  &   0.13\%  &   31\%  &   2.8\%  &  0.072  \\
\GCsp+WL+\GCph+XC                                   &   0.23\%  &   1.1\%  &   0.46\%  &   0.28\%  &   0.11\%  &   18\%  &   1.1\%  &  0.035   \\
\hline
\hline
\end{tabularx}
\label{tab:results}
\tablefoot{We show results for LIN and LOG models in the pessimistic and optimistic settings, using 
\Euclid observations \GCsp, WL+\GCph+XC, and their combination.}
\end{table*}

\begin{figure*}
\centering
\includegraphics[width=0.49\textwidth]{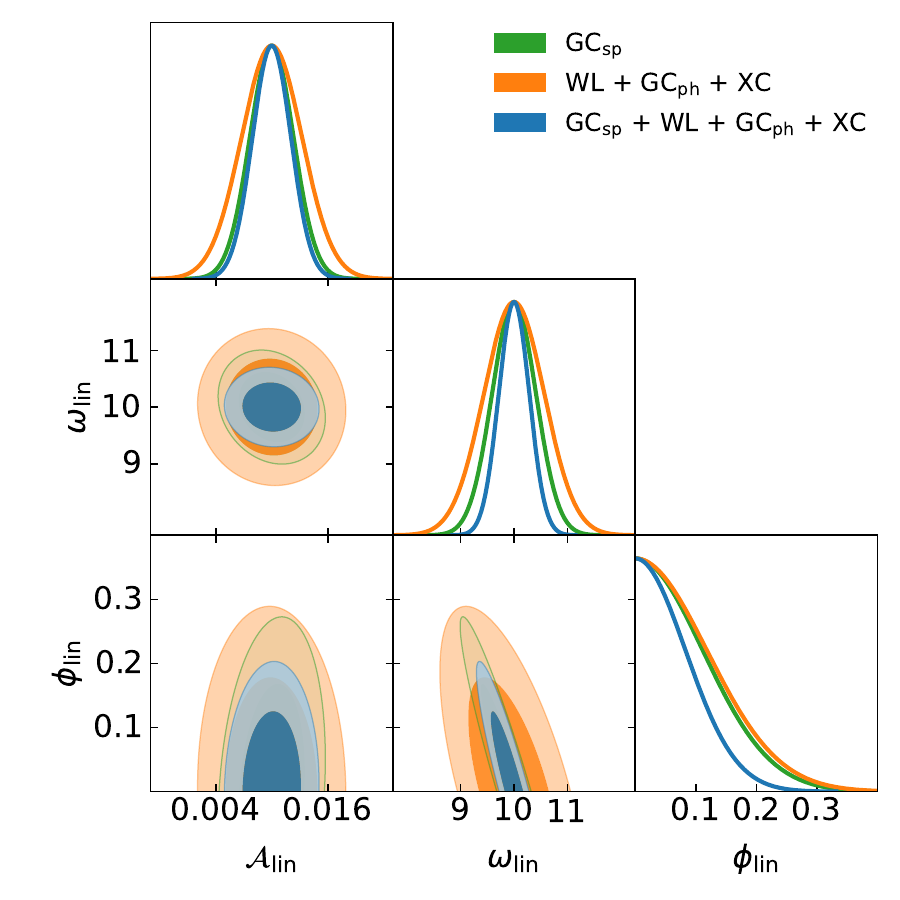}
\includegraphics[width=0.49\textwidth]{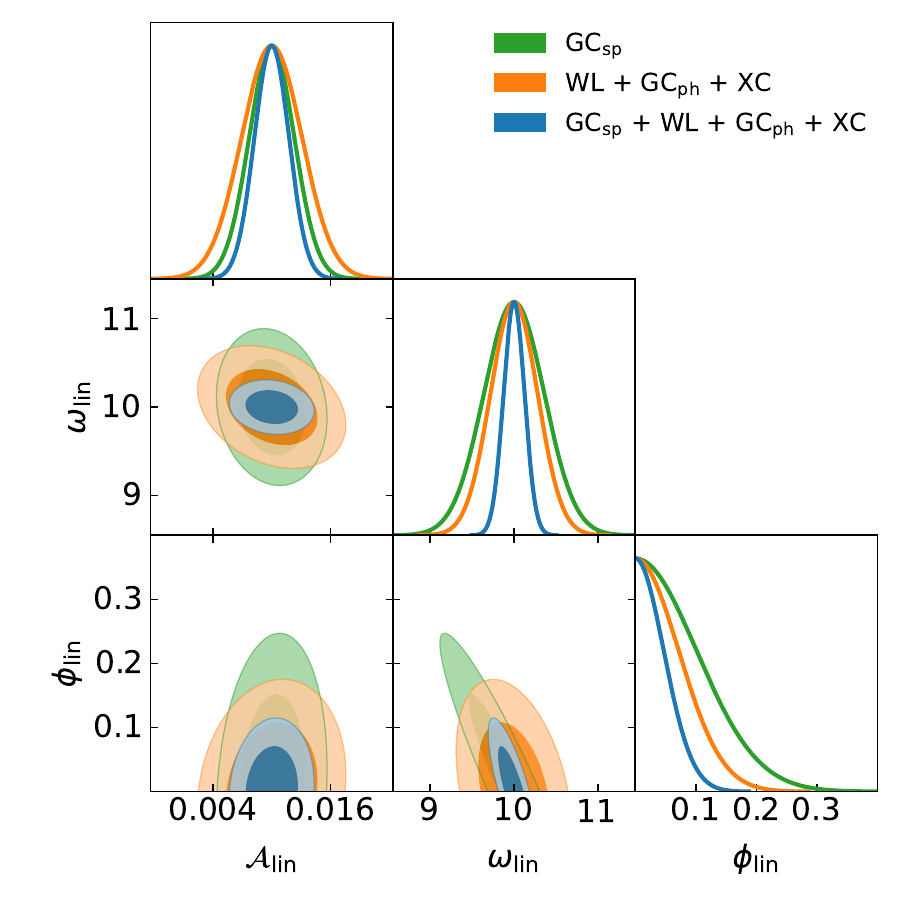}
\includegraphics[width=0.49\textwidth]{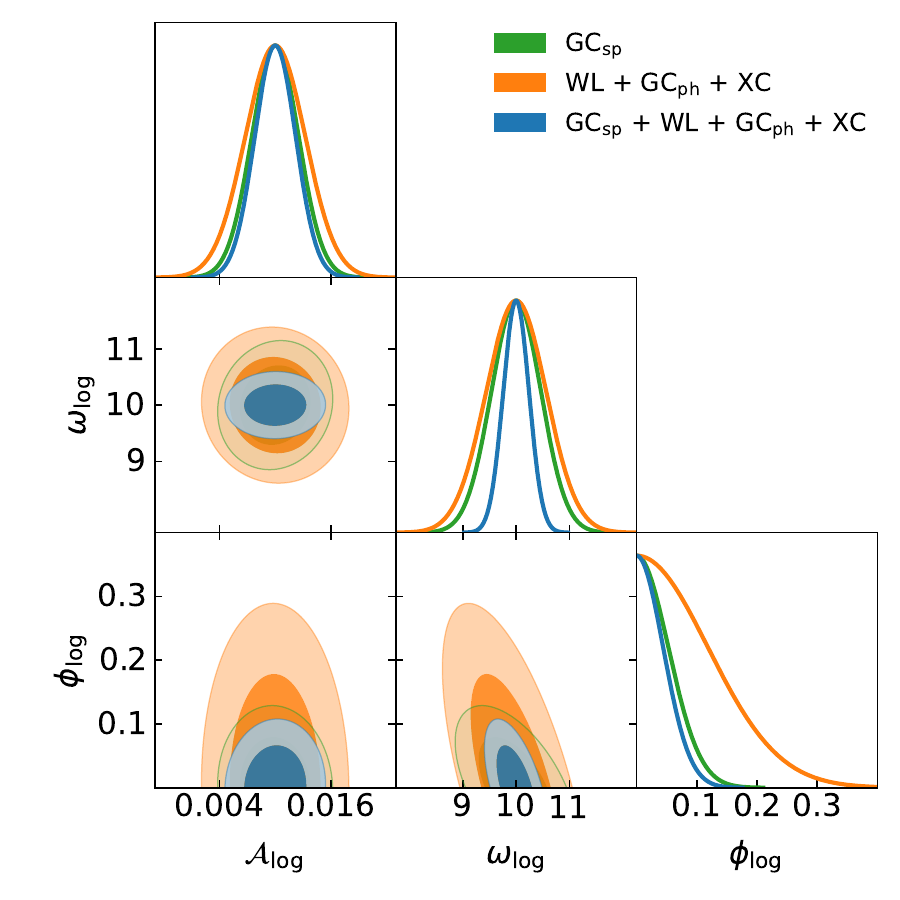}
\includegraphics[width=0.49\textwidth]{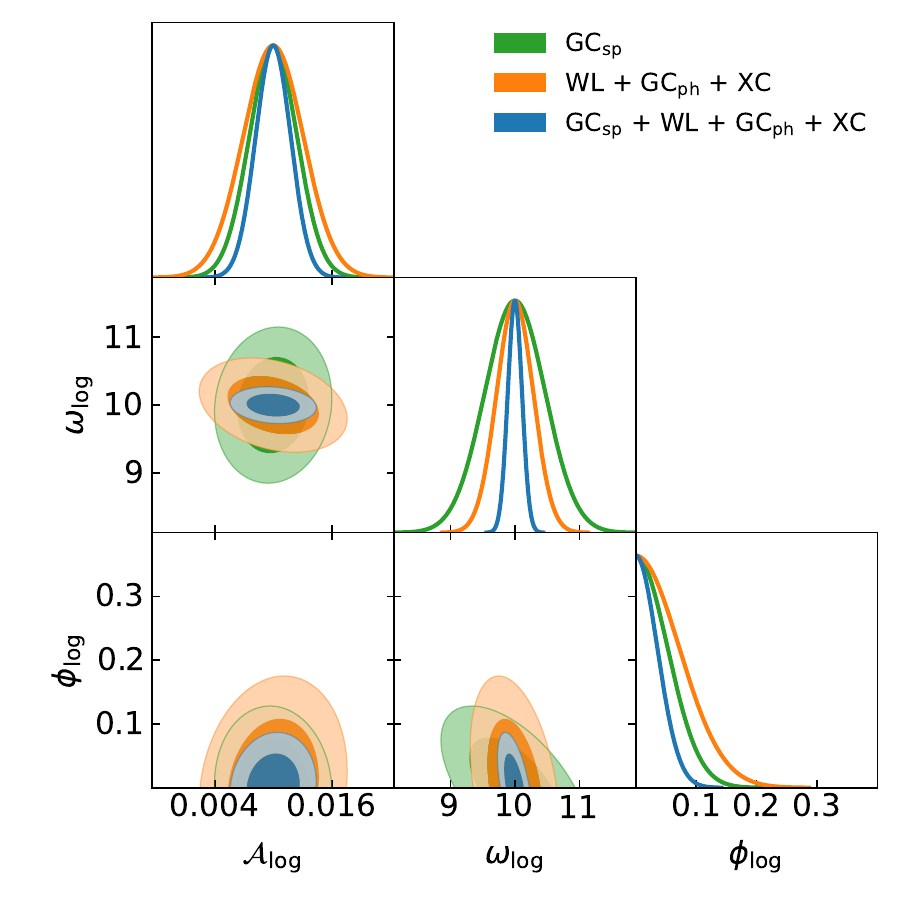}
\caption{Fisher-forecast marginalised two-dimensional contours and one-dimensional probability distribution functions 
from \Euclid on the primordial feature parameters for the LIN model with $\omega_{\rm lin} = 10$ (top panels) and the 
LOG model with $\omega_{\rm log} = 10$ (bottom panels). Left (right) panels correspond to the pessimistic (optimistic) 
setting for \GCsp\ (green), WL+\GCph+XC (orange), and their combination (blue).}
\label{fig:triangle}
\end{figure*}

We start by presenting results for a fiducial scenario with ${\cal A}_{\rm X} = 0.01$, $\omega_{\rm X} = 10$, and 
$\phi_{\rm X} = 0$. The marginalised uncertainties on the cosmological parameters and primordial feature parameters, 
percentages relative to the corresponding fiducial values, are collected in Table~\ref{tab:results} for the 
pessimistic (top panel) and optimistic (bottom panel) settings. The marginalised 68.3\% and 95.5\% confidence level (CL) 
contours for the primordial feature parameters are shown in Fig.~\ref{fig:triangle}.
For the fiducial value parameters, uncertainties on the amplitude of the primordial feature oscillations are dominated 
by \GCsp\ measurements resulting in ${\cal A}_{\rm lin} = 0.0100 \pm 0.0023$ ($\pm 0.0023$) and 
${\cal A}_{\rm log} = 0.0100 \pm 0.0025$ ($\pm 0.0025$) at a 68.3\% CL for the pessimistic (optimistic) setting. 
Combining \GCsp\ with the combination of photometric information, i.e. WL+\GCph+XC, we find 
${\cal A}_{\rm lin} = 0.0100 \pm 0.0021$ ($\pm 0.0018$) and ${\cal A}_{\rm log} = 0.0100 \pm 0.0022$ ($\pm 0.0018$) at a 
68.3\% CL for the pessimistic (optimistic) setting. 

For \GCsp, bounds on the primordial amplitude parameter ${\cal A}_X$ strongly depend on the primordial frequency 
value $\omega_X$; see \citet{Huang:2012mr,Ballardini:2016hpi,Slosar:2019gvt,Beutler:2019ojk,Ballardini:2019tuc}.
Low frequencies, lower than the BAO one, leave a broad modification on the matter power spectrum and are expected 
to be better constrained by CMB measurements. On the other hand, the imprint from high-frequency primordial oscillations 
is smoothed by projection effects on the observed matter power spectrum. Moreover, for the linear model the uncertainties 
for frequencies around the BAO frequency, i.e. $\log_{10} \omega_{\rm lin} \sim 0.87$, are degraded.
For the combination of photometric probes, bounds on the primordial amplitude parameter ${\cal A}_X$ 
are smoothed by projection effects in angular space and high-frequency primordial oscillations are severely 
washed out on the observed matter power spectrum. On the other hand, uncertainties for frequencies around the BAO 
scale and lower frequencies are tighter compared to the results from \GCsp. This represents an important result 
considering the lower accuracy of the PT predictions for low frequencies.

\begin{figure}
\centering
\includegraphics[width=0.49\textwidth]{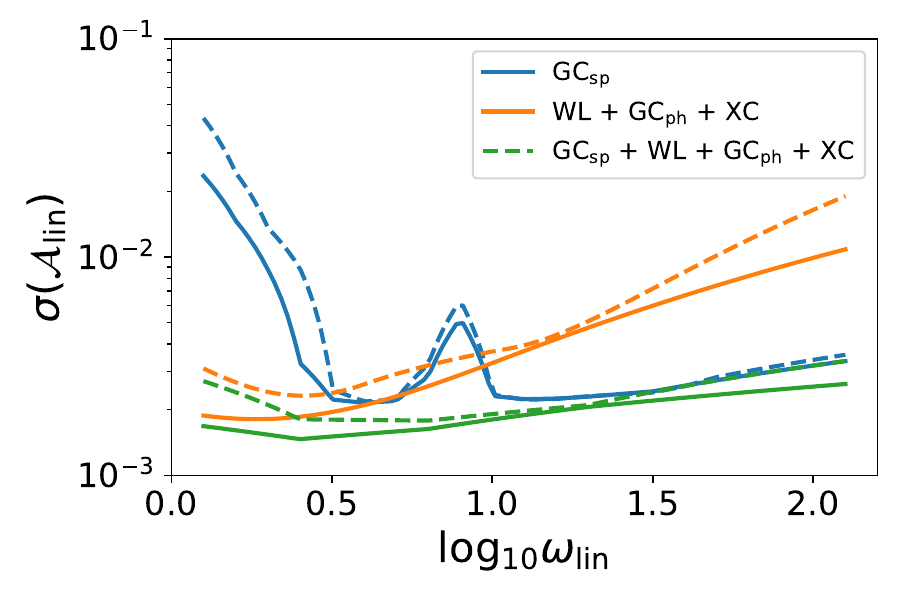}
\includegraphics[width=0.49\textwidth]{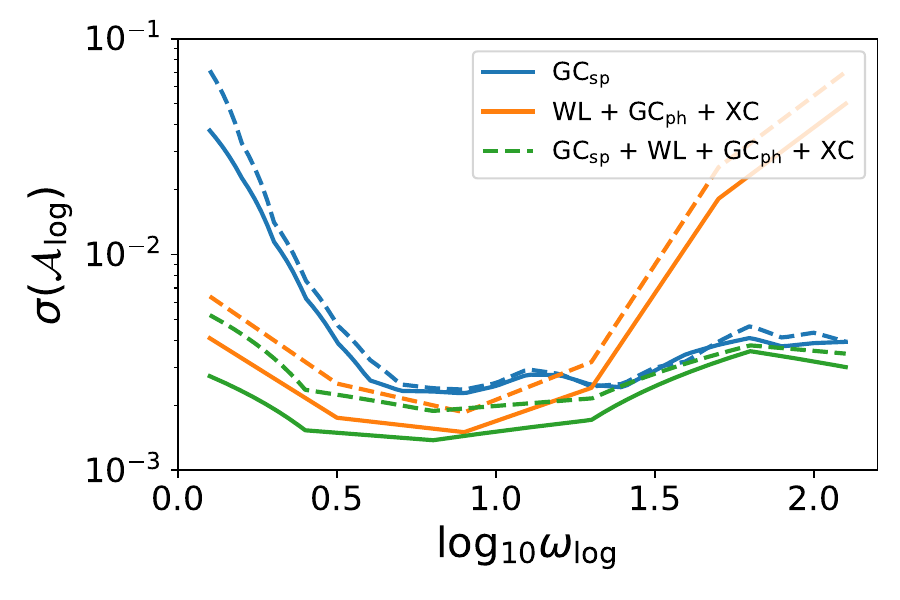}
\caption{Marginalised uncertainties on ${\cal A}_X$ as a function of the primordial frequency $\omega_X$ for the 
LIN model (top panel) and the LOG model (bottom panel). We show uncertainties for \GCsp\ (blue), WL+\GCph+XC (orange), 
and their combination (green). Dashed (solid) curves correspond to the pessimistic (optimistic) setting.}
\label{fig:error}
\end{figure}
In Fig.~\ref{fig:error}, we show the uncertainties at a 68.3\% CL for different values of the primordial frequency 
$\log_{10} \omega_{\rm X}$ within the range $(0.1,\,2.1)$. The combination of the spectroscopic and photometric probes 
allows us to reach uncertainties of 0.002--0.003 on the primordial feature amplitude for the entire range of frequencies.

The nonlinear reconstruction described in Sect.~\ref{sec:reconstruction} applied to \GCsp\ reduces significantly 
the uncertainties on the feature amplitude, going from $\sigma({\cal A}_{\rm lin}) = 0.0023$ ($0.0023$) at a 68.3\% CL 
for the pessimistic (optimistic) setting to $\sigma({\cal A}_{\rm lin}) = 0.0014$ ($0.0013$), and going from 
$\sigma({\cal A}_{\rm log}) = 0.0025$ ($0.0025$) to $\sigma({\cal A}_{\rm log}) = 0.0016$ ($0.0012$). Combining with 
the photometric probes, we find $\sigma({\cal A}_{\rm lin}) = 0.0014$ ($0.0012$) at a 68.3\% CL and 
$\sigma({\cal A}_{\rm log}) = 0.0015$ ($0.0012$) for the pessimistic (optimistic) setting, compared with 
$\sigma({\cal A}_{\rm lin}) = 0.0021$ ($0.0018$) at a 68.3\% CL and $\sigma({\cal A}_{\rm log}) = 0.0022$ ($0.0018$) 
for the same settings without reconstruction. We also note that in the analysis here $k_{\rm max}$ is set to be 
$0.25$--$0.30~h~{\rm Mpc}^{-1}$, while in \citet{Li:2021jvz} $k_{\rm max}=0.50~h~{\rm Mpc}^{-1}$ is used.

\section{\label{sec:bispectrum}Power spectrum and bispectrum combination}
In this section, we present the forecast results on primordial features by combining the power spectrum and 
bispectrum signals. The presence of a sharp feature in the inflaton potential violates the slow-roll evolution, 
generating a linearly spaced oscillating primordial bispectrum. In this case the primordial bispectrum can be 
characterised by the standard amplitude parameter $f_{\rm NL}^{\rm lin}$, the phase $\phi_{\rm lin}^B$, and the 
frequency $\omega_{\rm lin}$, and it is given by \citep{Chen:2006xjb,Chen:2008wn}
\begin{equation} \label{eq:Bklinosc}
    B_{\Phi}^{\rm lin}(k_1,k_2,k_3) = f_{\rm NL}^{\rm lin} \frac{6 A^2}{k_1^2k_2^2k_3^2} \sin\left[\omega_{\rm lin}\frac{K}{k_*} + 2\pi\phi_{\rm lin}^B \right] \,,
\end{equation}
where $K = k_1+k_2+k_3$ and $A = 9/25 (2\pi^2) k_*^{1-n_{\rm s}}A_{\rm s}$ is the normalisation parameter of 
$P_\Phi=A/k^{(4-n_{\rm s})}$.

A periodically modulated inflaton potential leads to resonances in the inflationary fluctuations with 
logarithmically-spaced oscillations \citep{Chen:2008wn}. This generates oscillatory features in the primordial power 
spectrum (see Sect.~\ref{sec:modelling}) and bispectrum \citep{Flauger:2009ab,Hannestad:2009yx,Barnaby:2011qe}. 
In this case, the primordial bispectrum is given by \citep{Chen:2010bka}
\begin{equation} \label{eq:Bklogar}
    B_\Phi^{\rm log}(k_1,k_2,k_3) = f_{\rm NL}^{\rm log}\frac{6A^2}{k_1^2k_2^2k_3^2}\sin\left[\omega_{\rm log}\ln\left(\frac{K}{k_*}\right)
    +2\pi\phi_{\rm log}^B\right] \,,
\end{equation}
where $\phi_{\rm log}^B$ and $f_{\rm NL}^{\rm log}$ are the phase and the amplitude, respectively, of the generated 
logarithmic features on the primordial bispectrum. Note that here we assume the frequency of the primordial feature 
bispectrum to be the same as in the case of the power spectrum (see Sect.~\ref{sec:features})

The redshift space model, for the galaxy bispectrum, is given after considering non-Gaussian initial conditions, as 
dictated by the oscillatory features, up to second order terms in RSD, bias, and matter expansions. The tree-level 
modelling can be written as  
  \begin{align}
   &P_g(\bk,z)=D^P(\bk,z)Z_1^2(\bk,z)P^{\rm lin}(k,z)+P_\veps(z) \,, \label{eq:Pgs} \\
   \nonumber \\ 
   &B_g(\bk_1,\bk_2,\bk_3,z)=D^B(\bk_1,\bk_2,\bk_3,z) \nonumber \\ 
   &\times\bigg[Z_1(\bk_1,z)Z_1(\bk_2,z)Z_1(\bk_3,z)B_{I}(\bk_1,\bk_2,\bk_3,z) \nonumber \\
   &+2Z_1(\bk_1,z)Z_1(\bk_2,z)Z_2(\bk_1,\bk_2,z)P_{\rm lin}(k_1,z)P_{\rm lin}(k_2,z)+2\text{ perm}\bigg] \nonumber \\
   &+2P_{\veps\veps_{\delta}}\big[Z_1(\bk_1,z)P_{\rm lin}(k_1,z)+2\text{ perm}\big]+B_{\veps} \,, \label{eq:Bgs}
  \end{align}
where the linearly propagated primordial oscillatory bispectrum is
\begin{equation}
    B_I(\bk_1,\bk_2,\bk_3,z)=M(k_1,z)M(k_2,z)M(k_3,z)B_\Phi^X(k_1,k_2,k_3,z)
\end{equation}
with $B_\Phi^X$ given by Eqs.~\eqref{eq:Bklinosc} and \eqref{eq:Bklogar} for the linear and logarithmic oscillations, 
respectively. The redshift space kernels are given by
\begin{align}
   &Z_1(\bk_i,z)=b_1(z)+f\mu_i^2\,, \label{eq:Z1}\\
   &Z_2(\bk_i,\bk_j,z)=b_1(z)F_2(\bk_i,\bk_j)+f(z)\mu_{ij}^2G_2(\bk_i,\bk_j)+\frac{b_2(z)}{2} \nonumber \\
   & +\frac{b_{s^2}(z)}{2}S_2(\bk_i,\bk_j)+\frac{f(z)\mu_{ij}k_{ij}}{2}\left[\frac{\mu_i}{k_i}Z_1(\bk_j,z)+\frac{\mu_j}{k_j}Z_1(\bk_i,z)\right] \label{eq:Z2} \,,
\end{align}  
where $\mu_i=\bk_i\cdot\hat{z}/k_i$ is the cosine of the angle between the wave vector $\bk_i$ and the line of sight, 
$\mu=\mu_1+\ldots\mu_n$, and $k=|\bk_1+\ldots\bk_n|$. Specifically $\mu_{12} = (\mu_1k_1+\mu_2k_2)/k_{12}$ and 
$k_{12}^2 = (\bk_1+\bk_2)^2$. 
The kernels $F_2(\bk_i,\bk_j)$ and $G_2(\bk_i,\bk_j)$ are the second-order symmetric PT kernels \citep{Bernardeau:2001qr}, 
while $S_2(\bk_1,\bk_2) = (\bk_1\cdot\bk_2)^2/(k_1^2 k_2^2)-1/3$ is the  tidal kernel \citep{McDonald:2009dh,Baldauf:2012hs}. 
The second-order tidal field bias term, following the convention of \citep{Baldauf:2012hs}, is given by 
$b_{s^2}(z)=-4[b_1(z)-1]/7$. The redshift-space bispectrum is characterised by five variables: three to define the 
triangle shape (the sides $k_1$, $k_2$, $k_3$) and  two that characterise the orientation of the  triangle 
relative to the line of sight, i.e. $B_{\rm g}(\bk_1,\bk_2,\bk_3,z)=B_{\rm g}(k_1,k_2,k_3,\mu_1,\phi,z)$. The forecasts 
presented here use the information of the bispectrum monopole obtained after taking the average over all angles.

Note that the presence of an oscillatory primordial bispectrum generates a scale-dependent correction to the linear 
bias, in a similar manner to the local primordial non-Gaussian case \cite[see][for discussions]{Karagiannis:2018jdt}. 
\citet{Cabass:2018roz} show that the scale-dependent correction oscillates with scale with an envelope similar to 
that of equilateral primordial non-Gaussianity (PNG) 
\cite[see][for a discussion]{Schmidt2010,Scoccimarro2011,Schmidt2012,Assassi2015}, while their findings indicate 
that the amplitude of such a term is very small to be detected by upcoming surveys. Therefore, we do not consider the 
scale-dependent terms in our analysis and exclude them from the expressions of the redshift kernels presented above. 
This means that the degeneracy between the scale-dependent bias coefficient and the primordial bispectrum amplitude 
($f_{\rm NL}^{\rm X}$), shown in \citet{Barreira:2020ekm}, does not affect our forecasts.

The multiplicative factors $D^P(\bk,z)$ and $D^B(\bk_1,\bk_2,\bk_3,z)$, which incorporate errors on redshift measurements 
and FOG effect, are $D^P(\bk,z)=\exp[-k\mu(\sigma_{\rm p}^2(z)+\sigma_{\rm r}^2(z))]$ and 
$D^B(\bk_1,\bk_2,\bk_3,z) = \exp[-(k_1^2\mu_1^2+k_2^2\mu_2^2+k_3^2\mu_3^2)(\sigma_{\rm p}^2(z)+\sigma_{\rm r}^2(z))/2]$ 
for the power spectrum and bispectrum, respectively (see Sect.~\ref{sec:GCsp} for details). The fiducial values of the 
stochastic terms in Eqs.~\eqref{eq:Pgs} and \eqref{eq:Bgs} are taken to be those predicted by Poisson statistics 
\citep{Schmidt:2015gwz,Desjacques:2016bnm}, i.e. $P_{\varepsilon} = 1/\bar{n}_{\rm g}$, 
$P_{\varepsilon\varepsilon_{\delta}}=b_1/(2\bar{n}_{\rm g})$, and $B_{\varepsilon} = 1/\bar{n}_{\rm g}^2$.
  
The AP effect is taken into account also for the bispectrum similar to the power spectrum case (see Sect.~\ref{sec:GCsp}). 
The observed bispectrum is thus given by \cite[see][]{Karagiannis:2022ylq} 
   \begin{align} \label{eq:BS_AP}
       B^{\rm obs}_{\rm g}(k_{{\rm ref,}1}&,k_{{\rm ref,}2},k_{{\rm ref,}3},\mu_{{\rm ref,}1},\phi;z) \nonumber \\
       &=\frac{1}{q_\parallel^2q_\perp^4}B_{\rm g}\left(k(k_{{\rm ref,}1}),k(k_{{\rm ref,}2}),k(k_{{\rm ref,}3}),\mu(\mu_{{\rm ref,}1}),\phi;z\right) \,. 
   \end{align}
 
Here we again use the Fisher matrix formalism to produce forecasts on the model parameters that control the primordial 
oscillations. The Fisher matrix of the redshift galaxy power spectrum is given by Eq.~\eqref{eq:fisher-gc}, while for 
the bispectrum the Fisher matrix for one redshift bin $z_i$ is given by 
\begin{align}\label{eq:fisherBs}
    F_{\alpha\beta}^B(z_i)&=\frac{1}{4\pi}\!\sum_{k_1,k_2,k_3}\!\int_{-1}^1\!\! {\rm d}\mu_1 \!\! \int_0^{2\pi}\!\! {\rm d} \phi \frac{1}{\Delta B^2(\bk_1,\bk_2,\bk_3,z_i)} \nonumber \\ 
    &\times\frac{\partial B_{\rm g}^{\rm obs}(\bk_1,\bk_2,\bk_3,z_i)}{\partial p_{\alpha}}\frac{\partial B_{\rm g}^{\rm obs}(\bk_1,\bk_2,\bk_3,z_i)}{\partial p_{\beta}} \,,
\end{align}
where the sum over triangles has $k_{\rm min}\le k_3\le k_2\le k_1 \le k_{\rm max}$, and $k_1$, $k_2$ and $k_3$ 
satisfy the triangle inequality. The bin size $\Delta k$ is taken to be the fundamental frequency of the survey, 
$k_{\rm f}=2\pi/L$, where for simplicity we approximate the survey volume as a cube, $L=V_{\rm survey}^{1/3}$. 
The $k$ mode is binned with a bin size of $\Delta k$, and here we consider $\Delta k=k_{\rm f}$, between its 
minimum and maximum values $k_{\rm min}=k_{\rm f}$ and $k_{\rm max}$. Gaussian approximation is used for the covariance 
of bispectrum, i.e. the off-diagonal terms are considered to be zero.\footnote{Neglecting the off-diagonal terms 
in the bispectrum covariance can have a significant impact on the constraints on primordial features, since these 
contributions are important even on large scales for squeezed triangles \citep{Gualdi:2020ymf,Biagetti:2021tua,Floss:2022wkq}. 
Considering the analytic expression of the full bispectrum covariance is left for future work.} The variance for 
the bispectrum is then \citep{Sefusatti:2006eu}
\begin{equation} \label{eq:deltaB2}
    \Delta B^2(\bk_1,\bk_2,\bk_3,z)=s_{123}\,\pi\, k_{\rm f}(z)^3\,\frac{P_{\rm g}(\bk_1,z)\,P_{\rm g}(\bk_2,z)\,P_{\rm g}(\bk_3,z)}{k_1k_2k_3\,\Delta k(z)^3} \,,
\end{equation}
where $s_{123}=6,2,1$ for equilateral, isosceles, and non-isosceles triangles, respectively. In addition, for 
degenerate configurations, i.e. $k_i=k_j+k_m$, the bispectrum variance should be multiplied by a factor of 2 
\citep{Chan:2016ehg,Desjacques:2016bnm}. We use the Fisher matrix of Eq.~\eqref{eq:fisherBs} to generate bispectrum 
forecasts on the parameters of interest
\begin{equation}
    \Theta_{\rm final}^B = \left\{\Omega_{\rm m,0}, \Omega_{\rm b,0}, h, n_{\rm s}, \sigma_8, A_X, \omega_X, \phi_X, f_{\rm NL}^X, \phi_X^B\right\} \,.
\end{equation}
The initial parameter vector considers all the parameters of the bispectrum tree-level modelling (i.e. cosmological 
parameters, biases, AP parameters, growth rate, and FOG amplitudes) to be free, and after marginalisation we derive 
the constraints for $\Theta_{\rm final}^B$. The forecasted marginalised $1\sigma$ errors on the feature model parameters, 
coming from the spectroscopic galaxy clustering, are presented in Table~\ref{table:results_PS_BS}. 

The galaxy bispectrum is capable of providing constraints not only on the feature parameters related to the primordial 
bispectrum but also on those controlling the primordial power spectrum feature model. 
The latter is due to the $P_{\rm lin}(k_i)P_{\rm lin}(k_j)$ terms that are present at tree level in the galaxy bispectrum 
\eqref{eq:Bgs} and can provide constraints equivalent to or even better than the galaxy power spectrum, especially in the 
case of a high $k_{\rm max}$. This can be seen in Table~\ref{table:results_PS_BS}, where adding the power spectrum to the 
bispectrum improves the forecasts by a few percent. This highlights the importance of a bispectrum analysis for the type 
of feature models discussed in this work. On the other hand, the sole contribution to the constraints on the 
$f_{\rm NL}^{\rm X}$ and $\phi^B_X$ feature parameters is the primordial component $B_I$ of the galaxy bispectrum, 
justifying the less rigorous forecasts presented here.    

\begin{table}[t]
\caption{Fisher-forecast 68.3\% CL marginalised uncertainties on primordial feature parameters, 
relative to their corresponding fiducial values, using \Euclid \GCsp\ bispectrum.}
\centerline{
\begin{tabular} { l c c c }
\noalign{\vskip 3pt}\hline\noalign{\vskip 1.5pt}\hline\noalign{\vskip 6pt}

  & {\bf P} & {\bf B} & {\bf P+B} \\ 
\noalign{\vskip 3pt}\hline\noalign{\vskip 6pt}
\multicolumn{4}{c}{Pessimistic setting ($k_{\rm max}=0.25\,h\,{\rm Mpc}^{-1}$)}  \\
\noalign{\vskip 3pt}\hline\noalign{\vskip 6pt}
$A_{\rm lin}$ & 23\% & 16\% & 13\%  \\  
$\omega_{\rm lin}$ & 4.1\% & 2.1\% & 1.7\%  \\  
$\phi_{\rm lin}$ & 0.11 & - & 0.049  \\  
$f_{\rm NL}^{\rm lin}$ & - & 31 & 31  \\  
$\phi_{\rm lin}^B$ & - & 4.9 & 4.9  \\ 
\noalign{\vskip 3pt}\hline\noalign{\vskip 6pt}
$A_{\rm log}$ & 25\% & 17\% & 13\%  \\  
$\omega_{\rm log}$ & 4.7\% &  2.9\% & 2.4\%  \\  
$\phi_{\rm log}$ &  0.053 & - & 0.026  \\  
$f_{\rm NL}^{\rm log}$ & - & 33 & 32  \\  
$\phi_{\rm log}^B$ & - & 5.2 & 5.1  \\ 
\noalign{\vskip 3pt}
\hline
\hline\noalign{\vskip 6pt}
\multicolumn{4}{l}{Optimistic setting ($k_{\rm max}=0.3\,h\,{\rm Mpc}^{-1}$)}  \\
\noalign{\vskip 3pt}\hline\noalign{\vskip 6pt}
$A_{\rm lin}$ & 23\% & 12\% & 11\%  \\  
$\omega_{\rm lin}$ & 3.6\% & 1.4\% & 1.2\%  \\  
$\phi_{\rm lin}^P$ & 0.10 & - & 0.039  \\  
$f_{\rm NL}^{\rm lin}$ & - & 29 & 29  \\  
$\phi_{\rm lin}^B$ & - & 4.6 & 4.6 \\ 
\noalign{\vskip 3pt}\hline\noalign{\vskip 6pt}
$A_{\rm log}$ & 25\% & 13\% & 11\%  \\  
$\omega_{\rm log}$ & 4.7\% & 2.2\% & 2.0\%  \\  
$\phi_{\rm log}$ & 0.052 & - & 0.023  \\  
$f_{\rm NL}^{\rm log}$ & - & 30 & 30  \\  
$\phi_{\rm log}^B$ & - & 4.8 & 4.8  \\ 
\noalign{\vskip 3pt}\hline\noalign{\vskip 1.5pt}\hline\noalign{\vskip 5pt}
\end{tabular}}
    \label{table:results_PS_BS}
\tablefoot{We show results for LIN and LOG models in the pessimistic and optimistic settings in the case, 
using the \Euclid \GCsp bispectrum and the joint power spectrum+bispectrum signal.}
\end{table}

\section{\label{sec:CMB}Combination with CMB data}
Future CMB polarisation data from ground-based experiments, such as Simons Observatory \citep{SimonsObservatory:2018koc} 
and CMB-S4 \citep{Abazajian:2019eic}, and satellites, such as LiteBIRD \citep{LiteBIRD:2022cnt}, will be able to reduce 
the uncertainties on the PPS, in particular in the presence of oscillatory signals. Indeed, E-mode polarisation is sourced 
by velocity gradient (scattering only) leading to sharper transfer functions compared to the temperature ones helping to 
characterise primordial features in the PPS; see 
\citet{Mortonson:2009qv,Miranda:2014fwa,Ballardini:2016hpi,Finelli:2016cyd,Hazra:2017joc,2023JCAP...03..014B}.

We consider the information available in the CMB data by using power spectra from temperature, E-mode polarisation, and 
CMB lensing in combination with \Euclid observations. We do not consider B-mode polarisation since it does not add 
information for the specific models examined here. We also neglect the cross-spectra between CMB fields and \Euclid 
primary probes; see \citet{Euclid:2021qvm} for a complete characterisation and quantification of the importance of 
cross-correlation between CMB and \Euclid measurements.
Following \citet{Euclid:2021qvm}, we consider three CMB experiments: {\em Planck}-like, Simons Observatory (SO), and 
CMB-S4. Noise curves correspond to isotropic noise deconvolved with instrumental beam \citep{Knox:1995dq}
\begin{equation}
    {\cal N}_\ell^X = w_X^{-1}b_\ell^{-2}\,,\qquad 
    b_\ell = \exp\left[-\ell(\ell+1)\theta_{\rm FWHM}^2/(16\ln 2)\right]\,,
\end{equation}
where $\theta_{\rm FWHM}$ is the full-width-at-half-maximum (FWHM) of the beam, $w_T$ and $w_E$ are the inverse square 
of the detector noise levels for temperature and polarisation, respectively. CMB lensing noise is reconstructed through 
minimum-variance estimator for the lensing potential using {\tt quicklens} 
code;\footnote{\url{https://github.com/dhanson/quicklens}} see \citet{Okamoto:2003zw}.

For the {\em Planck}-like experiment, we use an effective sensitivity in order to reproduce the {\em Planck} 2018 
results for the $\Lambda$CDM model \citep{Planck:2018vyg} avoiding the complexity of the real-data likelihood 
\citep{Planck:2019nip}. 
We use noise specifications corresponding to in-flight performances of the $143\,{\rm GHz}$ channel of the high-frequency 
instrument (HFI) \citep{Planck:2018nkj} with a sky fraction $f_{\rm sky} = 0.7$ and a multipole range for temperature and 
polarisation from $\ell_{\rm min}=2$ to $\ell_{\rm max}=1500$. The E-mode polarisation noise is inflated by a factor of 8 
to reproduce the uncertainty of the optical depth parameter $\tau$; see \citet{Bermejo-Climent:2021jxf}. Finally, CMB 
lensing is obtained by combining the $143\,{\rm GHz}$ and $217\,{\rm GHz}$ HFI channels assuming a conservative multipole 
range of 8--400.
For the CMB, we also consider the optical depth at reionisation $\tau$ and then we marginalise over it before combining 
the CMB Fisher matrix with the \Euclid ones.

For the SO-like experiment, we use noise curves provided by the SO collaboration in \citet{SimonsObservatory:2018koc} 
taking into account residuals and noises from component separation.\footnote{We use version 3.1.0 available at \\ \url{https://github.com/simonsobs/so\_noise\_models}.} 
We use spectra from $\ell_{\rm min}=40$ to $\ell_{\rm max}=3000$ for temperature and temperature-polarisation 
cross-correlation, and $\ell_{\rm max}=5000$ for E-mode polarisation. CMB lensing and temperature-lensing 
cross-correlation spectra cover the multipole range of 2--3000. The sky fraction considered is $f_{\rm sky} = 0.4$. 
We complement the SO-like information with the {\em Planck}-like large-scale data in the multipole range of 2--40 
in both temperature and polarisation. 

For CMB-S4, we use noise sensitivities of $1\,\mu{\rm K}\,{\rm arcmin}$ in temperature and 
$\sqrt{2}\,\mu{\rm K}\,{\rm arcmin}$ in polarisation, with resolution of $\theta_{\rm FWHM} = 1\,{\rm arcmin}$. 
We assume data over the same multipole ranges of SO and with the same sky coverage. 

\begin{table}[h]
\caption{Fisher-matrix-based forecasted 68.3\% CL marginalised uncertainties using \Euclid 
in combination with the CMB.}
\begin{tabular} {lccc}
\hline
\hline
 & \Euclid             & \Euclid    & \Euclid \\ 
 & + {\em Planck}-like & + SO-like  & + CMB-S4 \\ 
 &                     & + {\em Planck} low-$\ell$ & + {\em Planck} low-$\ell$ \\
\hline
\multicolumn{4}{l}{Pessimistic setting} \\
\hline
$A_{\rm lin}$      &  18\% & 14\% & 12\%  \\  
$\omega_{\rm lin}$ &  2.4\% & 1.2\% & 1.0\% \\  
$\phi_{\rm lin}$   &  0.066 & 0.046 & 0.043 \\  
\hline
$A_{\rm log}$      &  17\% & 11\% & 7.9\%  \\  
$\omega_{\rm log}$ &  1.9\% & 1.6\% & 1.5\% \\  
$\phi_{\rm log}$   &  0.048 & 0.036 & 0.033 \\ 
\hline
\hline
\multicolumn{4}{l}{Optimistic setting} \\
\hline
$A_{\rm lin}$      &  16\% & 13\% & 12\% \\  
$\omega_{\rm lin}$ &  1.0\% & 0.92\% & 0.85\%  \\  
$\phi_{\rm lin}$   & 0.042 & 0.038 & 0.037  \\  
\hline
$A_{\rm log}$      &  15\% & 10\% & 7.6\% \\  
$\omega_{\rm log}$ &  1.1\% & 1.0\% & 0.98\% \\  
$\phi_{\rm log}$   &  0.037 & 0.027 & 0.024 \\ 
\hline
\hline
\end{tabular}
\label{tab:results_CMB}
\tablefoot{We show results for LIN and LOG models in the pessimistic and optimistic settings in the case, 
relative to their corresponding fiducial values (${\cal A}_{\rm X} = 0.01$, $\omega_{\rm X} = 10$, using 
\Euclid (\GCsp+WL+\GCph+XC) in combination with the CMB.}
\end{table}
In Table~\ref{tab:results_CMB}, we show the marginalised uncertainties on the primordial feature parameters, 
percentage relative to the corresponding fiducial values, for the optimistic and pessimistic \Euclid+ CMB combination. 
Uncertainties on the primordial feature amplitude improve by 15\%--50\% depending on the choice of \Euclid settings 
and on the CMB experiment considered, for a frequency value of $\omega_{\rm X} = 10$.

\section{\label{sec:conclusions}Conclusions}
We have explored the constraining power of the future \Euclid space mission for oscillatory primordial features in 
the primordial power spectrum (PPS) of density perturbations. We have considered two templates with undamped oscillations, 
i.e. with constant amplitude all over the PPS, one linearly and one logarithmically spaced in $k$ space. We have used 
a common baseline set of parameters for both the templates with amplitude ${\cal A}_{\rm X} =0.01$, frequency 
$\omega_{\rm X} = 10$, and phase $\phi_{\rm X} = 0$, where X = \{lin, log\}.

Following previous studies for $\Lambda$CDM and simple extensions in \citet{Euclid:2019clj}, we have calculated 
Fisher-matrix-based uncertainties from \Euclid primary probes, i.e. spectroscopic galaxy clustering (\GCsp) and the 
combination of photometric galaxy clustering (\GCph) and photometric weak lensing (WL); see Sect.~\ref{sec:probes}.
We have considered two sets of \Euclid specifications: a pessimistic setting with $k_{\rm max} = 0.25\,h\,{\rm Mpc}^{-1}$ 
for \GCsp, $\ell_{\rm max} = 1500$ for WL, $\ell_{\rm max} = 750$ for \GCph\ and XC, and a cut in redshift of $z<0.9$ 
to \GCph; an optimistic setting with $k_{\rm max} = 0.30\,h\,{\rm Mpc}^{-1}$ for \GCsp, $\ell_{\rm max} = 5000$ for WL, 
and $\ell_{\rm max} = 3000$ for \GCph\ and XC. 

We have modelled the nonlinear matter power spectrum with time-sliced perturbation theory predictions at leading order 
\citep{Vasudevan:2019ewf,Beutler:2019ojk,Ballardini:2019tuc,Ballardini:2022vzh} calibrated on a set of N-body simulations 
from {\tt COLA} \citep{Tassev:2013pn,Howlett:2015hfa} and {\tt GADGET} 
\citep{Springel:2000qu,Springel:2005mi,Ballardini:2019tuc,Ballardini:2022vzh} for \GCsp; see Sect.~\ref{sec:PT}. For 
modelling of the photometric probes we rely on {\tt HMCODE} \citep{Mead:2016zqy} to describe the broadband small-scale 
behaviour and on PT to describe the smearing of BAO and primordial features; see Sect.~\ref{sec:PT}.

With the full set of probes, i.e. \GCsp+WL+\GCph+XC, we have found that \Euclid alone is able to constrain the amplitude 
of a primordial oscillatory feature with frequency $\omega_{\rm X} = 10$ to ${\cal A}_{\rm X} = 0.010\pm0.002$ at a 68.3\% 
CL both in the pessimistic and optimistic settings; see Sect.~\ref{sec:results} and Table~\ref{tab:results}. When we 
consider single \Euclid probes these uncertainties depend strongly on the frequency (see Fig.~\ref{fig:error}) and are 
weakened for low and high frequencies. However, we have found robust constraints of 0.002--0.003 from the full combination 
of \Euclid probes over the frequency range of $(1,\, 10^{2.1})$.

We have then considered further information available from \Euclid measurements including a numerical reconstruction of 
nonlinear spectroscopic galaxy clustering \citep{Beutler:2019ojk,Li:2021jvz} and the galaxy clustering bispectrum 
\citep{Karagiannis:2018jdt}; see Sect.~\ref{sec:reconstruction} and Sect.~\ref{sec:bispectrum}, respectively. 
We have found ${\cal A}_{\rm lin} = 0.0100\pm0.0014$ (${\cal A}_{\rm log} = 0.0100\pm0.0015$) in the pessimistic setting 
and ${\cal A}_{\rm lin} = 0.0100\pm0.0012$ (${\cal A}_{\rm log} = 0.0100\pm0.0012$) in the optimistic setting, both at 
68.3\% CL, from \GCsp\ (rec)+WL+\GCph+XC. This shows that reconstruction can potentially tighten the constraints on the 
primordial feature amplitude. Including the galaxy clustering bispectrum, we further tighten the uncertainties on the 
primordial feature amplitude to ${\cal A}_{\rm lin} = 0.0100\pm0.0009$ (${\cal A}_{\rm log} = 0.0100\pm0.0009$) in the 
pessimistic setting and ${\cal A}_{\rm lin} = 0.0100\pm0.0008$ (${\cal A}_{\rm log} = 0.0100\pm0.0008$) in the optimistic 
setting, both at 68.3\% CL, from \GCsp\ (rec)+WL+\GCph+XC+BS.

Finally, we have studied the combination of \Euclid probes and the expected information from CMB experiments by adding 
(without including the cross-correlation) forecasted results from {\em Planck}, Simons Observatory (SO), and CMB-S4 
following \citet{Euclid:2021qvm}; see Sect.~\ref{sec:CMB}. Combining the Fisher matrix information from a SO-like 
experiment (complemented with a {\em Planck}-like one at low-$\ell$) with the \Euclid primary probes, we have found 
${\cal A}_{\rm lin} = 0.0100\pm0.0014$ (${\cal A}_{\rm log} = 0.0100\pm0.0011$) in the pessimistic setting and 
${\cal A}_{\rm lin} = 0.0100\pm0.0013$ (${\cal A}_{\rm log} = 0.0100\pm0.0010$) in the optimistic setting, both at 68.3\% CL.

\begin{figure*}
\centering
\includegraphics[width=0.49\textwidth]{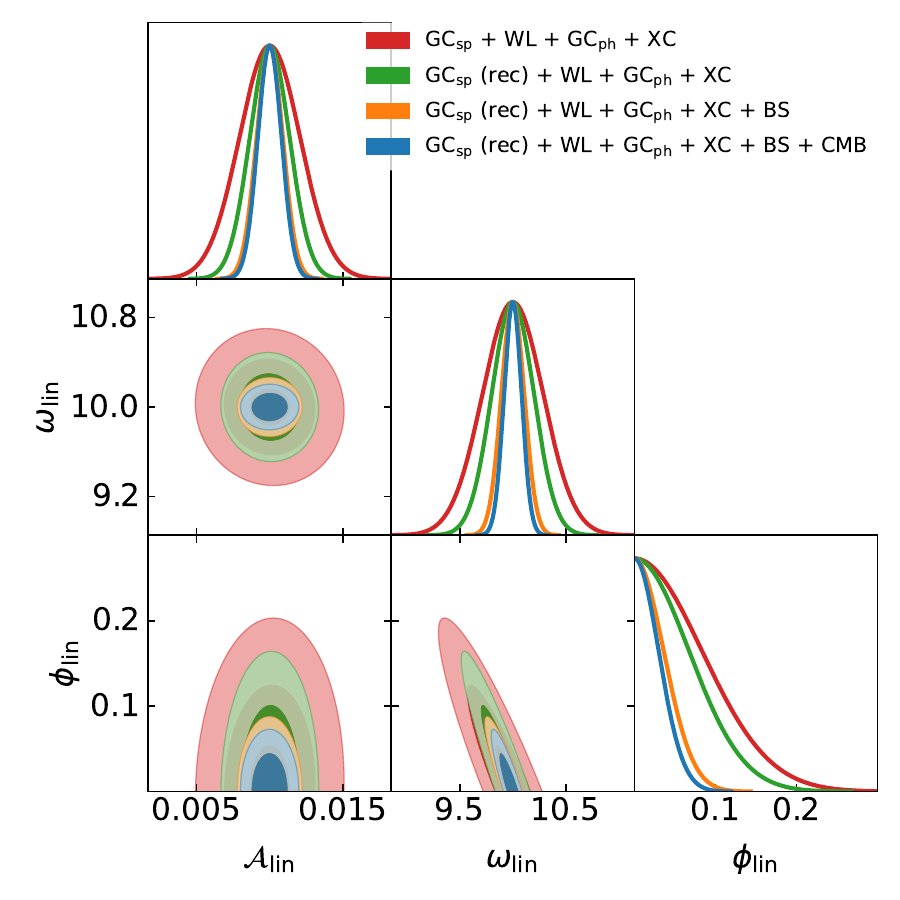}
\includegraphics[width=0.49\textwidth]{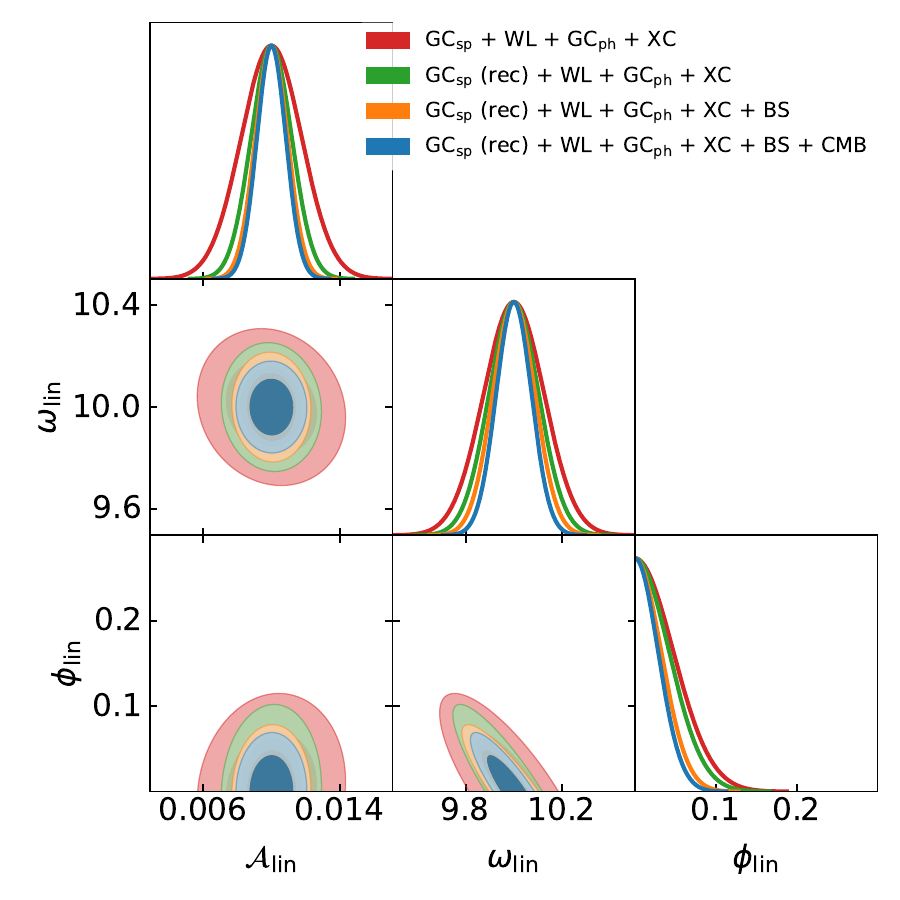}
\includegraphics[width=0.49\textwidth]{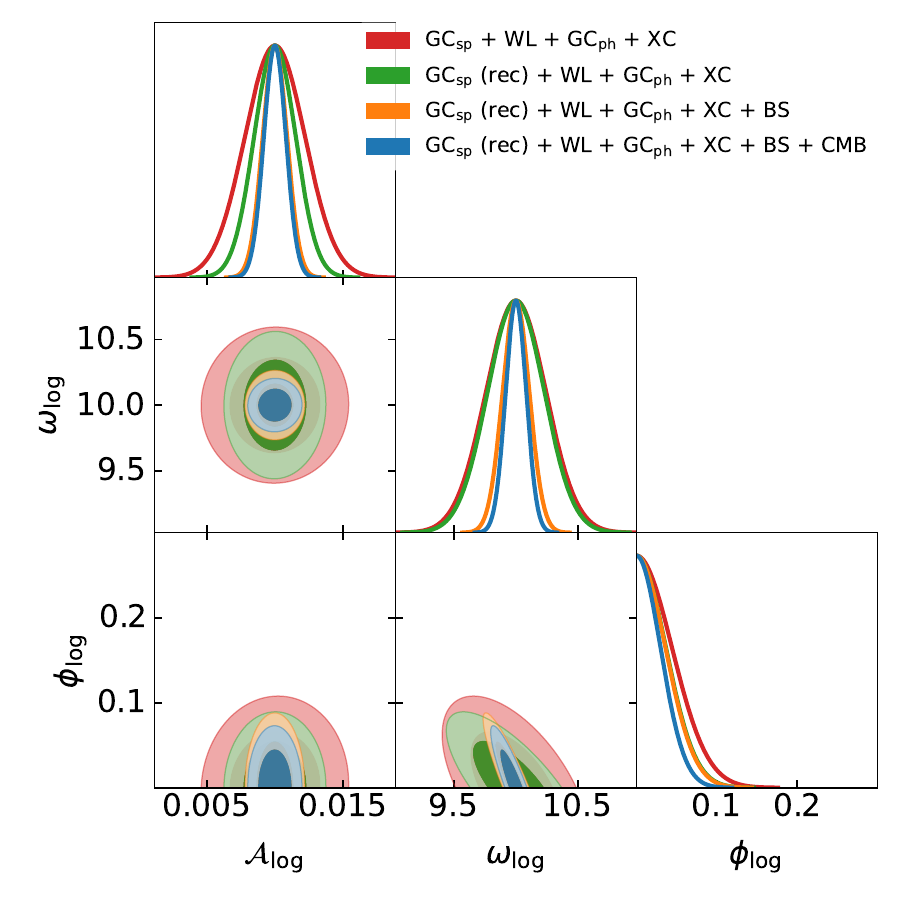}
\includegraphics[width=0.49\textwidth]{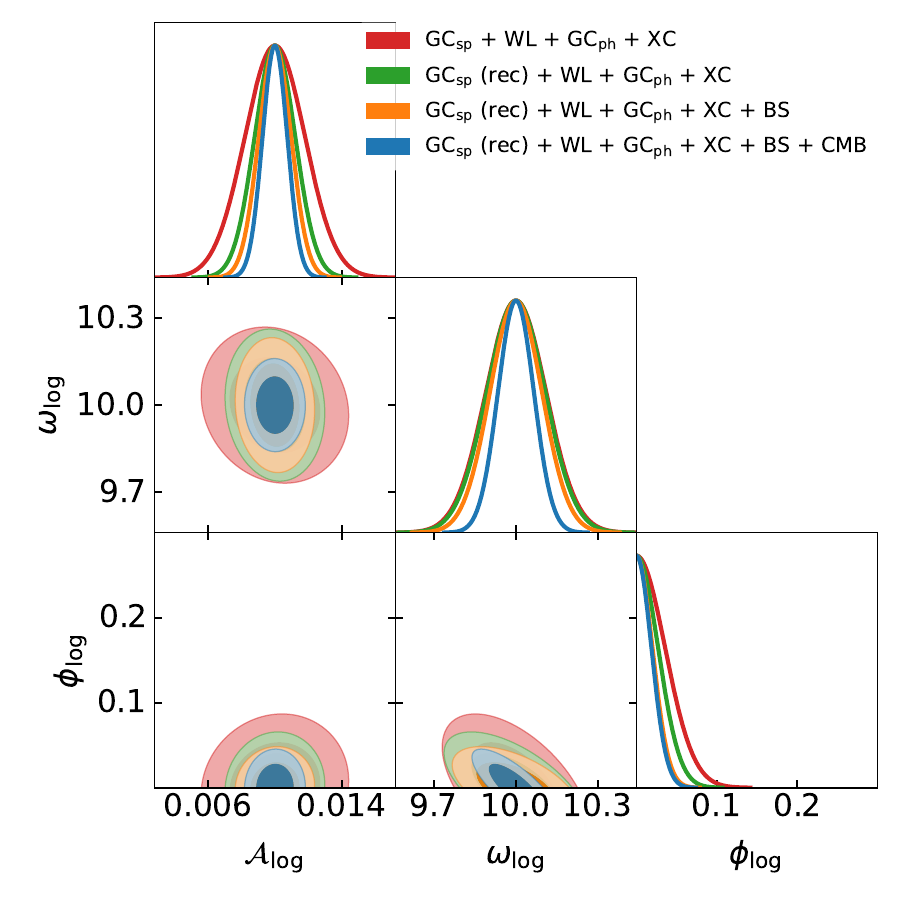}
\caption{Fisher-forecast marginalised two-dimensional contours and one-dimensional probability distribution functions 
from \Euclid on the primordial feature parameters for the LIN model with $\omega_{\rm lin} = 10$ (top panels) and the 
LOG model with $\omega_{\rm log} = 10$ (bottom panels). Left (right) panels correspond to the pessimistic (optimistic) 
setting for \GCsp+WL+\GCph+XC (red), \GCsp+WL+\GCph+XC with numerical reconstruction of \GCsp\ (green), 
\GCsp(rec)+WL+\GCph+XC in combination with \Euclid \GCsp\ bispectrum (orange), and their combination plus SO-like CMB (blue).}
\label{fig:triangle_final}
\end{figure*}

We summarise in Fig.~\ref{fig:triangle_final} the comparison of \Euclid measurements in combination with other 
non-primary sources of information. Our tightest uncertainties, combining all the sources of information expected from 
\Euclid in combination with future CMB experiments, e.g SO complemented with {\em Planck} at low-$\ell$, correspond to 
${\cal A}_{\rm lin} = 0.0100 \pm 0.0008$ at a 68.3\% CL and to ${\cal A}_{\rm log} = 0.0100 \pm 0.0008$ for \GCsp\ 
(PS rec + BS)+WL+\GCph+XC+SO-like for both the pessimistic and the optimistic settings.

Our results highlight the power of \Euclid measurements to constrain primordial feature signals helping us to reach 
a more complete picture of the physics of the early Universe and to improve over the current bounds on the primordial 
feature amplitude parameter ${\cal A}_{\rm X}$. In addition, the expected observations from \Euclid will allow us to 
scrutinise the primordial interpretation of some of the anomalies in the CMB temperature and polarisation angular power 
spectra corresponding to the following best-fits of {\em Planck} data \citep{Planck:2018vyg}: 
${\cal A}_{\rm lin} = 0.015$ and $\log_{10}\omega_{\rm lin} = 1.05$
for the linear oscillations, 
${\cal A}_{\rm log} = 0.014$ and $\log_{10}\omega_{\rm log} = 1.26$
for the logarithmic oscillations.

%
%
\begin{acknowledgements}
MBall acknowledges financial support from the INFN InDark initiative and from the COSMOS network ({\tt www.cosmosnet.it}) through the ASI (Italian Space Agency) 
Grants 2016-24-H.0 and 2016-24-H.1-2018, as well as 2020-9-HH.0 (participation in LiteBIRD phase A).
YA acknowledges support by the Spanish Research Agency (Agencia Estatal de Investigaci\'on)'s grant RYC2020-030193-I/AEI/10.13039/501100011033 and the European Social Fund (Fondo Social Europeo) through the  Ram\'{o}n y Cajal program within the State Plan for Scientific and Technical Research and Innovation (Plan Estatal de Investigaci\'on Cient\'ifica y T\'ecnica y de Innovaci\'on) 2017-2020, and by the Spanish Research Agency through the grant IFT Centro de Excelencia Severo Ochoa No CEX2020-001007-S funded by MCIN/AEI/10.13039/501100011033.
DK was supported by the South African Radio Astronomy Observatory and the National Research Foundation (Grant No. 75415). 
BL is supported by the European Research Council (ERC) through a starting Grant (ERC-StG-716532 PUNCA), and the UK Science and Technology Facilities Council 
(STFC) Consolidated Grant No. ST/I00162X/1 and ST/P000541/1.
ZS acknowledge ITP institute funding from DFG project 456622116 and IRAP lab support.
DS acknowledges financial support from the Fondecyt Regular project number 1200171.
MBald acknowledges support by the project `Combining Cosmic Microwave Background and Large Scale Structure data: an Integrated Approach for Addressing Fundamental Questions in Cosmology', funded by the MIUR Progetti di Ricerca di Rilevante Interesse Nazionale (PRIN) Bando 2017 - grant 2017YJYZAH.
GCH acknowledges support through the ESA research fellowship programme.
HAW acknowledges support from the Research Council of Norway through project 325113.
MV acknowledges partial financial support from the INFN PD51 INDARK grant and from the agreement ASI-INAF n.2017-14-H.0. Part of the N-body simulations presented in this work were obtained on the Ulysses supercomputer at SISSA.
JJ acknowledges support by the Swedish Research Council (VR) under the project 2020-05143 -- `Deciphering the Dynamics of Cosmic Structure'.
GL acknowledges support by the ANR BIG4 project, grant ANR-16-CE23-0002 of the French Agence Nationale de la Recherche.
AA, JJ, and GL acknowledge the support of the Simons Collaboration on ``Learning the Universe''. 
DP acknowledges financial support by agreement n. 2020-9-HH.0 ASI-UniRM2.
JV was supported by Ruth och Nils-Erik Stenb{\"a}cks Stiftelse.
\AckEC
\end{acknowledgements}

\bibliography{Biblio}

\begin{thebibliography}{170}
\expandafter\ifx\csname natexlab\endcsname\relax\def\natexlab#1{#1}\fi

\bibitem[{{Abazajian} {et~al.}(2019){Abazajian}, {Addison}, {Adshead}, {Ahmed},
  {Allen}, {Alonso}, {Alvarez}, {Anderson}, {Arnold}, {Baccigalupi}, {Bailey},
  {Barkats}, {Barron}, {Barry}, {Bartlett}, {Basu Thakur}, {Battaglia},
  {Baxter}, {Bean}, {Bebek}, {Bender}, {Benson}, {Berger}, {Bhimani},
  {Bischoff}, {Bleem}, {Bocquet}, {Boddy}, {Bonato}, {Bond}, {Borrill},
  {Bouchet}, {Brown}, {Bryan}, {Burkhart}, {Buza}, {Byrum}, {Calabrese},
  {Calafut}, {Caldwell}, {Carlstrom}, {Carron}, {Cecil}, {Challinor}, {Chang},
  {Chinone}, {Cho}, {Cooray}, {Crawford}, {Crites}, {Cukierman}, {Cyr-Racine},
  {de Haan}, {de Zotti}, {Delabrouille}, {Demarteau}, {Devlin}, {Di Valentino},
  {Dobbs}, {Duff}, {Duivenvoorden}, {Dvorkin}, {Edwards}, {Eimer}, {Errard},
  {Essinger-Hileman}, {Fabbian}, {Feng}, {Ferraro}, {Filippini}, {Flauger},
  {Flaugher}, {Fraisse}, {Frolov}, {Galitzki}, {Galli}, {Ganga}, {Gerbino},
  {Gilchriese}, {Gluscevic}, {Green}, {Grin}, {Grohs}, {Gualtieri}, {Guarino},
  {Gudmundsson}, {Habib}, {Haller}, {Halpern}, {Halverson}, {Hanany},
  {Harrington}, {Hasegawa}, {Hasselfield}, {Hazumi}, {Heitmann}, {Henderson},
  {Henning}, {Hill}, {Hlozek}, {Holder}, {Holzapfel}, {Hubmayr},
  {Huffenberger}, {Huffer}, {Hui}, {Irwin}, {Johnson}, {Johnstone}, {Jones},
  {Karkare}, {Katayama}, {Kerby}, {Kernovsky}, {Keskitalo}, {Kisner}, {Knox},
  {Kosowsky}, {Kovac}, {Kovetz}, {Kuhlmann}, {Kuo}, {Kurita}, {Kusaka},
  {Lahteenmaki}, {Lawrence}, {Lee}, {Lewis}, {Li}, {Linder}, {Loverde},
  {Lowitz}, {Madhavacheril}, {Mantz}, {Matsuda}, {Mauskopf}, {McMahon},
  {McQuinn}, {Meerburg}, {Melin}, {Meyers}, {Millea}, {Mohr}, {Moncelsi},
  {Mroczkowski}, {Mukherjee}, {M{\"u}nchmeyer}, {Nagai}, {Nagy}, {Namikawa},
  {Nati}, {Natoli}, {Negrello}, {Newburgh}, {Niemack}, {Nishino}, {Nordby},
  {Novosad}, {O'Connor}, {Obied}, {Padin}, {Pandey}, {Partridge}, {Pierpaoli},
  {Pogosian}, {Pryke}, {Puglisi}, {Racine}, {Raghunathan}, {Rahlin},
  {Rajagopalan}, {Raveri}, {Reichanadter}, {Reichardt}, {Remazeilles}, {Rocha},
  {Roe}, {Roy}, {Ruhl}, {Salatino}, {Saliwanchik}, {Schaan}, {Schillaci},
  {Schmittfull}, {Scott}, {Sehgal}, {Shandera}, {Sheehy}, {Sherwin},
  {Shirokoff}, {Simon}, {Slosar}, {Somerville}, {Spergel}, {Staggs}, {Stark},
  {Stompor}, {Story}, {Stoughton}, {Suzuki}, {Tajima}, {Teply}, {Thompson},
  {Timbie}, {Tomasi}, {Treu}, {Tristram}, {Tucker}, {Umilt{\`a}}, {van
  Engelen}, {Vieira}, {Vieregg}, {Vogelsberger}, {Wang}, {Watson}, {White},
  {Whitehorn}, {Wollack}, {Kimmy Wu}, {Xu}, {Yasini}, {Yeck}, {Yoon}, {Young},
  \& {Zonca}}]{Abazajian:2019eic}
{Abazajian}, K., {Addison}, G., {Adshead}, P., {et~al.} 2019, arXiv e-prints,
  arXiv:1907.04473

\bibitem[{{Ach{\'u}carro} {et~al.}(2014){Ach{\'u}carro}, {Atal}, {Hu}, {Ortiz},
  \& {Torrado}}]{Achucarro:2014msa}
{Ach{\'u}carro}, A., {Atal}, V., {Hu}, B., {Ortiz}, P., \& {Torrado}, J. 2014,
  \prd, 90, 023511

\bibitem[{{Ach{\'u}carro} {et~al.}(2022){Ach{\'u}carro}, {Biagetti}, {Braglia},
  {Cabass}, {Caldwell}, {Castorina}, {Chen}, {Coulton}, {Flauger}, {Fumagalli},
  {Ivanov}, {Lee}, {Maleknejad}, {Meerburg}, {Moradinezhad Dizgah}, {Palma},
  {Pimentel}, {Renaux-Petel}, {Wallisch}, {Wandelt}, {Witkowski}, \& {Kimmy
  Wu}}]{2022arXiv220308128A}
{Ach{\'u}carro}, A., {Biagetti}, M., {Braglia}, M., {et~al.} 2022, arXiv
  e-prints, arXiv:2203.08128

\bibitem[{{Ach{\'u}carro} {et~al.}(2011){Ach{\'u}carro}, {Gong}, {Hardeman},
  {Palma}, \& {Patil}}]{Achucarro:2010da}
{Ach{\'u}carro}, A., {Gong}, J.-O., {Hardeman}, S., {Palma}, G.~A., \& {Patil},
  S.~P. 2011, \jcap, 2011, 030

\bibitem[{{Ach{\'u}carro} {et~al.}(2012){Ach{\'u}carro}, {Gong}, {Hardeman},
  {Palma}, \& {Patil}}]{Achucarro:2012sm}
{Ach{\'u}carro}, A., {Gong}, J.-O., {Hardeman}, S., {Palma}, G.~A., \& {Patil},
  S.~P. 2012, Journal of High Energy Physics, 2012, 66

\bibitem[{{Adams} {et~al.}(2001){Adams}, {Cresswell}, \&
  {Easther}}]{Adams:2001vc}
{Adams}, J., {Cresswell}, B., \& {Easther}, R. 2001, \prd, 64, 123514

\bibitem[{{Ade} {et~al.}(2019){Ade}, {Aguirre}, {Ahmed}, {Aiola}, {Ali},
  {Alonso}, {Alvarez}, {Arnold}, {Ashton}, {Austermann}, {Awan}, {Baccigalupi},
  {Baildon}, {Barron}, {Battaglia}, {Battye}, {Baxter}, {Bazarko}, {Beall},
  {Bean}, {Beck}, {Beckman}, {Beringue}, {Bianchini}, {Boada}, {Boettger},
  {Bond}, {Borrill}, {Brown}, {Bruno}, {Bryan}, {Calabrese}, {Calafut},
  {Calisse}, {Carron}, {Challinor}, {Chesmore}, {Chinone}, {Chluba}, {Cho},
  {Choi}, {Coppi}, {Cothard}, {Coughlin}, {Crichton}, {Crowley}, {Crowley},
  {Cukierman}, {D'Ewart}, {D{\"u}nner}, {de Haan}, {Devlin}, {Dicker},
  {Didier}, {Dobbs}, {Dober}, {Duell}, {Duff}, {Duivenvoorden}, {Dunkley},
  {Dusatko}, {Errard}, {Fabbian}, {Feeney}, {Ferraro}, {Flux{\`a}}, {Freese},
  {Frisch}, {Frolov}, {Fuller}, {Fuzia}, {Galitzki}, {Gallardo}, {Tomas Galvez
  Ghersi}, {Gao}, {Gawiser}, {Gerbino}, {Gluscevic}, {Goeckner-Wald}, {Golec},
  {Gordon}, {Gralla}, {Green}, {Grigorian}, {Groh}, {Groppi}, {Guan},
  {Gudmundsson}, {Han}, {Hargrave}, {Hasegawa}, {Hasselfield}, {Hattori},
  {Haynes}, {Hazumi}, {He}, {Healy}, {Henderson}, {Hervias-Caimapo}, {Hill},
  {Hill}, {Hilton}, {Hilton}, {Hincks}, {Hinshaw}, {Hlo{\v{z}}ek}, {Ho}, {Ho},
  {Howe}, {Huang}, {Hubmayr}, {Huffenberger}, {Hughes}, {Ijjas}, {Ikape},
  {Irwin}, {Jaffe}, {Jain}, {Jeong}, {Kaneko}, {Karpel}, {Katayama}, {Keating},
  {Kernasovskiy}, {Keskitalo}, {Kisner}, {Kiuchi}, {Klein}, {Knowles},
  {Koopman}, {Kosowsky}, {Krachmalnicoff}, {Kuenstner}, {Kuo}, {Kusaka},
  {Lashner}, {Lee}, {Lee}, {Leon}, {Leung}, {Lewis}, {Li}, {Li}, {Limon},
  {Linder}, {Lopez-Caraballo}, {Louis}, {Lowry}, {Lungu}, {Madhavacheril},
  {Mak}, {Maldonado}, {Mani}, {Mates}, {Matsuda}, {Maurin}, {Mauskopf}, {May},
  {McCallum}, {McKenney}, {McMahon}, {Meerburg}, {Meyers}, {Miller},
  {Mirmelstein}, {Moodley}, {Munchmeyer}, {Munson}, {Naess}, {Nati},
  {Navaroli}, {Newburgh}, {Nguyen}, {Niemack}, {Nishino}, {Orlowski-Scherer},
  {Page}, {Partridge}, {Peloton}, {Perrotta}, {Piccirillo}, {Pisano},
  {Poletti}, {Puddu}, {Puglisi}, {Raum}, {Reichardt}, {Remazeilles},
  {Rephaeli}, {Riechers}, {Rojas}, {Roy}, {Sadeh}, {Sakurai}, {Salatino},
  {Sathyanarayana Rao}, {Schaan}, {Schmittfull}, {Sehgal}, {Seibert}, {Seljak},
  {Sherwin}, {Shimon}, {Sierra}, {Sievers}, {Sikhosana}, {Silva-Feaver},
  {Simon}, {Sinclair}, {Siritanasak}, {Smith}, {Smith}, {Spergel}, {Staggs},
  {Stein}, {Stevens}, {Stompor}, {Suzuki}, {Tajima}, {Takakura}, {Teply},
  {Thomas}, {Thorne}, {Thornton}, {Trac}, {Tsai}, {Tucker}, {Ullom},
  {Vagnozzi}, {van Engelen}, {Van Lanen}, {Van Winkle}, {Vavagiakis},
  {Verg{\`e}s}, {Vissers}, {Wagoner}, {Walker}, {Ward}, {Westbrook},
  {Whitehorn}, {Williams}, {Williams}, {Wollack}, {Xu}, {Yu}, {Yu}, {Zago},
  {Zhang}, {Zhu}, \& {Simons Observatory
  Collaboration}}]{SimonsObservatory:2018koc}
{Ade}, P., {Aguirre}, J., {Ahmed}, Z., {et~al.} 2019, \jcap, 2019, 056

\bibitem[{{Albrecht} \& {Steinhardt}(1982)}]{Albrecht:1982wi}
{Albrecht}, A. \& {Steinhardt}, P.~J. 1982, \prl, 48, 1220

\bibitem[{{Alcock} \& {Paczynski}(1979)}]{Alcock:1979mp}
{Alcock}, C. \& {Paczynski}, B. 1979, \nat, 281, 358

\bibitem[{{Amendola} {et~al.}(2018){Amendola}, {Appleby}, {Avgoustidis},
  {Bacon}, {Baker}, {Baldi}, {Bartolo}, {Blanchard}, {Bonvin}, {Borgani},
  {Branchini}, {Burrage}, {Camera}, {Carbone}, {Casarini}, {Cropper}, {de
  Rham}, {Dietrich}, {Di Porto}, {Durrer}, {Ealet}, {Ferreira}, {Finelli},
  {Garc{\'\i}a-Bellido}, {Giannantonio}, {Guzzo}, {Heavens}, {Heisenberg},
  {Heymans}, {Hoekstra}, {Hollenstein}, {Holmes}, {Hwang}, {Jahnke},
  {Kitching}, {Koivisto}, {Kunz}, {La Vacca}, {Linder}, {March}, {Marra},
  {Martins}, {Majerotto}, {Markovic}, {Marsh}, {Marulli}, {Massey}, {Mellier},
  {Montanari}, {Mota}, {Nunes}, {Percival}, {Pettorino}, {Porciani},
  {Quercellini}, {Read}, {Rinaldi}, {Sapone}, {Sawicki}, {Scaramella},
  {Skordis}, {Simpson}, {Taylor}, {Thomas}, {Trotta}, {Verde}, {Vernizzi},
  {Vollmer}, {Wang}, {Weller}, \& {Zlosnik}}]{Amendola:2016saw}
{Amendola}, L., {Appleby}, S., {Avgoustidis}, A., {et~al.} 2018, Living Reviews
  in Relativity, 21, 2

\bibitem[{{Ansari Fard} \& {Baghram}(2018)}]{Fard:2017oex}
{Ansari Fard}, M. \& {Baghram}, S. 2018, \jcap, 2018, 051

\bibitem[{{Antony} {et~al.}(2023){Antony}, {Finelli}, {Hazra}, \&
  {Shafieloo}}]{Antony:2022ert}
{Antony}, A., {Finelli}, F., {Hazra}, D.~K., \& {Shafieloo}, A. 2023, \prl,
  130, 111001

\bibitem[{{Assassi} {et~al.}(2015){Assassi}, {Baumann}, \&
  {Schmidt}}]{Assassi2015}
{Assassi}, V., {Baumann}, D., \& {Schmidt}, F. 2015, \jcap, 12, 043

\bibitem[{{Baldauf} {et~al.}(2015){Baldauf}, {Mirbabayi}, {Simonovi{\'{c}}}, \&
  {Zaldarriaga}}]{Baldauf:2015xfa}
{Baldauf}, T., {Mirbabayi}, M., {Simonovi{\'{c}}}, M., \& {Zaldarriaga}, M.
  2015, \prd, 92, 043514

\bibitem[{{Baldauf} {et~al.}(2012){Baldauf}, {Seljak}, {Desjacques}, \&
  {McDonald}}]{Baldauf:2012hs}
{Baldauf}, T., {Seljak}, U., {Desjacques}, V., \& {McDonald}, P. 2012, \prd,
  86, 083540

\bibitem[{{Ballardini}(2019)}]{Ballardini:2018noo}
{Ballardini}, M. 2019, Physics of the Dark Universe, 23, 100245

\bibitem[{{Ballardini} \& {Finelli}(2022)}]{Ballardini:2022vzh}
{Ballardini}, M. \& {Finelli}, F. 2022, \jcap, 2022, 083

\bibitem[{{Ballardini} {et~al.}(2016){Ballardini}, {Finelli}, {Fedeli}, \&
  {Moscardini}}]{Ballardini:2016hpi}
{Ballardini}, M., {Finelli}, F., {Fedeli}, C., \& {Moscardini}, L. 2016, \jcap,
  2016, 041

\bibitem[{{Ballardini} {et~al.}(2018){Ballardini}, {Finelli}, {Maartens}, \&
  {Moscardini}}]{Ballardini:2017qwq}
{Ballardini}, M., {Finelli}, F., {Maartens}, R., \& {Moscardini}, L. 2018,
  \jcap, 2018, 044

\bibitem[{{Ballardini} {et~al.}(2023){Ballardini}, {Finelli}, {Marulli},
  {Moscardini}, \& {Veropalumbo}}]{Ballardini:2022wzu}
{Ballardini}, M., {Finelli}, F., {Marulli}, F., {Moscardini}, L., \&
  {Veropalumbo}, A. 2023, \prd, 107, 043532

\bibitem[{{Ballardini} {et~al.}(2020){Ballardini}, {Murgia}, {Baldi},
  {Finelli}, \& {Viel}}]{Ballardini:2019tuc}
{Ballardini}, M., {Murgia}, R., {Baldi}, M., {Finelli}, F., \& {Viel}, M. 2020,
  \jcap, 2020, 030

\bibitem[{{Barnaby} {et~al.}(2012){Barnaby}, {Pajer}, \&
  {Peloso}}]{Barnaby:2011qe}
{Barnaby}, N., {Pajer}, E., \& {Peloso}, M. 2012, \prd, 85, 023525

\bibitem[{{Barreira}(2020)}]{Barreira:2020ekm}
{Barreira}, A. 2020, \jcap, 2020, 031

\bibitem[{{Bartolo} {et~al.}(2013){Bartolo}, {Cannone}, \&
  {Matarrese}}]{Bartolo:2013exa}
{Bartolo}, N., {Cannone}, D., \& {Matarrese}, S. 2013, \jcap, 2013, 038

\bibitem[{{Benetti}(2013)}]{Benetti:2013cja}
{Benetti}, M. 2013, \prd, 88, 087302

\bibitem[{{Bermejo-Climent} {et~al.}(2021){Bermejo-Climent}, {Ballardini},
  {Finelli}, {Paoletti}, {Maartens}, {Rubi{\~n}o-Mart{\'\i}n}, \&
  {Valenziano}}]{Bermejo-Climent:2021jxf}
{Bermejo-Climent}, J.~R., {Ballardini}, M., {Finelli}, F., {et~al.} 2021, \prd,
  103, 103502

\bibitem[{{Bernardeau} {et~al.}(2002){Bernardeau}, {Colombi}, {Gazta{\~n}aga},
  \& {Scoccimarro}}]{Bernardeau:2001qr}
{Bernardeau}, F., {Colombi}, S., {Gazta{\~n}aga}, E., \& {Scoccimarro}, R.
  2002, \physrep, 367, 1

\bibitem[{{Beutler} {et~al.}(2019){Beutler}, {Biagetti}, {Green}, {Slosar}, \&
  {Wallisch}}]{Beutler:2019ojk}
{Beutler}, F., {Biagetti}, M., {Green}, D., {Slosar}, A., \& {Wallisch}, B.
  2019, Physical Review Research, 1, 033209

\bibitem[{{Bezrukov} \& {Shaposhnikov}(2008)}]{Bezrukov:2007ep}
{Bezrukov}, F. \& {Shaposhnikov}, M. 2008, Physics Letters B, 659, 703

\bibitem[{{Biagetti} {et~al.}(2022){Biagetti}, {Castiblanco}, {Nore{\~n}a}, \&
  {Sefusatti}}]{Biagetti:2021tua}
{Biagetti}, M., {Castiblanco}, L., {Nore{\~n}a}, J., \& {Sefusatti}, E. 2022,
  \jcap, 2022, 009

\bibitem[{{Birkin} {et~al.}(2019){Birkin}, {Li}, {Cautun}, \&
  {Shi}}]{Birkin:2018nag}
{Birkin}, J., {Li}, B., {Cautun}, M., \& {Shi}, Y. 2019, \mnras, 483, 5267

\bibitem[{{Blas} {et~al.}(2016{\natexlab{a}}){Blas}, {Garny}, {Ivanov}, \&
  {Sibiryakov}}]{Blas:2015qsi}
{Blas}, D., {Garny}, M., {Ivanov}, M.~M., \& {Sibiryakov}, S.
  2016{\natexlab{a}}, \jcap, 2016, 052

\bibitem[{{Blas} {et~al.}(2016{\natexlab{b}}){Blas}, {Garny}, {Ivanov}, \&
  {Sibiryakov}}]{Blas:2016sfa}
{Blas}, D., {Garny}, M., {Ivanov}, M.~M., \& {Sibiryakov}, S.
  2016{\natexlab{b}}, \jcap, 2016, 028

\bibitem[{{Boyle} \& {Komatsu}(2018)}]{Boyle:2017lzt}
{Boyle}, A. \& {Komatsu}, E. 2018, \jcap, 2018, 035

\bibitem[{{Braglia} {et~al.}(2021){Braglia}, {Chen}, \&
  {Hazra}}]{Braglia:2021ckn}
{Braglia}, M., {Chen}, X., \& {Hazra}, D.~K. 2021, \jcap, 2021, 005

\bibitem[{{Braglia} {et~al.}(2022{\natexlab{a}}){Braglia}, {Chen}, \&
  {Hazra}}]{Braglia:2021rej}
{Braglia}, M., {Chen}, X., \& {Hazra}, D.~K. 2022{\natexlab{a}}, \prd, 105,
  103523

\bibitem[{{Braglia} {et~al.}(2022{\natexlab{b}}){Braglia}, {Chen}, \&
  {Hazra}}]{Braglia:2021sun}
{Braglia}, M., {Chen}, X., \& {Hazra}, D.~K. 2022{\natexlab{b}}, European
  Physical Journal C, 82, 498

\bibitem[{{Braglia} {et~al.}(2023){Braglia}, {Chen}, {Kumar Hazra}, \&
  {Pinol}}]{2023JCAP...03..014B}
{Braglia}, M., {Chen}, X., {Kumar Hazra}, D., \& {Pinol}, L. 2023, \jcap, 2023,
  014

\bibitem[{{Brenier} {et~al.}(2003){Brenier}, {Frisch}, {H{\'e}non}, {Loeper},
  {Matarrese}, {Mohayaee}, \& {Sobolevski{\u{i}}}}]{Brenier:2003mn}
{Brenier}, Y., {Frisch}, U., {H{\'e}non}, M., {et~al.} 2003, \mnras, 346, 501

\bibitem[{{Burden} {et~al.}(2014){Burden}, {Percival}, {Manera}, {Cuesta},
  {Vargas Magana}, \& {Ho}}]{Burden:2014cwa}
{Burden}, A., {Percival}, W.~J., {Manera}, M., {et~al.} 2014, \mnras, 445, 3152

\bibitem[{{Ca{\~n}as-Herrera} {et~al.}(2021){Ca{\~n}as-Herrera}, {Torrado}, \&
  {Ach{\'u}carro}}]{Canas-Herrera:2020mme}
{Ca{\~n}as-Herrera}, G., {Torrado}, J., \& {Ach{\'u}carro}, A. 2021, \prd, 103,
  123531

\bibitem[{{Cabass} {et~al.}(2018){Cabass}, {Pajer}, \&
  {Schmidt}}]{Cabass:2018roz}
{Cabass}, G., {Pajer}, E., \& {Schmidt}, F. 2018, \jcap, 2018, 003

\bibitem[{{Cannone} {et~al.}(2014){Cannone}, {Bartolo}, \&
  {Matarrese}}]{Cannone:2014qna}
{Cannone}, D., {Bartolo}, N., \& {Matarrese}, S. 2014, \prd, 89, 127301

\bibitem[{{Cautun} \& {van de Weygaert}(2011)}]{Cautun:2011dtfe}
{Cautun}, M.~C. \& {van de Weygaert}, R. 2011, arXiv e-prints
  [\eprint[arXiv]{1105.0370}]

\bibitem[{{Chan} \& {Blot}(2017)}]{Chan:2016ehg}
{Chan}, K.~C. \& {Blot}, L. 2017, \prd, 96, 023528

\bibitem[{{Chen} {et~al.}(2020){Chen}, {Vlah}, \& {White}}]{Chen:2020ckc}
{Chen}, S.-F., {Vlah}, Z., \& {White}, M. 2020, \jcap, 2020, 035

\bibitem[{{Chen}(2010)}]{Chen:2010bka}
{Chen}, X. 2010, \jcap, 2010, 003

\bibitem[{{Chen}(2012)}]{Chen:2011zf}
{Chen}, X. 2012, \jcap, 2012, 038

\bibitem[{{Chen} {et~al.}(2016{\natexlab{a}}){Chen}, {Dvorkin}, {Huang},
  {Namjoo}, \& {Verde}}]{Chen:2016vvw}
{Chen}, X., {Dvorkin}, C., {Huang}, Z., {Namjoo}, M.~H., \& {Verde}, L.
  2016{\natexlab{a}}, \jcap, 2016, 014

\bibitem[{{Chen} {et~al.}(2007){Chen}, {Easther}, \& {Lim}}]{Chen:2006xjb}
{Chen}, X., {Easther}, R., \& {Lim}, E.~A. 2007, \jcap, 2007, 023

\bibitem[{{Chen} {et~al.}(2008){Chen}, {Easther}, \& {Lim}}]{Chen:2008wn}
{Chen}, X., {Easther}, R., \& {Lim}, E.~A. 2008, \jcap, 2008, 010

\bibitem[{{Chen} {et~al.}(2016{\natexlab{b}}){Chen}, {Meerburg}, \&
  {M{\"u}nchmeyer}}]{Chen:2016zuu}
{Chen}, X., {Meerburg}, P.~D., \& {M{\"u}nchmeyer}, M. 2016{\natexlab{b}},
  \jcap, 2016, 023

\bibitem[{{Chen} \& {Namjoo}(2014)}]{Chen:2014joa}
{Chen}, X. \& {Namjoo}, M.~H. 2014, Physics Letters B, 739, 285

\bibitem[{{Cheung} {et~al.}(2008){Cheung}, {Fitzpatrick}, {Kaplan}, {Senatore},
  \& {Creminelli}}]{2008JHEP...03..014C}
{Cheung}, C., {Fitzpatrick}, A.~L., {Kaplan}, J., {Senatore}, L., \&
  {Creminelli}, P. 2008, Journal of High Energy Physics, 3, 014

\bibitem[{{Chluba} {et~al.}(2015){Chluba}, {Hamann}, \&
  {Patil}}]{Chluba:2015bqa}
{Chluba}, J., {Hamann}, J., \& {Patil}, S.~P. 2015, International Journal of
  Modern Physics D, 24, 1530023

\bibitem[{{Covi} {et~al.}(2006){Covi}, {Hamann}, {Melchiorri}, {Slosar}, \&
  {Sorbera}}]{Covi:2006ci}
{Covi}, L., {Hamann}, J., {Melchiorri}, A., {Slosar}, A., \& {Sorbera}, I.
  2006, \prd, 74, 083509

\bibitem[{{Creminelli} {et~al.}(2014){Creminelli}, {Gleyzes}, {Simonovi{\'c}},
  \& {Vernizzi}}]{Creminelli:2013poa}
{Creminelli}, P., {Gleyzes}, J., {Simonovi{\'c}}, M., \& {Vernizzi}, F. 2014,
  \jcap, 2014, 051

\bibitem[{{Crocce} {et~al.}(2006){Crocce}, {Pueblas}, \&
  {Scoccimarro}}]{Crocce:2006ve}
{Crocce}, M., {Pueblas}, S., \& {Scoccimarro}, R. 2006, \mnras, 373, 369

\bibitem[{{Crocce} \& {Scoccimarro}(2008)}]{Crocce:2007dt}
{Crocce}, M. \& {Scoccimarro}, R. 2008, \prd, 77, 023533

\bibitem[{{Croft} \& {Gaztanaga}(1997)}]{Croft:1997mn}
{Croft}, R. A.~C. \& {Gaztanaga}, E. 1997, \mnras, 285, 793

\bibitem[{{Debono} {et~al.}(2020){Debono}, {Hazra}, {Shafieloo}, {Smoot}, \&
  {Starobinsky}}]{Debono:2020emh}
{Debono}, I., {Hazra}, D.~K., {Shafieloo}, A., {Smoot}, G.~F., \&
  {Starobinsky}, A.~A. 2020, \mnras, 496, 3448

\bibitem[{{Desjacques} {et~al.}(2018){Desjacques}, {Jeong}, \&
  {Schmidt}}]{Desjacques:2016bnm}
{Desjacques}, V., {Jeong}, D., \& {Schmidt}, F. 2018, \physrep, 733, 1

\bibitem[{{Dom{\`e}nech} {et~al.}(2020){Dom{\`e}nech}, {Chen}, {Kamionkowski},
  \& {Loeb}}]{Domenech:2020qay}
{Dom{\`e}nech}, G., {Chen}, X., {Kamionkowski}, M., \& {Loeb}, A. 2020, \jcap,
  2020, 005

\bibitem[{{Dom{\`e}nech} \& {Kamionkowski}(2019)}]{Domenech:2019cyh}
{Dom{\`e}nech}, G. \& {Kamionkowski}, M. 2019, \jcap, 2019, 040

\bibitem[{{Easther} \& {Flauger}(2014)}]{Easther:2013kla}
{Easther}, R. \& {Flauger}, R. 2014, \jcap, 2014, 037

\bibitem[{{Eisenstein} {et~al.}(2007){Eisenstein}, {Seo}, {Sirko}, \&
  {Spergel}}]{Eisenstein:2006nk}
{Eisenstein}, D.~J., {Seo}, H.-J., {Sirko}, E., \& {Spergel}, D.~N. 2007, \apj,
  664, 675

\bibitem[{{Euclid Collaboration: Blanchard} {et~al.}(2020){Euclid
  Collaboration: Blanchard}, {Camera}, {Carbone}, {Cardone}, {Casas}, {Clesse},
  {Ili{\'c}}, {Kilbinger}, {Kitching}, {Kunz}, {Lacasa}, {Linder}, {Majerotto},
  {Markovi{\v{c}}}, {Martinelli}, {Pettorino}, {Pourtsidou}, {Sakr},
  {S{\'a}nchez}, {Sapone}, {Tutusaus}, {Yahia-Cherif}, {Yankelevich},
  {Andreon}, {Aussel}, {Balaguera-Antol{\'\i}nez}, {Baldi}, {Bardelli},
  {Bender}, {Biviano}, {Bonino}, {Boucaud}, {Bozzo}, {Branchini}, {Brau-Nogue},
  {Brescia}, {Brinchmann}, {Burigana}, {Cabanac}, {Capobianco}, {Cappi},
  {Carretero}, {Carvalho}, {Casas}, {Castander}, {Castellano}, {Cavuoti},
  {Cimatti}, {Cledassou}, {Colodro-Conde}, {Congedo}, {Conselice}, {Conversi},
  {Copin}, {Corcione}, {Coupon}, {Courtois}, {Cropper}, {Da Silva}, {de la
  Torre}, {Di Ferdinando}, {Dubath}, {Ducret}, {Duncan}, {Dupac}, {Dusini},
  {Fabbian}, {Fabricius}, {Farrens}, {Fosalba}, {Fotopoulou}, {Fourmanoit},
  {Frailis}, {Franceschi}, {Franzetti}, {Fumana}, {Galeotta}, {Gillard},
  {Gillis}, {Giocoli}, {G{\'o}mez-Alvarez}, {Graci{\'a}-Carpio}, {Grupp},
  {Guzzo}, {Hoekstra}, {Hormuth}, {Israel}, {Jahnke}, {Keihanen}, {Kermiche},
  {Kirkpatrick}, {Kohley}, {Kubik}, {Kurki-Suonio}, {Ligori}, {Lilje}, {Lloro},
  {Maino}, {Maiorano}, {Marggraf}, {Martinet}, {Marulli}, {Massey},
  {Medinaceli}, {Mei}, {Mellier}, {Metcalf}, {Metge}, {Meylan}, {Moresco},
  {Moscardini}, {Munari}, {Nichol}, {Niemi}, {Nucita}, {Padilla}, {Paltani},
  {Pasian}, {Percival}, {Pires}, {Polenta}, {Poncet}, {Pozzetti}, {Racca},
  {Raison}, {Renzi}, {Rhodes}, {Romelli}, {Roncarelli}, {Rossetti}, {Saglia},
  {Schneider}, {Scottez}, {Secroun}, {Sirri}, {Stanco}, {Starck}, {Sureau},
  {Tallada-Cresp{\'\i}}, {Tavagnacco}, {Taylor}, {Tenti}, {Tereno},
  {Toledo-Moreo}, {Torradeflot}, {Valenziano}, {Vassallo}, {Verdoes Kleijn},
  {Viel}, {Wang}, {Zacchei}, {Zoubian}, \& {Zucca}}]{Euclid:2019clj}
{Euclid Collaboration: Blanchard}, A., {Camera}, S., {Carbone}, C., {et~al.}
  2020, \aap, 642, A191

\bibitem[{{Euclid Collaboration: Ili{\'c}} {et~al.}(2022){Euclid Collaboration:
  Ili{\'c}}, {Aghanim}, {Baccigalupi}, {Bermejo-Climent}, {Fabbian}, {Legrand},
  {Paoletti}, {Ballardini}, {Archidiacono}, {Douspis}, {Finelli}, {Ganga},
  {Hern{\'a}ndez-Monteagudo}, {Lattanzi}, {Marinucci}, {Migliaccio}, {Carbone},
  {Casas}, {Martinelli}, {Tutusaus}, {Natoli}, {Ntelis}, {Pagano}, {Wenzl},
  {Gruppuso}, {Kitching}, {Langer}, {Mauri}, {Patrizii}, {Renzi}, {Sirri},
  {Stanco}, {Tenti}, {Vielzeuf}, {Lacasa}, {Polenta}, {Yankelevich},
  {Blanchard}, {Sakr}, {Pourtsidou}, {Camera}, {Cardone}, {Kilbinger}, {Kunz},
  {Markovic}, {Pettorino}, {S{\'a}nchez}, {Sapone}, {Amara}, {Auricchio},
  {Bender}, {Bodendorf}, {Bonino}, {Branchini}, {Brescia}, {Brinchmann},
  {Capobianco}, {Carretero}, {Castander}, {Castellano}, {Cavuoti}, {Cimatti},
  {Cledassou}, {Congedo}, {Conselice}, {Conversi}, {Copin}, {Corcione},
  {Costille}, {Cropper}, {Da Silva}, {Degaudenzi}, {Dubath}, {Duncan}, {Dupac},
  {Dusini}, {Ealet}, {Farrens}, {Fosalba}, {Frailis}, {Franceschi},
  {Franzetti}, {Fumana}, {Garilli}, {Gillard}, {Gillis}, {Giocoli}, {Grazian},
  {Grupp}, {Guzzo}, {Haugan}, {Hoekstra}, {Holmes}, {Hormuth}, {Hudelot},
  {Jahnke}, {Kermiche}, {Kiessling}, {Kohley}, {Kubik}, {K{\"u}mmel},
  {Kurki-Suonio}, {Laureijs}, {Ligori}, {Lilje}, {Lloro}, {Mansutti},
  {Marggraf}, {Marulli}, {Massey}, {Maurogordato}, {Meneghetti}, {Merlin},
  {Meylan}, {Moresco}, {Morin}, {Moscardini}, {Munari}, {Niemi}, {Padilla},
  {Paltani}, {Pasian}, {Pedersen}, {Percival}, {Pires}, {Poncet}, {Popa},
  {Pozzetti}, {Raison}, {Rebolo}, {Rhodes}, {Roncarelli}, {Rossetti}, {Saglia},
  {Scaramella}, {Schneider}, {Secroun}, {Seidel}, {Serrano}, {Sirignano},
  {Starck}, {Tallada-Cresp{\'\i}}, {Taylor}, {Tereno}, {Toledo-Moreo},
  {Torradeflot}, {Valentijn}, {Valenziano}, {Verdoes Kleijn}, {Wang},
  {Welikala}, {Weller}, {Zamorani}, {Zoubian}, {Medinaceli}, {Mei}, {Rosset},
  {Sureau}, {Vassallo}, {Zacchei}, {Andreon}, {Balaguera-Antol{\'\i}nez},
  {Baldi}, {Bardelli}, {Biviano}, {Borgani}, {Bozzo}, {Burigana}, {Cabanac},
  {Cappi}, {Carvalho}, {Castignani}, {Colodro-Conde}, {Coupon}, {Courtois},
  {Cuby}, {de la Torre}, {Di Ferdinando}, {Dole}, {Farina}, {Ferreira},
  {Flose-Reimberg}, {Galeotta}, {Gozaliasl}, {Graci{\'a}-Carpio}, {Keihanen},
  {Kirkpatrick}, {Lindholm}, {Mainetti}, {Maino}, {Martinet}, {Maturi},
  {Metcalf}, {Morgante}, {Neissner}, {Nightingale}, {Nucita}, {Potter},
  {Riccio}, {Romelli}, {Schirmer}, {Schultheis}, {Scottez}, {Teyssier},
  {Tramacere}, {Valiviita}, {Viel}, {Whittaker}, \& {Zucca}}]{Euclid:2021qvm}
{Euclid Collaboration: Ili{\'c}}, S., {Aghanim}, N., {Baccigalupi}, C.,
  {et~al.} 2022, \aap, 657, A91

\bibitem[{{Euclid Collaboration: Scaramella} {et~al.}(2022){Euclid
  Collaboration: Scaramella}, {Amiaux}, {Mellier}, {Burigana}, {Carvalho},
  {Cuillandre}, {Da Silva}, {Derosa}, {Dinis}, {Maiorano}, {Maris}, {Tereno},
  {Laureijs}, {Boenke}, {Buenadicha}, {Dupac}, {Gaspar Venancio},
  {G{\'o}mez-{\'A}lvarez}, {Hoar}, {Lorenzo Alvarez}, {Racca},
  {Saavedra-Criado}, {Schwartz}, {Vavrek}, {Schirmer}, {Aussel}, {Azzollini},
  {Cardone}, {Cropper}, {Ealet}, {Garilli}, {Gillard}, {Granett}, {Guzzo},
  {Hoekstra}, {Jahnke}, {Kitching}, {Maciaszek}, {Meneghetti}, {Miller},
  {Nakajima}, {Niemi}, {Pasian}, {Percival}, {Pottinger}, {Sauvage},
  {Scodeggio}, {Wachter}, {Zacchei}, {Aghanim}, {Amara}, {Auphan}, {Auricchio},
  {Awan}, {Balestra}, {Bender}, {Bodendorf}, {Bonino}, {Branchini},
  {Brau-Nogue}, {Brescia}, {Candini}, {Capobianco}, {Carbone}, {Carlberg},
  {Carretero}, {Casas}, {Castander}, {Castellano}, {Cavuoti}, {Cimatti},
  {Cledassou}, {Congedo}, {Conselice}, {Conversi}, {Copin}, {Corcione},
  {Costille}, {Courbin}, {Degaudenzi}, {Douspis}, {Dubath}, {Duncan}, {Dusini},
  {Farrens}, {Ferriol}, {Fosalba}, {Fourmanoit}, {Frailis}, {Franceschi},
  {Franzetti}, {Fumana}, {Gillis}, {Giocoli}, {Grazian}, {Grupp}, {Haugan},
  {Holmes}, {Hormuth}, {Hudelot}, {Kermiche}, {Kiessling}, {Kilbinger},
  {Kohley}, {Kubik}, {K{\"u}mmel}, {Kunz}, {Kurki-Suonio}, {Lahav}, {Ligori},
  {Lilje}, {Lloro}, {Mansutti}, {Marggraf}, {Markovic}, {Marulli}, {Massey},
  {Maurogordato}, {Melchior}, {Merlin}, {Meylan}, {Mohr}, {Moresco}, {Morin},
  {Moscardini}, {Munari}, {Nichol}, {Padilla}, {Paltani}, {Peacock},
  {Pedersen}, {Pettorino}, {Pires}, {Poncet}, {Popa}, {Pozzetti}, {Raison},
  {Rebolo}, {Rhodes}, {Rix}, {Roncarelli}, {Rossetti}, {Saglia}, {Schneider},
  {Schrabback}, {Secroun}, {Seidel}, {Serrano}, {Sirignano}, {Sirri},
  {Skottfelt}, {Stanco}, {Starck}, {Tallada-Cresp{\'\i}}, {Tavagnacco},
  {Taylor}, {Teplitz}, {Toledo-Moreo}, {Torradeflot}, {Trifoglio}, {Valentijn},
  {Valenziano}, {Verdoes Kleijn}, {Wang}, {Welikala}, {Weller}, {Wetzstein},
  {Zamorani}, {Zoubian}, {Andreon}, {Baldi}, {Bardelli}, {Boucaud}, {Camera},
  {Di Ferdinando}, {Fabbian}, {Farinelli}, {Galeotta}, {Graci{\'a}-Carpio},
  {Maino}, {Medinaceli}, {Mei}, {Neissner}, {Polenta}, {Renzi}, {Romelli},
  {Rosset}, {Sureau}, {Tenti}, {Vassallo}, {Zucca}, {Baccigalupi},
  {Balaguera-Antol{\'\i}nez}, {Battaglia}, {Biviano}, {Borgani}, {Bozzo},
  {Cabanac}, {Cappi}, {Casas}, {Castignani}, {Colodro-Conde}, {Coupon},
  {Courtois}, {Cuby}, {de la Torre}, {Desai}, {Dole}, {Fabricius}, {Farina},
  {Ferreira}, {Finelli}, {Flose-Reimberg}, {Fotopoulou}, {Ganga}, {Gozaliasl},
  {Hook}, {Keihanen}, {Kirkpatrick}, {Liebing}, {Lindholm}, {Mainetti},
  {Martinelli}, {Martinet}, {Maturi}, {McCracken}, {Metcalf}, {Morgante},
  {Nightingale}, {Nucita}, {Patrizii}, {Potter}, {Riccio}, {S{\'a}nchez},
  {Sapone}, {Schewtschenko}, {Schultheis}, {Scottez}, {Teyssier}, {Tutusaus},
  {Valiviita}, {Viel}, {Vriend}, \& {Whittaker}}]{Euclid:2021icp}
{Euclid Collaboration: Scaramella}, R., {Amiaux}, J., {Mellier}, Y., {et~al.}
  2022, \aap, 662, A112

\bibitem[{{Fang} {et~al.}(2017){Fang}, {Blazek}, {McEwen}, \&
  {Hirata}}]{Fang:2016wcf}
{Fang}, X., {Blazek}, J.~A., {McEwen}, J.~E., \& {Hirata}, C.~M. 2017, \jcap,
  2017, 030

\bibitem[{{Fergusson} {et~al.}(2015){Fergusson}, {Gruetjen}, {Shellard}, \&
  {Wallisch}}]{Fergusson:2014tza}
{Fergusson}, J.~R., {Gruetjen}, H.~F., {Shellard}, E.~P.~S., \& {Wallisch}, B.
  2015, \prd, 91, 123506

\bibitem[{{Finelli} {et~al.}(2018){Finelli}, {Bucher}, {Ach{\'u}carro},
  {Ballardini}, {Bartolo}, {Baumann}, {Clesse}, {Errard}, {Handley},
  {Hindmarsh}, {Kiiveri}, {Kunz}, {Lasenby}, {Liguori}, {Paoletti}, {Ringeval},
  {V{\"a}liviita}, {van Tent}, {Vennin}, {Ade}, {Allison}, {Arroja}, {Ashdown},
  {Banday}, {Banerji}, {Bartlett}, {Basak}, {de Bernardis}, {Bersanelli},
  {Bonaldi}, {Borril}, {Bouchet}, {Boulanger}, {Brinckmann}, {Burigana},
  {Buzzelli}, {Cai}, {Calvo}, {Carvalho}, {Castellano}, {Challinor}, {Chluba},
  {Colantoni}, {Coppolecchia}, {Crook}, {D'Alessandro}, {D'Amico},
  {Delabrouille}, {Desjacques}, {De Zotti}, {Diego}, {Di Valentino}, {Feeney},
  {Fergusson}, {Fernandez-Cobos}, {Ferraro}, {Forastieri}, {Galli},
  {Garc{\'\i}a-Bellido}, {de Gasperis}, {G{\'e}nova-Santos}, {Gerbino},
  {Gonz{\'a}lez-Nuevo}, {Grandis}, {Greenslade}, {Hagstotz}, {Hanany}, {Hazra},
  {Hern{\'a}ndez-Monteagudo}, {Hervias-Caimapo}, {Hills}, {Hivon}, {Hu},
  {Kisner}, {Kitching}, {Kovetz}, {Kurki-Suonio}, {Lamagna}, {Lattanzi},
  {Lesgourgues}, {Lewis}, {Lindholm}, {Lizarraga}, {L{\'o}pez-Caniego},
  {Luzzi}, {Maffei}, {Mandolesi}, {Mart{\'\i}nez-Gonz{\'a}lez}, {Martins},
  {Masi}, {McCarthy}, {Matarrese}, {Melchiorri}, {Melin}, {Molinari},
  {Monfardini}, {Natoli}, {Negrello}, {Notari}, {Oppizzi}, {Paiella}, {Pajer},
  {Patanchon}, {Patil}, {Piat}, {Pisano}, {Polastri}, {Polenta}, {Pollo},
  {Poulin}, {Quartin}, {Ravenni}, {Remazeilles}, {Renzi}, {Roest}, {Roman},
  {Rubi{\~n}o-Martin}, {Salvati}, {Starobinsky}, {Tartari}, {Tasinato},
  {Tomasi}, {Torrado}, {Trappe}, {Trombetti}, {Tucci}, {Tucker}, {Urrestilla},
  {van de Weygaert}, {Vielva}, {Vittorio}, {Young}, \&
  {Zannoni}}]{Finelli:2016cyd}
{Finelli}, F., {Bucher}, M., {Ach{\'u}carro}, A., {et~al.} 2018, \jcap, 2018,
  016

\bibitem[{{Flauger} {et~al.}(2010){Flauger}, {McAllister}, {Pajer}, {Westphal},
  \& {Xu}}]{Flauger:2009ab}
{Flauger}, R., {McAllister}, L., {Pajer}, E., {Westphal}, A., \& {Xu}, G. 2010,
  \jcap, 2010, 009

\bibitem[{{Fl{\"o}ss} {et~al.}(2023){Fl{\"o}ss}, {Biagetti}, \&
  {Meerburg}}]{Floss:2022wkq}
{Fl{\"o}ss}, T., {Biagetti}, M., \& {Meerburg}, P.~D. 2023, \prd, 107, 023528

\bibitem[{{Freese} {et~al.}(1990){Freese}, {Frieman}, \&
  {Olinto}}]{Freese:1990rb}
{Freese}, K., {Frieman}, J.~A., \& {Olinto}, A.~V. 1990, \prl, 65, 3233

\bibitem[{{Gruppuso} {et~al.}(2016){Gruppuso}, {Kitazawa}, {Mandolesi},
  {Natoli}, \& {Sagnotti}}]{Gruppuso:2015xqa}
{Gruppuso}, A., {Kitazawa}, N., {Mandolesi}, N., {Natoli}, P., \& {Sagnotti},
  A. 2016, Physics of the Dark Universe, 11, 68

\bibitem[{{Gruppuso} \& {Sagnotti}(2015)}]{Gruppuso:2015zia}
{Gruppuso}, A. \& {Sagnotti}, A. 2015, International Journal of Modern Physics
  D, 24, 1544008

\bibitem[{{Gualdi} \& {Verde}(2020)}]{Gualdi:2020ymf}
{Gualdi}, D. \& {Verde}, L. 2020, \jcap, 2020, 041

\bibitem[{{Guth}(1981)}]{Guth:1980zm}
{Guth}, A.~H. 1981, \prd, 23, 347

\bibitem[{{Hada} \& {Eisenstein}(2018)}]{Hada:2018fde}
{Hada}, R. \& {Eisenstein}, D.~J. 2018, \mnras, 478, 1866

\bibitem[{{Hada} \& {Eisenstein}(2019)}]{Hada:2018ziy}
{Hada}, R. \& {Eisenstein}, D.~J. 2019, \mnras, 482, 5685

\bibitem[{{Hamann} {et~al.}(2007){Hamann}, {Covi}, {Melchiorri}, \&
  {Slosar}}]{Hamann:2007pa}
{Hamann}, J., {Covi}, L., {Melchiorri}, A., \& {Slosar}, A. 2007, \prd, 76,
  023503

\bibitem[{{Hamann} \& {Wons}(2022)}]{Hamann:2021eyw}
{Hamann}, J. \& {Wons}, J. 2022, \jcap, 2022, 036

\bibitem[{{Hamilton}(1998)}]{Hamilton:1997zq}
{Hamilton}, A.~J.~S. 1998, in Astrophysics and Space Science Library, Vol. 231,
  The Evolving Universe, ed. D.~{Hamilton}, 185

\bibitem[{{Hannestad} {et~al.}(2010){Hannestad}, {Haugb{\o}lle}, {Jarnhus}, \&
  {Sloth}}]{Hannestad:2009yx}
{Hannestad}, S., {Haugb{\o}lle}, T., {Jarnhus}, P.~R., \& {Sloth}, M.~S. 2010,
  \jcap, 2010, 001

\bibitem[{{Hawking} {et~al.}(1982){Hawking}, {Moss}, \&
  {Stewart}}]{Hawking:1982ga}
{Hawking}, S.~W., {Moss}, I.~G., \& {Stewart}, J.~M. 1982, \prd, 26, 2681

\bibitem[{{Hazra} {et~al.}(2022){Hazra}, {Antony}, \&
  {Shafieloo}}]{Hazra:2022rdl}
{Hazra}, D.~K., {Antony}, A., \& {Shafieloo}, A. 2022, \jcap, 2022, 063

\bibitem[{{Hazra} {et~al.}(2018){Hazra}, {Paoletti}, {Ballardini}, {Finelli},
  {Shafieloo}, {Smoot}, \& {Starobinsky}}]{Hazra:2017joc}
{Hazra}, D.~K., {Paoletti}, D., {Ballardini}, M., {et~al.} 2018, \jcap, 2018,
  017

\bibitem[{{Hazra} {et~al.}(2014{\natexlab{a}}){Hazra}, {Shafieloo}, {Smoot}, \&
  {Starobinsky}}]{Hazra:2014goa}
{Hazra}, D.~K., {Shafieloo}, A., {Smoot}, G.~F., \& {Starobinsky}, A.~A.
  2014{\natexlab{a}}, \jcap, 2014, 048

\bibitem[{{Hazra} {et~al.}(2016){Hazra}, {Shafieloo}, {Smoot}, \&
  {Starobinsky}}]{Hazra:2016fkm}
{Hazra}, D.~K., {Shafieloo}, A., {Smoot}, G.~F., \& {Starobinsky}, A.~A. 2016,
  \jcap, 2016, 009

\bibitem[{{Hazra} {et~al.}(2014{\natexlab{b}}){Hazra}, {Shafieloo}, \&
  {Souradeep}}]{Hazra:2014jwa}
{Hazra}, D.~K., {Shafieloo}, A., \& {Souradeep}, T. 2014{\natexlab{b}}, \jcap,
  2014, 011

\bibitem[{{Howlett} {et~al.}(2015){Howlett}, {Manera}, \&
  {Percival}}]{Howlett:2015hfa}
{Howlett}, C., {Manera}, M., \& {Percival}, W.~J. 2015, Astronomy and
  Computing, 12, 109

\bibitem[{{Hu} \& {Torrado}(2015)}]{Hu:2014hra}
{Hu}, B. \& {Torrado}, J. 2015, \prd, 91, 064039

\bibitem[{{Huang} {et~al.}(2012){Huang}, {Verde}, \& {Vernizzi}}]{Huang:2012mr}
{Huang}, Z., {Verde}, L., \& {Vernizzi}, F. 2012, \jcap, 2012, 005

\bibitem[{{Kaiser}(1987)}]{Kaiser:1987qv}
{Kaiser}, N. 1987, \mnras, 227, 1

\bibitem[{{Karagiannis} {et~al.}(2018){Karagiannis}, {Lazanu}, {Liguori},
  {Raccanelli}, {Bartolo}, \& {Verde}}]{Karagiannis:2018jdt}
{Karagiannis}, D., {Lazanu}, A., {Liguori}, M., {et~al.} 2018, \mnras, 478,
  1341

\bibitem[{{Karagiannis} {et~al.}(2022){Karagiannis}, {Maartens}, \&
  {Randrianjanahary}}]{Karagiannis:2022ylq}
{Karagiannis}, D., {Maartens}, R., \& {Randrianjanahary}, L.~F. 2022, \jcap,
  2022, 003

\bibitem[{{Keeley} {et~al.}(2020){Keeley}, {Shafieloo}, {Hazra}, \&
  {Souradeep}}]{Keeley:2020rmo}
{Keeley}, R.~E., {Shafieloo}, A., {Hazra}, D.~K., \& {Souradeep}, T. 2020,
  \jcap, 2020, 055

\bibitem[{{Knox}(1995)}]{Knox:1995dq}
{Knox}, L. 1995, \prd, 52, 4307

\bibitem[{{Laureijs} {et~al.}(2011){Laureijs}, {Amiaux}, {Arduini},
  {Augu{\`e}res}, {Brinchmann}, {Cole}, {Cropper}, {Dabin}, {Duvet}, {Ealet},
  {Garilli}, {Gondoin}, {Guzzo}, {Hoar}, {Hoekstra}, {Holmes}, {Kitching},
  {Maciaszek}, {Mellier}, {Pasian}, {Percival}, {Rhodes}, {Saavedra Criado},
  {Sauvage}, {Scaramella}, {Valenziano}, {Warren}, {Bender}, {Castander},
  {Cimatti}, {Le F{\`e}vre}, {Kurki-Suonio}, {Levi}, {Lilje}, {Meylan},
  {Nichol}, {Pedersen}, {Popa}, {Rebolo Lopez}, {Rix}, {Rottgering},
  {Zeilinger}, {Grupp}, {Hudelot}, {Massey}, {Meneghetti}, {Miller}, {Paltani},
  {Paulin-Henriksson}, {Pires}, {Saxton}, {Schrabback}, {Seidel}, {Walsh},
  {Aghanim}, {Amendola}, {Bartlett}, {Baccigalupi}, {Beaulieu}, {Benabed},
  {Cuby}, {Elbaz}, {Fosalba}, {Gavazzi}, {Helmi}, {Hook}, {Irwin}, {Kneib},
  {Kunz}, {Mannucci}, {Moscardini}, {Tao}, {Teyssier}, {Weller}, {Zamorani},
  {Zapatero Osorio}, {Boulade}, {Foumond}, {Di Giorgio}, {Guttridge}, {James},
  {Kemp}, {Martignac}, {Spencer}, {Walton}, {Bl{\"u}mchen}, {Bonoli},
  {Bortoletto}, {Cerna}, {Corcione}, {Fabron}, {Jahnke}, {Ligori}, {Madrid},
  {Martin}, {Morgante}, {Pamplona}, {Prieto}, {Riva}, {Toledo}, {Trifoglio},
  {Zerbi}, {Abdalla}, {Douspis}, {Grenet}, {Borgani}, {Bouwens}, {Courbin},
  {Delouis}, {Dubath}, {Fontana}, {Frailis}, {Grazian}, {Koppenh{\"o}fer},
  {Mansutti}, {Melchior}, {Mignoli}, {Mohr}, {Neissner}, {Noddle}, {Poncet},
  {Scodeggio}, {Serrano}, {Shane}, {Starck}, {Surace}, {Taylor},
  {Verdoes-Kleijn}, {Vuerli}, {Williams}, {Zacchei}, {Altieri}, {Escudero
  Sanz}, {Kohley}, {Oosterbroek}, {Astier}, {Bacon}, {Bardelli}, {Baugh},
  {Bellagamba}, {Benoist}, {Bianchi}, {Biviano}, {Branchini}, {Carbone},
  {Cardone}, {Clements}, {Colombi}, {Conselice}, {Cresci}, {Deacon}, {Dunlop},
  {Fedeli}, {Fontanot}, {Franzetti}, {Giocoli}, {Garcia-Bellido}, {Gow},
  {Heavens}, {Hewett}, {Heymans}, {Holland}, {Huang}, {Ilbert}, {Joachimi},
  {Jennins}, {Kerins}, {Kiessling}, {Kirk}, {Kotak}, {Krause}, {Lahav}, {van
  Leeuwen}, {Lesgourgues}, {Lombardi}, {Magliocchetti}, {Maguire}, {Majerotto},
  {Maoli}, {Marulli}, {Maurogordato}, {McCracken}, {McLure}, {Melchiorri},
  {Merson}, {Moresco}, {Nonino}, {Norberg}, {Peacock}, {Pello}, {Penny},
  {Pettorino}, {Di Porto}, {Pozzetti}, {Quercellini}, {Radovich}, {Rassat},
  {Roche}, {Ronayette}, {Rossetti}, {Sartoris}, {Schneider}, {Semboloni},
  {Serjeant}, {Simpson}, {Skordis}, {Smadja}, {Smartt}, {Spano}, {Spiro},
  {Sullivan}, {Tilquin}, {Trotta}, {Verde}, {Wang}, {Williger}, {Zhao},
  {Zoubian}, \& {Zucca}}]{EUCLID:2011zbd}
{Laureijs}, R., {Amiaux}, J., {Arduini}, S., {et~al.} 2011, arXiv e-prints,
  arXiv:1110.3193

\bibitem[{{Li} {et~al.}(2022){Li}, {Zhu}, \& {Li}}]{Li:2021jvz}
{Li}, Y., {Zhu}, H.-M., \& {Li}, B. 2022, \mnras, 514, 4363

\bibitem[{{Linde}(1982)}]{Linde:1981mu}
{Linde}, A.~D. 1982, Physics Letters B, 108, 389

\bibitem[{{Linde}(1983)}]{Linde:1983gd}
{Linde}, A.~D. 1983, Physics Letters B, 129, 177

\bibitem[{{LiteBIRD Collaboration: Allys} {et~al.}(2023){LiteBIRD
  Collaboration: Allys}, {Arnold}, {Aumont}, {Aurlien}, {Azzoni},
  {Baccigalupi}, {Banday}, {Banerji}, {Barreiro}, {Bartolo}, {Bautista},
  {Beck}, {Beckman}, {Bersanelli}, {Boulanger}, {Brilenkov}, {Bucher},
  {Calabrese}, {Campeti}, {Carones}, {Casas}, {Catalano}, {Chan}, {Cheung},
  {Chinone}, {Clark}, {Columbro}, {D'Alessandro}, {de Bernardis}, {de Haan},
  {de la Hoz}, {De Petris}, {Torre}, {Diego-Palazuelos}, {Dobbs}, {Dotani},
  {Duval}, {Elleflot}, {Eriksen}, {Errard}, {Essinger-Hileman}, {Finelli},
  {Flauger}, {Franceschet}, {Fuskeland}, {Galloway}, {Ganga}, {Gerbino},
  {Gervasi}, {G{\'e}nova-Santos}, {Ghigna}, {Giardiello}, {Gjerl{\o}w},
  {Grain}, {Grupp}, {Gruppuso}, {Gudmundsson}, {Halverson}, {Hargrave},
  {Hasebe}, {Hasegawa}, {Hazumi}, {Henrot-Versill{\'e}}, {Hensley}, {Hergt},
  {Herman}, {Hivon}, {Hlozek}, {Hornsby}, {Hoshino}, {Hubmayr}, {Ichiki},
  {Iida}, {Imada}, {Ishino}, {Jaehnig}, {Katayama}, {Kato}, {Keskitalo},
  {Kisner}, {Kobayashi}, {Kogut}, {Kohri}, {Komatsu}, {Komatsu}, {Konishi},
  {Krachmalnicoff}, {Kuo}, {Lamagna}, {Lattanzi}, {Lee}, {Leloup}, {Levrier},
  {Linder}, {Luzzi}, {Macias-Perez}, {Maciaszek}, {Maffei}, {Maino},
  {Mandelli}, {Mart{\'\i}nez-Gonz{\'a}lez}, {Masi}, {Massa}, {Matarrese},
  {Matsuda}, {Matsumura}, {Mele}, {Migliaccio}, {Minami}, {Moggi},
  {Montgomery}, {Montier}, {Morgante}, {Mot}, {Nagano}, {Nagasaki}, {Nagata},
  {Nakano}, {Namikawa}, {Nati}, {Natoli}, {Nerval}, {Noviello}, {Odagiri},
  {Oguri}, {Ohsaki}, {Pagano}, {Paiella}, {Paoletti}, {Passerini}, {Patanchon},
  {Piacentini}, {Piat}, {Pisano}, {Polenta}, {Poletti}, {Prouv{\'e}},
  {Puglisi}, {Rambaud}, {Raum}, {Realini}, {Reinecke}, {Remazeilles},
  {Ritacco}, {Roudil}, {Rubino-Martin}, {Russell}, {Sakurai}, {Sakurai},
  {Sasaki}, {Scott}, {Sekimoto}, {Shinozaki}, {Shiraishi}, {Shirron},
  {Signorelli}, {Spinella}, {Stever}, {Stompor}, {Sugiyama}, {Sullivan},
  {Suzuki}, {Svalheim}, {Switzer}, {Takaku}, {Takakura}, {Takase}, {Tartari},
  {Terao}, {Thermeau}, {Thommesen}, {Thompson}, {Tomasi}, {Tominaga},
  {Tristram}, {Tsuji}, {Tsujimoto}, {Vacher}, {Vielva}, {Vittorio}, {Wang},
  {Watanuki}, {Wehus}, {Weller}, {Westbrook}, {Wilms}, {Winter}, {Wollack},
  {Yumoto}, {Zannoni}, \& {Collaboration LiteB I R D}}]{LiteBIRD:2022cnt}
{LiteBIRD Collaboration: Allys}, E., {Arnold}, K., {Aumont}, J., {et~al.} 2023,
  Progress of Theoretical and Experimental Physics, 2023, 042F01

\bibitem[{{Liu} \& {Huang}(2020)}]{Liu:2019dxr}
{Liu}, M. \& {Huang}, Z. 2020, \apj, 897, 166

\bibitem[{{Mao} {et~al.}(2021){Mao}, {Wang}, {Li}, {Cai}, {Falck}, {Neyrinck},
  \& {Szalay}}]{Mao:2020vdp}
{Mao}, T.-X., {Wang}, J., {Li}, B., {et~al.} 2021, \mnras, 501, 1499

\bibitem[{{McDonald} \& {Roy}(2009)}]{McDonald:2009dh}
{McDonald}, P. \& {Roy}, A. 2009, \jcap, 2009, 020

\bibitem[{{McEwen} {et~al.}(2016){McEwen}, {Fang}, {Hirata}, \&
  {Blazek}}]{McEwen:2016fjn}
{McEwen}, J.~E., {Fang}, X., {Hirata}, C.~M., \& {Blazek}, J.~A. 2016, \jcap,
  2016, 015

\bibitem[{{Mead} {et~al.}(2016){Mead}, {Heymans}, {Lombriser}, {Peacock},
  {Steele}, \& {Winther}}]{Mead:2016zqy}
{Mead}, A.~J., {Heymans}, C., {Lombriser}, L., {et~al.} 2016, \mnras, 459, 1468

\bibitem[{{Meerburg} {et~al.}(2016){Meerburg}, {M{\"u}nchmeyer}, \&
  {Wandelt}}]{Meerburg:2015owa}
{Meerburg}, P.~D., {M{\"u}nchmeyer}, M., \& {Wandelt}, B. 2016, \prd, 93,
  043536

\bibitem[{{Meerburg} {et~al.}(2014){Meerburg}, {Spergel}, \&
  {Wandelt}}]{Meerburg:2013dla}
{Meerburg}, P.~D., {Spergel}, D.~N., \& {Wandelt}, B.~D. 2014, \prd, 89, 063537

\bibitem[{{Meerburg} {et~al.}(2012){Meerburg}, {Wijers}, \& {van der
  Schaar}}]{Meerburg:2011gd}
{Meerburg}, P.~D., {Wijers}, R. A.~M.~J., \& {van der Schaar}, J.~P. 2012,
  \mnras, 421, 369

\bibitem[{{Mergulh{\~a}o} {et~al.}(2023){Mergulh{\~a}o}, {Beutler}, \&
  {Peacock}}]{Mergulhao:2023ukp}
{Mergulh{\~a}o}, T., {Beutler}, F., \& {Peacock}, J.~A. 2023, \jcap, 2023, 012

\bibitem[{{Miranda} \& {Hu}(2014)}]{Miranda:2013wxa}
{Miranda}, V. \& {Hu}, W. 2014, \prd, 89, 083529

\bibitem[{{Miranda} {et~al.}(2015){Miranda}, {Hu}, \&
  {Dvorkin}}]{Miranda:2014fwa}
{Miranda}, V., {Hu}, W., \& {Dvorkin}, C. 2015, \prd, 91, 063514

\bibitem[{{Mohayaee} {et~al.}(2003){Mohayaee}, {Frisch}, {Matarrese}, \&
  {Sobolevskii}}]{Mohayaee:2006aap}
{Mohayaee}, R., {Frisch}, U., {Matarrese}, S., \& {Sobolevskii}, A. 2003, \aap,
  406, 393

\bibitem[{{Monaco} \& {Efstathiou}(1999)}]{Monaco:1999mn}
{Monaco}, P. \& {Efstathiou}, G. 1999, \mnras, 308, 763

\bibitem[{{Mortonson} {et~al.}(2009){Mortonson}, {Dvorkin}, {Peiris}, \&
  {Hu}}]{Mortonson:2009qv}
{Mortonson}, M.~J., {Dvorkin}, C., {Peiris}, H.~V., \& {Hu}, W. 2009, \prd, 79,
  103519

\bibitem[{{Mukhanov} \& {Chibisov}(1981)}]{Mukhanov:1981xt}
{Mukhanov}, V.~F. \& {Chibisov}, G.~V. 1981, Soviet Journal of Experimental and
  Theoretical Physics Letters, 33, 532

\bibitem[{{Mukherjee} \& {Wang}(2003)}]{Mukherjee:2003ag}
{Mukherjee}, P. \& {Wang}, Y. 2003, \apj, 599, 1

\bibitem[{{Naik} {et~al.}(2022){Naik}, {Furuuchi}, \&
  {Chingangbam}}]{Naik:2022mxn}
{Naik}, S.~S., {Furuuchi}, K., \& {Chingangbam}, P. 2022, \jcap, 2022, 016

\bibitem[{{Nusser} \& {Branchini}(2000)}]{Nusser:1999hm}
{Nusser}, A. \& {Branchini}, E. 2000, \mnras, 313, 587

\bibitem[{{Okamoto} \& {Hu}(2003)}]{Okamoto:2003zw}
{Okamoto}, T. \& {Hu}, W. 2003, \prd, 67, 083002

\bibitem[{{Palma} {et~al.}(2018){Palma}, {Sapone}, \& {Sypsas}}]{Palma:2017wxu}
{Palma}, G.~A., {Sapone}, D., \& {Sypsas}, S. 2018, \jcap, 2018, 004

\bibitem[{{Peebles}(1989)}]{Peebles:1989apj}
{Peebles}, P.~J.~E. 1989, \apjl, 344, L53

\bibitem[{{Peiris} {et~al.}(2003){Peiris}, {Komatsu}, {Verde}, {Spergel},
  {Bennett}, {Halpern}, {Hinshaw}, {Jarosik}, {Kogut}, {Limon}, {Meyer},
  {Page}, {Tucker}, {Wollack}, \& {Wright}}]{Peiris:2003ff}
{Peiris}, H.~V., {Komatsu}, E., {Verde}, L., {et~al.} 2003, \apjs, 148, 213

\bibitem[{{Planck Collaboration: Ade} {et~al.}(2014{\natexlab{a}}){Planck
  Collaboration: Ade}, {Aghanim}, {Armitage-Caplan}, {Arnaud}, {Ashdown},
  {Atrio-Barandela}, {Aumont}, {Baccigalupi}, {Banday}, {Barreiro}, {Bartlett},
  {Bartolo}, {Battaner}, {Benabed}, {Beno{\^\i}t}, {Benoit-L{\'e}vy},
  {Bernard}, {Bersanelli}, {Bielewicz}, {Bobin}, {Bock}, {Bonaldi}, {Bonavera},
  {Bond}, {Borrill}, {Bouchet}, {Bridges}, {Bucher}, {Burigana}, {Butler},
  {Cardoso}, {Catalano}, {Challinor}, {Chamballu}, {Chiang}, {Chiang},
  {Christensen}, {Church}, {Clements}, {Colombi}, {Colombo}, {Couchot},
  {Coulais}, {Crill}, {Curto}, {Cuttaia}, {Danese}, {Davies}, {Davis}, {de
  Bernardis}, {de Rosa}, {de Zotti}, {Delabrouille}, {Delouis}, {D{\'e}sert},
  {Diego}, {Dole}, {Donzelli}, {Dor{\'e}}, {Douspis}, {Ducout}, {Dunkley},
  {Dupac}, {Efstathiou}, {Elsner}, {En{\ss}lin}, {Eriksen}, {Fergusson},
  {Finelli}, {Forni}, {Frailis}, {Franceschi}, {Galeotta}, {Ganga}, {Giard},
  {Giraud-H{\'e}raud}, {Gonz{\'a}lez-Nuevo}, {G{\'o}rski}, {Gratton},
  {Gregorio}, {Gruppuso}, {Hansen}, {Hanson}, {Harrison}, {Heavens},
  {Henrot-Versill{\'e}}, {Hern{\'a}ndez-Monteagudo}, {Herranz}, {Hildebrandt},
  {Hivon}, {Hobson}, {Holmes}, {Hornstrup}, {Hovest}, {Huffenberger}, {Jaffe},
  {Jaffe}, {Jones}, {Juvela}, {Keih{\"a}nen}, {Keskitalo}, {Kisner}, {Knoche},
  {Knox}, {Kunz}, {Kurki-Suonio}, {Lacasa}, {Lagache}, {L{\"a}hteenm{\"a}ki},
  {Lamarre}, {Lasenby}, {Laureijs}, {Lawrence}, {Leahy}, {Leonardi},
  {Lesgourgues}, {Lewis}, {Liguori}, {Lilje}, {Linden-V{\o}rnle},
  {L{\'o}pez-Caniego}, {Lubin}, {Mac{\'\i}as-P{\'e}rez}, {Maffei}, {Maino},
  {Mandolesi}, {Mangilli}, {Marinucci}, {Maris}, {Marshall}, {Martin},
  {Mart{\'\i}nez-Gonz{\'a}lez}, {Masi}, {Massardi}, {Matarrese}, {Matthai},
  {Mazzotta}, {Meinhold}, {Melchiorri}, {Mendes}, {Mennella}, {Migliaccio},
  {Mitra}, {Miville-Desch{\^e}nes}, {Moneti}, {Montier}, {Morgante},
  {Mortlock}, {Moss}, {Munshi}, {Murphy}, {Naselsky}, {Natoli}, {Netterfield},
  {N{\o}rgaard-Nielsen}, {Noviello}, {Novikov}, {Novikov}, {Osborne},
  {Oxborrow}, {Paci}, {Pagano}, {Pajot}, {Paoletti}, {Pasian}, {Patanchon},
  {Peiris}, {Perdereau}, {Perotto}, {Perrotta}, {Piacentini}, {Piat},
  {Pierpaoli}, {Pietrobon}, {Plaszczynski}, {Pointecouteau}, {Polenta},
  {Ponthieu}, {Popa}, {Poutanen}, {Pratt}, {Pr{\'e}zeau}, {Prunet}, {Puget},
  {Rachen}, {Racine}, {Rebolo}, {Reinecke}, {Remazeilles}, {Renault}, {Renzi},
  {Ricciardi}, {Riller}, {Ristorcelli}, {Rocha}, {Rosset}, {Roudier},
  {Rubi{\~n}o-Mart{\'\i}n}, {Rusholme}, {Sandri}, {Santos}, {Savini}, {Scott},
  {Seiffert}, {Shellard}, {Smith}, {Spencer}, {Starck}, {Stolyarov}, {Stompor},
  {Sudiwala}, {Sunyaev}, {Sureau}, {Sutter}, {Sutton}, {Suur-Uski}, {Sygnet},
  {Tauber}, {Tavagnacco}, {Terenzi}, {Toffolatti}, {Tomasi}, {Tristram},
  {Tucci}, {Tuovinen}, {Valenziano}, {Valiviita}, {Van Tent}, {Varis},
  {Vielva}, {Villa}, {Vittorio}, {Wade}, {Wandelt}, {White}, {White}, {Yvon},
  {Zacchei}, \& {Zonca}}]{Planck:2013wtn}
{Planck Collaboration: Ade}, P.~A.~R., {Aghanim}, N., {Armitage-Caplan}, C.,
  {et~al.} 2014{\natexlab{a}}, \aap, 571, A24

\bibitem[{{Planck Collaboration: Ade} {et~al.}(2014{\natexlab{b}}){Planck
  Collaboration: Ade}, {Aghanim}, {Armitage-Caplan}, {Arnaud}, {Ashdown},
  {Atrio-Barandela}, {Aumont}, {Baccigalupi}, {Banday}, {Barreiro}, {Bartlett},
  {Bartolo}, {Battaner}, {Benabed}, {Beno{\^\i}t}, {Benoit-L{\'e}vy},
  {Bernard}, {Bersanelli}, {Bielewicz}, {Bobin}, {Bock}, {Bonaldi}, {Bond},
  {Borrill}, {Bouchet}, {Bridges}, {Bucher}, {Burigana}, {Butler}, {Calabrese},
  {Cardoso}, {Catalano}, {Challinor}, {Chamballu}, {Chiang}, {Chiang},
  {Christensen}, {Church}, {Clements}, {Colombi}, {Colombo}, {Couchot},
  {Coulais}, {Crill}, {Curto}, {Cuttaia}, {Danese}, {Davies}, {Davis}, {de
  Bernardis}, {de Rosa}, {de Zotti}, {Delabrouille}, {Delouis}, {D{\'e}sert},
  {Dickinson}, {Diego}, {Dole}, {Donzelli}, {Dor{\'e}}, {Douspis}, {Dunkley},
  {Dupac}, {Efstathiou}, {En{\ss}lin}, {Eriksen}, {Finelli}, {Forni},
  {Frailis}, {Franceschi}, {Galeotta}, {Ganga}, {Gauthier}, {Giard},
  {Giardino}, {Giraud-H{\'e}raud}, {Gonz{\'a}lez-Nuevo}, {G{\'o}rski},
  {Gratton}, {Gregorio}, {Gruppuso}, {Hamann}, {Hansen}, {Hanson}, {Harrison},
  {Henrot-Versill{\'e}}, {Hern{\'a}ndez-Monteagudo}, {Herranz}, {Hildebrandt},
  {Hivon}, {Hobson}, {Holmes}, {Hornstrup}, {Hovest}, {Huffenberger}, {Jaffe},
  {Jaffe}, {Jones}, {Juvela}, {Keih{\"a}nen}, {Keskitalo}, {Kisner}, {Kneissl},
  {Knoche}, {Knox}, {Kunz}, {Kurki-Suonio}, {Lagache}, {L{\"a}hteenm{\"a}ki},
  {Lamarre}, {Lasenby}, {Laureijs}, {Lawrence}, {Leach}, {Leahy}, {Leonardi},
  {Lesgourgues}, {Lewis}, {Liguori}, {Lilje}, {Linden-V{\o}rnle},
  {L{\'o}pez-Caniego}, {Lubin}, {Mac{\'\i}as-P{\'e}rez}, {Maffei}, {Maino},
  {Mandolesi}, {Maris}, {Marshall}, {Martin}, {Mart{\'\i}nez-Gonz{\'a}lez},
  {Masi}, {Massardi}, {Matarrese}, {Matthai}, {Mazzotta}, {Meinhold},
  {Melchiorri}, {Mendes}, {Mennella}, {Migliaccio}, {Mitra},
  {Miville-Desch{\^e}nes}, {Moneti}, {Montier}, {Morgante}, {Mortlock}, {Moss},
  {Munshi}, {Murphy}, {Naselsky}, {Nati}, {Natoli}, {Netterfield},
  {N{\o}rgaard-Nielsen}, {Noviello}, {Novikov}, {Novikov}, {O'Dwyer},
  {Osborne}, {Oxborrow}, {Paci}, {Pagano}, {Pajot}, {Paladini}, {Pandolfi},
  {Paoletti}, {Partridge}, {Pasian}, {Patanchon}, {Peiris}, {Perdereau},
  {Perotto}, {Perrotta}, {Piacentini}, {Piat}, {Pierpaoli}, {Pietrobon},
  {Plaszczynski}, {Pointecouteau}, {Polenta}, {Ponthieu}, {Popa}, {Poutanen},
  {Pratt}, {Pr{\'e}zeau}, {Prunet}, {Puget}, {Rachen}, {Rebolo}, {Reinecke},
  {Remazeilles}, {Renault}, {Ricciardi}, {Riller}, {Ristorcelli}, {Rocha},
  {Rosset}, {Roudier}, {Rowan-Robinson}, {Rubi{\~n}o-Mart{\'\i}n}, {Rusholme},
  {Sandri}, {Santos}, {Savelainen}, {Savini}, {Scott}, {Seiffert}, {Shellard},
  {Spencer}, {Starck}, {Stolyarov}, {Stompor}, {Sudiwala}, {Sunyaev}, {Sureau},
  {Sutton}, {Suur-Uski}, {Sygnet}, {Tauber}, {Tavagnacco}, {Terenzi},
  {Toffolatti}, {Tomasi}, {Tr{\'e}guer-Goudineau}, {Tristram}, {Tucci},
  {Tuovinen}, {Valenziano}, {Valiviita}, {Van Tent}, {Varis}, {Vielva},
  {Villa}, {Vittorio}, {Wade}, {Wandelt}, {White}, {Wilkinson}, {Yvon},
  {Zacchei}, {Zibin}, \& {Zonca}}]{Planck:2013jfk}
{Planck Collaboration: Ade}, P.~A.~R., {Aghanim}, N., {Armitage-Caplan}, C.,
  {et~al.} 2014{\natexlab{b}}, \aap, 571, A22

\bibitem[{{Planck Collaboration: Ade} {et~al.}(2016{\natexlab{a}}){Planck
  Collaboration: Ade}, {Aghanim}, {Arnaud}, {Arroja}, {Ashdown}, {Aumont},
  {Baccigalupi}, {Ballardini}, {Banday}, {Barreiro}, {Bartolo}, {Basak},
  {Battaner}, {Benabed}, {Beno{\^\i}t}, {Benoit-L{\'e}vy}, {Bernard},
  {Bersanelli}, {Bielewicz}, {Bock}, {Bonaldi}, {Bonavera}, {Bond}, {Borrill},
  {Bouchet}, {Boulanger}, {Bucher}, {Burigana}, {Butler}, {Calabrese},
  {Cardoso}, {Catalano}, {Challinor}, {Chamballu}, {Chiang}, {Christensen},
  {Church}, {Clements}, {Colombi}, {Colombo}, {Combet}, {Couchot}, {Coulais},
  {Crill}, {Curto}, {Cuttaia}, {Danese}, {Davies}, {Davis}, {de Bernardis}, {de
  Rosa}, {de Zotti}, {Delabrouille}, {D{\'e}sert}, {Diego}, {Dole}, {Donzelli},
  {Dor{\'e}}, {Douspis}, {Ducout}, {Dupac}, {Efstathiou}, {Elsner},
  {En{\ss}lin}, {Eriksen}, {Fergusson}, {Finelli}, {Forni}, {Frailis},
  {Fraisse}, {Franceschi}, {Frejsel}, {Galeotta}, {Galli}, {Ganga}, {Gauthier},
  {Ghosh}, {Giard}, {Giraud-H{\'e}raud}, {Gjerl{\o}w}, {Gonz{\'a}lez-Nuevo},
  {G{\'o}rski}, {Gratton}, {Gregorio}, {Gruppuso}, {Gudmundsson}, {Hamann},
  {Hansen}, {Hanson}, {Harrison}, {Heavens}, {Helou}, {Henrot-Versill{\'e}},
  {Hern{\'a}ndez-Monteagudo}, {Herranz}, {Hildebrandt}, {Hivon}, {Hobson},
  {Holmes}, {Hornstrup}, {Hovest}, {Huang}, {Huffenberger}, {Hurier}, {Jaffe},
  {Jaffe}, {Jones}, {Juvela}, {Keih{\"a}nen}, {Keskitalo}, {Kim}, {Kisner},
  {Knoche}, {Kunz}, {Kurki-Suonio}, {Lacasa}, {Lagache}, {L{\"a}hteenm{\"a}ki},
  {Lamarre}, {Lasenby}, {Lattanzi}, {Lawrence}, {Leonardi}, {Lesgourgues},
  {Levrier}, {Lewis}, {Liguori}, {Lilje}, {Linden-V{\o}rnle},
  {L{\'o}pez-Caniego}, {Lubin}, {Mac{\'\i}as-P{\'e}rez}, {Maggio}, {Maino},
  {Mandolesi}, {Mangilli}, {Marinucci}, {Maris}, {Martin},
  {Mart{\'\i}nez-Gonz{\'a}lez}, {Masi}, {Matarrese}, {McGehee}, {Meinhold},
  {Melchiorri}, {Mendes}, {Mennella}, {Migliaccio}, {Mitra},
  {Miville-Desch{\^e}nes}, {Moneti}, {Montier}, {Morgante}, {Mortlock}, {Moss},
  {M{\"u}nchmeyer}, {Munshi}, {Murphy}, {Naselsky}, {Nati}, {Natoli},
  {Netterfield}, {N{\o}rgaard-Nielsen}, {Noviello}, {Novikov}, {Novikov},
  {Oxborrow}, {Paci}, {Pagano}, {Pajot}, {Paoletti}, {Pasian}, {Patanchon},
  {Peiris}, {Perdereau}, {Perotto}, {Perrotta}, {Pettorino}, {Piacentini},
  {Piat}, {Pierpaoli}, {Pietrobon}, {Plaszczynski}, {Pointecouteau}, {Polenta},
  {Popa}, {Pratt}, {Pr{\'e}zeau}, {Prunet}, {Puget}, {Rachen}, {Racine},
  {Rebolo}, {Reinecke}, {Remazeilles}, {Renault}, {Renzi}, {Ristorcelli},
  {Rocha}, {Rosset}, {Rossetti}, {Roudier}, {Rubi{\~n}o-Mart{\'\i}n},
  {Rusholme}, {Sandri}, {Santos}, {Savelainen}, {Savini}, {Scott}, {Seiffert},
  {Shellard}, {Shiraishi}, {Smith}, {Spencer}, {Stolyarov}, {Stompor},
  {Sudiwala}, {Sunyaev}, {Sutter}, {Sutton}, {Suur-Uski}, {Sygnet}, {Tauber},
  {Terenzi}, {Toffolatti}, {Tomasi}, {Tristram}, {Troja}, {Tucci}, {Tuovinen},
  {Valenziano}, {Valiviita}, {Van Tent}, {Vielva}, {Villa}, {Wade}, {Wandelt},
  {Wehus}, {Yvon}, {Zacchei}, \& {Zonca}}]{Planck:2015zfm}
{Planck Collaboration: Ade}, P.~A.~R., {Aghanim}, N., {Arnaud}, M., {et~al.}
  2016{\natexlab{a}}, \aap, 594, A17

\bibitem[{{Planck Collaboration: Ade} {et~al.}(2016{\natexlab{b}}){Planck
  Collaboration: Ade}, {Aghanim}, {Arnaud}, {Arroja}, {Ashdown}, {Aumont},
  {Baccigalupi}, {Ballardini}, {Banday}, {Barreiro}, {Bartolo}, {Battaner},
  {Benabed}, {Beno{\^\i}t}, {Benoit-L{\'e}vy}, {Bernard}, {Bersanelli},
  {Bielewicz}, {Bock}, {Bonaldi}, {Bonavera}, {Bond}, {Borrill}, {Bouchet},
  {Boulanger}, {Bucher}, {Burigana}, {Butler}, {Calabrese}, {Cardoso},
  {Catalano}, {Challinor}, {Chamballu}, {Chary}, {Chiang}, {Christensen},
  {Church}, {Clements}, {Colombi}, {Colombo}, {Combet}, {Contreras}, {Couchot},
  {Coulais}, {Crill}, {Curto}, {Cuttaia}, {Danese}, {Davies}, {Davis}, {de
  Bernardis}, {de Rosa}, {de Zotti}, {Delabrouille}, {D{\'e}sert}, {Diego},
  {Dole}, {Donzelli}, {Dor{\'e}}, {Douspis}, {Ducout}, {Dupac}, {Efstathiou},
  {Elsner}, {En{\ss}lin}, {Eriksen}, {Fergusson}, {Finelli}, {Forni},
  {Frailis}, {Fraisse}, {Franceschi}, {Frejsel}, {Frolov}, {Galeotta}, {Galli},
  {Ganga}, {Gauthier}, {Giard}, {Giraud-H{\'e}raud}, {Gjerl{\o}w},
  {Gonz{\'a}lez-Nuevo}, {G{\'o}rski}, {Gratton}, {Gregorio}, {Gruppuso},
  {Gudmundsson}, {Hamann}, {Handley}, {Hansen}, {Hanson}, {Harrison},
  {Henrot-Versill{\'e}}, {Hern{\'a}ndez-Monteagudo}, {Herranz}, {Hildebrandt},
  {Hivon}, {Hobson}, {Holmes}, {Hornstrup}, {Hovest}, {Huang}, {Huffenberger},
  {Hurier}, {Jaffe}, {Jaffe}, {Jones}, {Juvela}, {Keih{\"a}nen}, {Keskitalo},
  {Kim}, {Kisner}, {Kneissl}, {Knoche}, {Kunz}, {Kurki-Suonio}, {Lagache},
  {L{\"a}hteenm{\"a}ki}, {Lamarre}, {Lasenby}, {Lattanzi}, {Lawrence},
  {Leonardi}, {Lesgourgues}, {Levrier}, {Lewis}, {Liguori}, {Lilje},
  {Linden-V{\o}rnle}, {L{\'o}pez-Caniego}, {Lubin}, {Ma},
  {Mac{\'\i}as-P{\'e}rez}, {Maggio}, {Maino}, {Mandolesi}, {Mangilli}, {Maris},
  {Martin}, {Mart{\'\i}nez-Gonz{\'a}lez}, {Masi}, {Matarrese}, {McGehee},
  {Meinhold}, {Melchiorri}, {Mendes}, {Mennella}, {Migliaccio}, {Mitra},
  {Miville-Desch{\^e}nes}, {Molinari}, {Moneti}, {Montier}, {Morgante},
  {Mortlock}, {Moss}, {M{\"u}nchmeyer}, {Munshi}, {Murphy}, {Naselsky}, {Nati},
  {Natoli}, {Netterfield}, {N{\o}rgaard-Nielsen}, {Noviello}, {Novikov},
  {Novikov}, {Oxborrow}, {Paci}, {Pagano}, {Pajot}, {Paladini}, {Pandolfi},
  {Paoletti}, {Pasian}, {Patanchon}, {Pearson}, {Peiris}, {Perdereau},
  {Perotto}, {Perrotta}, {Pettorino}, {Piacentini}, {Piat}, {Pierpaoli},
  {Pietrobon}, {Plaszczynski}, {Pointecouteau}, {Polenta}, {Popa}, {Pratt},
  {Pr{\'e}zeau}, {Prunet}, {Puget}, {Rachen}, {Reach}, {Rebolo}, {Reinecke},
  {Remazeilles}, {Renault}, {Renzi}, {Ristorcelli}, {Rocha}, {Rosset},
  {Rossetti}, {Roudier}, {Rowan-Robinson}, {Rubi{\~n}o-Mart{\'\i}n},
  {Rusholme}, {Sandri}, {Santos}, {Savelainen}, {Savini}, {Scott}, {Seiffert},
  {Shellard}, {Shiraishi}, {Spencer}, {Stolyarov}, {Stompor}, {Sudiwala},
  {Sunyaev}, {Sutton}, {Suur-Uski}, {Sygnet}, {Tauber}, {Terenzi},
  {Toffolatti}, {Tomasi}, {Tristram}, {Trombetti}, {Tucci}, {Tuovinen},
  {Valenziano}, {Valiviita}, {Van Tent}, {Vielva}, {Villa}, {Wade}, {Wandelt},
  {Wehus}, {White}, {Yvon}, {Zacchei}, {Zibin}, \& {Zonca}}]{Ade:2015lrj}
{Planck Collaboration: Ade}, P.~A.~R., {Aghanim}, N., {Arnaud}, M., {et~al.}
  2016{\natexlab{b}}, \aap, 594, A20

\bibitem[{{Planck Collaboration: Aghanim} {et~al.}(2020{\natexlab{a}}){Planck
  Collaboration: Aghanim}, {Akrami}, {Arroja}, {Ashdown}, {Aumont},
  {Baccigalupi}, {Ballardini}, {Banday}, {Barreiro}, {Bartolo}, {Basak},
  {Battye}, {Benabed}, {Bernard}, {Bersanelli}, {Bielewicz}, {Bock}, {Bond},
  {Borrill}, {Bouchet}, {Boulanger}, {Bucher}, {Burigana}, {Butler},
  {Calabrese}, {Cardoso}, {Carron}, {Casaponsa}, {Challinor}, {Chiang},
  {Colombo}, {Combet}, {Contreras}, {Crill}, {Cuttaia}, {de Bernardis}, {de
  Zotti}, {Delabrouille}, {Delouis}, {D{\'e}sert}, {Di Valentino}, {Dickinson},
  {Diego}, {Donzelli}, {Dor{\'e}}, {Douspis}, {Ducout}, {Dupac}, {Efstathiou},
  {Elsner}, {En{\ss}lin}, {Eriksen}, {Falgarone}, {Fantaye}, {Fergusson},
  {Fernandez-Cobos}, {Finelli}, {Forastieri}, {Frailis}, {Franceschi},
  {Frolov}, {Galeotta}, {Galli}, {Ganga}, {G{\'e}nova-Santos}, {Gerbino},
  {Ghosh}, {Gonz{\'a}lez-Nuevo}, {G{\'o}rski}, {Gratton}, {Gruppuso},
  {Gudmundsson}, {Hamann}, {Handley}, {Hansen}, {Helou}, {Herranz},
  {Hildebrandt}, {Hivon}, {Huang}, {Jaffe}, {Jones}, {Karakci}, {Keih{\"a}nen},
  {Keskitalo}, {Kiiveri}, {Kim}, {Kisner}, {Knox}, {Krachmalnicoff}, {Kunz},
  {Kurki-Suonio}, {Lagache}, {Lamarre}, {Langer}, {Lasenby}, {Lattanzi},
  {Lawrence}, {Le Jeune}, {Leahy}, {Lesgourgues}, {Levrier}, {Lewis},
  {Liguori}, {Lilje}, {Lilley}, {Lindholm}, {L{\'o}pez-Caniego}, {Lubin}, {Ma},
  {Mac{\'\i}as-P{\'e}rez}, {Maggio}, {Maino}, {Mandolesi}, {Mangilli},
  {Marcos-Caballero}, {Maris}, {Martin}, {Martinelli},
  {Mart{\'\i}nez-Gonz{\'a}lez}, {Matarrese}, {Mauri}, {McEwen}, {Meerburg},
  {Meinhold}, {Melchiorri}, {Mennella}, {Migliaccio}, {Millea}, {Mitra},
  {Miville-Desch{\^e}nes}, {Molinari}, {Moneti}, {Montier}, {Morgante}, {Moss},
  {Mottet}, {M{\"u}nchmeyer}, {Natoli}, {N{\o}rgaard-Nielsen}, {Oxborrow},
  {Pagano}, {Paoletti}, {Partridge}, {Patanchon}, {Pearson}, {Peel}, {Peiris},
  {Perrotta}, {Pettorino}, {Piacentini}, {Polastri}, {Polenta}, {Puget},
  {Rachen}, {Reinecke}, {Remazeilles}, {Renault}, {Renzi}, {Rocha}, {Rosset},
  {Roudier}, {Rubi{\~n}o-Mart{\'\i}n}, {Ruiz-Granados}, {Salvati}, {Sandri},
  {Savelainen}, {Scott}, {Shellard}, {Shiraishi}, {Sirignano}, {Sirri},
  {Spencer}, {Sunyaev}, {Suur-Uski}, {Tauber}, {Tavagnacco}, {Tenti},
  {Terenzi}, {Toffolatti}, {Tomasi}, {Trombetti}, {Valiviita}, {Van Tent},
  {Vibert}, {Vielva}, {Villa}, {Vittorio}, {Wandelt}, {Wehus}, {White},
  {White}, {Zacchei}, \& {Zonca}}]{Planck:2018nkj}
{Planck Collaboration: Aghanim}, N., {Akrami}, Y., {Arroja}, F., {et~al.}
  2020{\natexlab{a}}, \aap, 641, A1

\bibitem[{{Planck Collaboration: Aghanim} {et~al.}(2020{\natexlab{b}}){Planck
  Collaboration: Aghanim}, {Akrami}, {Ashdown}, {Aumont}, {Baccigalupi},
  {Ballardini}, {Banday}, {Barreiro}, {Bartolo}, {Basak}, {Battye}, {Benabed},
  {Bernard}, {Bersanelli}, {Bielewicz}, {Bock}, {Bond}, {Borrill}, {Bouchet},
  {Boulanger}, {Bucher}, {Burigana}, {Butler}, {Calabrese}, {Cardoso},
  {Carron}, {Challinor}, {Chiang}, {Chluba}, {Colombo}, {Combet}, {Contreras},
  {Crill}, {Cuttaia}, {de Bernardis}, {de Zotti}, {Delabrouille}, {Delouis},
  {Di Valentino}, {Diego}, {Dor{\'e}}, {Douspis}, {Ducout}, {Dupac}, {Dusini},
  {Efstathiou}, {Elsner}, {En{\ss}lin}, {Eriksen}, {Fantaye}, {Farhang},
  {Fergusson}, {Fernandez-Cobos}, {Finelli}, {Forastieri}, {Frailis},
  {Fraisse}, {Franceschi}, {Frolov}, {Galeotta}, {Galli}, {Ganga},
  {G{\'e}nova-Santos}, {Gerbino}, {Ghosh}, {Gonz{\'a}lez-Nuevo}, {G{\'o}rski},
  {Gratton}, {Gruppuso}, {Gudmundsson}, {Hamann}, {Handley}, {Hansen},
  {Herranz}, {Hildebrandt}, {Hivon}, {Huang}, {Jaffe}, {Jones}, {Karakci},
  {Keih{\"a}nen}, {Keskitalo}, {Kiiveri}, {Kim}, {Kisner}, {Knox},
  {Krachmalnicoff}, {Kunz}, {Kurki-Suonio}, {Lagache}, {Lamarre}, {Lasenby},
  {Lattanzi}, {Lawrence}, {Le Jeune}, {Lemos}, {Lesgourgues}, {Levrier},
  {Lewis}, {Liguori}, {Lilje}, {Lilley}, {Lindholm}, {L{\'o}pez-Caniego},
  {Lubin}, {Ma}, {Mac{\'\i}as-P{\'e}rez}, {Maggio}, {Maino}, {Mandolesi},
  {Mangilli}, {Marcos-Caballero}, {Maris}, {Martin}, {Martinelli},
  {Mart{\'\i}nez-Gonz{\'a}lez}, {Matarrese}, {Mauri}, {McEwen}, {Meinhold},
  {Melchiorri}, {Mennella}, {Migliaccio}, {Millea}, {Mitra},
  {Miville-Desch{\^e}nes}, {Molinari}, {Montier}, {Morgante}, {Moss}, {Natoli},
  {N{\o}rgaard-Nielsen}, {Pagano}, {Paoletti}, {Partridge}, {Patanchon},
  {Peiris}, {Perrotta}, {Pettorino}, {Piacentini}, {Polastri}, {Polenta},
  {Puget}, {Rachen}, {Reinecke}, {Remazeilles}, {Renzi}, {Rocha}, {Rosset},
  {Roudier}, {Rubi{\~n}o-Mart{\'\i}n}, {Ruiz-Granados}, {Salvati}, {Sandri},
  {Savelainen}, {Scott}, {Shellard}, {Sirignano}, {Sirri}, {Spencer},
  {Sunyaev}, {Suur-Uski}, {Tauber}, {Tavagnacco}, {Tenti}, {Toffolatti},
  {Tomasi}, {Trombetti}, {Valenziano}, {Valiviita}, {Van Tent}, {Vibert},
  {Vielva}, {Villa}, {Vittorio}, {Wandelt}, {Wehus}, {White}, {White},
  {Zacchei}, \& {Zonca}}]{Planck:2018vyg}
{Planck Collaboration: Aghanim}, N., {Akrami}, Y., {Ashdown}, M., {et~al.}
  2020{\natexlab{b}}, \aap, 641, A6

\bibitem[{{Planck Collaboration: Aghanim} {et~al.}(2020{\natexlab{c}}){Planck
  Collaboration: Aghanim}, {Akrami}, {Ashdown}, {Aumont}, {Baccigalupi},
  {Ballardini}, {Banday}, {Barreiro}, {Bartolo}, {Basak}, {Benabed}, {Bernard},
  {Bersanelli}, {Bielewicz}, {Bock}, {Bond}, {Borrill}, {Bouchet}, {Boulanger},
  {Bucher}, {Burigana}, {Butler}, {Calabrese}, {Cardoso}, {Carron},
  {Casaponsa}, {Challinor}, {Chiang}, {Colombo}, {Combet}, {Crill}, {Cuttaia},
  {de Bernardis}, {de Rosa}, {de Zotti}, {Delabrouille}, {Delouis}, {Di
  Valentino}, {Diego}, {Dor{\'e}}, {Douspis}, {Ducout}, {Dupac}, {Dusini},
  {Efstathiou}, {Elsner}, {En{\ss}lin}, {Eriksen}, {Fantaye},
  {Fernandez-Cobos}, {Finelli}, {Frailis}, {Fraisse}, {Franceschi}, {Frolov},
  {Galeotta}, {Galli}, {Ganga}, {G{\'e}nova-Santos}, {Gerbino}, {Ghosh},
  {Giraud-H{\'e}raud}, {Gonz{\'a}lez-Nuevo}, {G{\'o}rski}, {Gratton},
  {Gruppuso}, {Gudmundsson}, {Hamann}, {Handley}, {Hansen}, {Herranz}, {Hivon},
  {Huang}, {Jaffe}, {Jones}, {Keih{\"a}nen}, {Keskitalo}, {Kiiveri}, {Kim},
  {Kisner}, {Krachmalnicoff}, {Kunz}, {Kurki-Suonio}, {Lagache}, {Lamarre},
  {Lasenby}, {Lattanzi}, {Lawrence}, {Le Jeune}, {Levrier}, {Lewis}, {Liguori},
  {Lilje}, {Lilley}, {Lindholm}, {L{\'o}pez-Caniego}, {Lubin}, {Ma},
  {Mac{\'\i}as-P{\'e}rez}, {Maggio}, {Maino}, {Mandolesi}, {Mangilli},
  {Marcos-Caballero}, {Maris}, {Martin}, {Mart{\'\i}nez-Gonz{\'a}lez},
  {Matarrese}, {Mauri}, {McEwen}, {Meinhold}, {Melchiorri}, {Mennella},
  {Migliaccio}, {Millea}, {Miville-Desch{\^e}nes}, {Molinari}, {Moneti},
  {Montier}, {Morgante}, {Moss}, {Natoli}, {N{\o}rgaard-Nielsen}, {Pagano},
  {Paoletti}, {Partridge}, {Patanchon}, {Peiris}, {Perrotta}, {Pettorino},
  {Piacentini}, {Polenta}, {Puget}, {Rachen}, {Reinecke}, {Remazeilles},
  {Renzi}, {Rocha}, {Rosset}, {Roudier}, {Rubi{\~n}o-Mart{\'\i}n},
  {Ruiz-Granados}, {Salvati}, {Sandri}, {Savelainen}, {Scott}, {Shellard},
  {Sirignano}, {Sirri}, {Spencer}, {Sunyaev}, {Suur-Uski}, {Tauber},
  {Tavagnacco}, {Tenti}, {Toffolatti}, {Tomasi}, {Trombetti}, {Valiviita}, {Van
  Tent}, {Vielva}, {Villa}, {Vittorio}, {Wandelt}, {Wehus}, {Zacchei}, \&
  {Zonca}}]{Planck:2019nip}
{Planck Collaboration: Aghanim}, N., {Akrami}, Y., {Ashdown}, M., {et~al.}
  2020{\natexlab{c}}, \aap, 641, A5

\bibitem[{{Planck Collaboration: Akrami} {et~al.}(2020{\natexlab{a}}){Planck
  Collaboration: Akrami}, {Arroja}, {Ashdown}, {Aumont}, {Baccigalupi},
  {Ballardini}, {Banday}, {Barreiro}, {Bartolo}, {Basak}, {Benabed}, {Bernard},
  {Bersanelli}, {Bielewicz}, {Bock}, {Bond}, {Borrill}, {Bouchet}, {Boulanger},
  {Bucher}, {Burigana}, {Butler}, {Calabrese}, {Cardoso}, {Carron},
  {Challinor}, {Chiang}, {Colombo}, {Combet}, {Contreras}, {Crill}, {Cuttaia},
  {de Bernardis}, {de Zotti}, {Delabrouille}, {Delouis}, {Di Valentino},
  {Diego}, {Donzelli}, {Dor{\'e}}, {Douspis}, {Ducout}, {Dupac}, {Dusini},
  {Efstathiou}, {Elsner}, {En{\ss}lin}, {Eriksen}, {Fantaye}, {Fergusson},
  {Fernandez-Cobos}, {Finelli}, {Forastieri}, {Frailis}, {Franceschi},
  {Frolov}, {Galeotta}, {Galli}, {Ganga}, {Gauthier}, {G{\'e}nova-Santos},
  {Gerbino}, {Ghosh}, {Gonz{\'a}lez-Nuevo}, {G{\'o}rski}, {Gratton},
  {Gruppuso}, {Gudmundsson}, {Hamann}, {Handley}, {Hansen}, {Herranz}, {Hivon},
  {Hooper}, {Huang}, {Jaffe}, {Jones}, {Keih{\"a}nen}, {Keskitalo}, {Kiiveri},
  {Kim}, {Kisner}, {Krachmalnicoff}, {Kunz}, {Kurki-Suonio}, {Lagache},
  {Lamarre}, {Lasenby}, {Lattanzi}, {Lawrence}, {Le Jeune}, {Lesgourgues},
  {Levrier}, {Lewis}, {Liguori}, {Lilje}, {Lindholm}, {L{\'o}pez-Caniego},
  {Lubin}, {Ma}, {Mac{\'\i}as-P{\'e}rez}, {Maggio}, {Maino}, {Mandolesi},
  {Mangilli}, {Marcos-Caballero}, {Maris}, {Martin},
  {Mart{\'\i}nez-Gonz{\'a}lez}, {Matarrese}, {Mauri}, {McEwen}, {Meerburg},
  {Meinhold}, {Melchiorri}, {Mennella}, {Migliaccio}, {Mitra},
  {Miville-Desch{\^e}nes}, {Molinari}, {Moneti}, {Montier}, {Morgante}, {Moss},
  {M{\"u}nchmeyer}, {Natoli}, {N{\o}rgaard-Nielsen}, {Pagano}, {Paoletti},
  {Partridge}, {Patanchon}, {Peiris}, {Perrotta}, {Pettorino}, {Piacentini},
  {Polastri}, {Polenta}, {Puget}, {Rachen}, {Reinecke}, {Remazeilles}, {Renzi},
  {Rocha}, {Rosset}, {Roudier}, {Rubi{\~n}o-Mart{\'\i}n}, {Ruiz-Granados},
  {Salvati}, {Sandri}, {Savelainen}, {Scott}, {Shellard}, {Shiraishi},
  {Sirignano}, {Sirri}, {Spencer}, {Sunyaev}, {Suur-Uski}, {Tauber},
  {Tavagnacco}, {Tenti}, {Toffolatti}, {Tomasi}, {Trombetti}, {Valiviita}, {Van
  Tent}, {Vielva}, {Villa}, {Vittorio}, {Wandelt}, {Wehus}, {White}, {Zacchei},
  {Zibin}, \& {Zonca}}]{Akrami:2018odb}
{Planck Collaboration: Akrami}, Y., {Arroja}, F., {Ashdown}, M., {et~al.}
  2020{\natexlab{a}}, \aap, 641, A10

\bibitem[{{Planck Collaboration: Akrami} {et~al.}(2020{\natexlab{b}}){Planck
  Collaboration: Akrami}, {Arroja}, {Ashdown}, {Aumont}, {Baccigalupi},
  {Ballardini}, {Banday}, {Barreiro}, {Bartolo}, {Basak}, {Benabed}, {Bernard},
  {Bersanelli}, {Bielewicz}, {Bond}, {Borrill}, {Bouchet}, {Bucher},
  {Burigana}, {Butler}, {Calabrese}, {Cardoso}, {Casaponsa}, {Challinor},
  {Chiang}, {Colombo}, {Combet}, {Crill}, {Cuttaia}, {de Bernardis}, {de Rosa},
  {de Zotti}, {Delabrouille}, {Delouis}, {Di Valentino}, {Diego}, {Dor{\'e}},
  {Douspis}, {Ducout}, {Dupac}, {Dusini}, {Efstathiou}, {Elsner}, {En{\ss}lin},
  {Eriksen}, {Fantaye}, {Fergusson}, {Fernandez-Cobos}, {Finelli}, {Frailis},
  {Fraisse}, {Franceschi}, {Frolov}, {Galeotta}, {Galli}, {Ganga},
  {G{\'e}nova-Santos}, {Gerbino}, {Gonz{\'a}lez-Nuevo}, {G{\'o}rski},
  {Gratton}, {Gruppuso}, {Gudmundsson}, {Hamann}, {Handley}, {Hansen},
  {Herranz}, {Hivon}, {Huang}, {Jaffe}, {Jones}, {Jung}, {Keih{\"a}nen},
  {Keskitalo}, {Kiiveri}, {Kim}, {Krachmalnicoff}, {Kunz}, {Kurki-Suonio},
  {Lamarre}, {Lasenby}, {Lattanzi}, {Lawrence}, {Le Jeune}, {Levrier}, {Lewis},
  {Liguori}, {Lilje}, {Lindholm}, {L{\'o}pez-Caniego}, {Ma},
  {Mac{\'\i}as-P{\'e}rez}, {Maggio}, {Maino}, {Mandolesi}, {Marcos-Caballero},
  {Maris}, {Martin}, {Mart{\'\i}nez-Gonz{\'a}lez}, {Matarrese}, {Mauri},
  {McEwen}, {Meerburg}, {Meinhold}, {Melchiorri}, {Mennella}, {Migliaccio},
  {Miville-Desch{\^e}nes}, {Molinari}, {Moneti}, {Montier}, {Morgante}, {Moss},
  {M{\"u}nchmeyer}, {Natoli}, {Oppizzi}, {Pagano}, {Paoletti}, {Partridge},
  {Patanchon}, {Perrotta}, {Pettorino}, {Piacentini}, {Polenta}, {Puget},
  {Rachen}, {Racine}, {Reinecke}, {Remazeilles}, {Renzi}, {Rocha},
  {Rubi{\~n}o-Mart{\'\i}n}, {Ruiz-Granados}, {Salvati}, {Savelainen}, {Scott},
  {Shellard}, {Shiraishi}, {Sirignano}, {Sirri}, {Smith}, {Spencer}, {Stanco},
  {Sunyaev}, {Suur-Uski}, {Tauber}, {Tavagnacco}, {Tenti}, {Toffolatti},
  {Tomasi}, {Trombetti}, {Valiviita}, {Van Tent}, {Vielva}, {Villa},
  {Vittorio}, {Wandelt}, {Wehus}, {Zacchei}, \& {Zonca}}]{Planck:2019izv}
{Planck Collaboration: Akrami}, Y., {Arroja}, F., {Ashdown}, M., {et~al.}
  2020{\natexlab{b}}, \aap, 641, A9

\bibitem[{{Sarpa} {et~al.}(2019){Sarpa}, {Schimd}, {Branchini}, \&
  {Matarrese}}]{Sarpa:2018ucb}
{Sarpa}, E., {Schimd}, C., {Branchini}, E., \& {Matarrese}, S. 2019, \mnras,
  484, 3818

\bibitem[{{Sato}(1981)}]{Sato:1980yn}
{Sato}, K. 1981, \mnras, 195, 467

\bibitem[{{Schaap} \& {van de Weygaert}(2000)}]{Schaap:2000aap}
{Schaap}, W.~E. \& {van de Weygaert}, R. 2000, \aap, 363, L29

\bibitem[{{Schmid} \& {Hui}(2013)}]{Schmidt2012}
{Schmid}, F. \& {Hui}, L. 2013, Physical Review Letters, 110, 011301

\bibitem[{{Schmidt}(2016)}]{Schmidt:2015gwz}
{Schmidt}, F. 2016, \prd, 93, 063512

\bibitem[{{Schmidt} \& {Kamionkowski}(2010)}]{Schmidt2010}
{Schmidt}, F. \& {Kamionkowski}, M. 2010, \prd, 82, 103002

\bibitem[{{Schmittfull} {et~al.}(2017){Schmittfull}, {Baldauf}, \&
  {Zaldarriaga}}]{Schmittfull:2017prd}
{Schmittfull}, M., {Baldauf}, T., \& {Zaldarriaga}, M. 2017, \prd, 96, 023505

\bibitem[{{Scoccimarro} {et~al.}(2012){Scoccimarro}, {Hui}, {Manera}, \&
  {Chan}}]{Scoccimarro2011}
{Scoccimarro}, R., {Hui}, L., {Manera}, M., \& {Chan}, K.~C. 2012, \prd, 85,
  083002

\bibitem[{{Sefusatti} {et~al.}(2007){Sefusatti}, {Vale}, {Kadota}, \&
  {Frieman}}]{Sefusatti:2006eu}
{Sefusatti}, E., {Vale}, C., {Kadota}, K., \& {Frieman}, J. 2007, \apj, 658,
  669

\bibitem[{{Senatore} \& {Trevisan}(2018)}]{Senatore:2017pbn}
{Senatore}, L. \& {Trevisan}, G. 2018, \jcap, 2018, 019

\bibitem[{{Shi} {et~al.}(2018){Shi}, {Cautun}, \& {Li}}]{Shi:2017gqs}
{Shi}, Y., {Cautun}, M., \& {Li}, B. 2018, \prd, 97, 023505

\bibitem[{{Slosar} {et~al.}(2019){Slosar}, {Chen}, {Dvorkin}, {Meerburg},
  {Wallisch}, {Green}, \& {Silverstein}}]{Slosar:2019gvt}
{Slosar}, A., {Chen}, X., {Dvorkin}, C., {et~al.} 2019, \baas, 51, 98

\bibitem[{{Springel}(2005)}]{Springel:2005mi}
{Springel}, V. 2005, \mnras, 364, 1105

\bibitem[{{Springel} {et~al.}(2001){Springel}, {White}, {Tormen}, \&
  {Kauffmann}}]{Springel:2000qu}
{Springel}, V., {White}, S. D.~M., {Tormen}, G., \& {Kauffmann}, G. 2001,
  \mnras, 328, 726

\bibitem[{{Starobinsky}(1980)}]{Starobinsky:1980te}
{Starobinsky}, A.~A. 1980, Physics Letters B, 91, 99

\bibitem[{{Starobinsky}(1992)}]{Starobinsky:1992ts}
{Starobinsky}, A.~A. 1992, Soviet Journal of Experimental and Theoretical
  Physics Letters, 55, 489

\bibitem[{{Tassev} {et~al.}(2015){Tassev}, {Eisenstein}, {Wandelt}, \&
  {Zaldarriaga}}]{Tassev:2015mia}
{Tassev}, S., {Eisenstein}, D.~J., {Wandelt}, B.~D., \& {Zaldarriaga}, M. 2015,
  arXiv e-prints, arXiv:1502.07751

\bibitem[{{Tassev} {et~al.}(2013){Tassev}, {Zaldarriaga}, \&
  {Eisenstein}}]{Tassev:2013pn}
{Tassev}, S., {Zaldarriaga}, M., \& {Eisenstein}, D.~J. 2013, \jcap, 2013, 036

\bibitem[{{Teyssier}(2002)}]{teyssier2002}
{Teyssier}, R. 2002, \aap, 385, 337

\bibitem[{{Torrado} {et~al.}(2017){Torrado}, {Hu}, \&
  {Ach{\'u}carro}}]{Torrado:2016sls}
{Torrado}, J., {Hu}, B., \& {Ach{\'u}carro}, A. 2017, \prd, 96, 083515

\bibitem[{{van de Weygaert} \& {Schaap}(2009)}]{vandeWeygaert:2007ze}
{van de Weygaert}, R. \& {Schaap}, W. 2009, {The Cosmic Web: Geometric
  Analysis}, ed. V.~J. {Mart{\'\i}nez}, E.~{Saar},
  E.~{Mart{\'\i}nez-Gonz{\'a}lez}, \& M.~J. {Pons-Border{\'\i}a}, Vol. 665,
  291--413

\bibitem[{{Vasudevan} {et~al.}(2019){Vasudevan}, {Ivanov}, {Sibiryakov}, \&
  {Lesgourgues}}]{Vasudevan:2019ewf}
{Vasudevan}, A., {Ivanov}, M.~M., {Sibiryakov}, S., \& {Lesgourgues}, J. 2019,
  \jcap, 2019, 037

\bibitem[{{Viel} {et~al.}(2010){Viel}, {Haehnelt}, \& {Springel}}]{Viel:2010bn}
{Viel}, M., {Haehnelt}, M.~G., \& {Springel}, V. 2010, \jcap, 2010, 015

\bibitem[{{Villaescusa-Navarro} {et~al.}(2018){Villaescusa-Navarro}, {Naess},
  {Genel}, {Pontzen}, {Wandelt}, {Anderson}, {Font-Ribera}, {Battaglia}, \&
  {Spergel}}]{Villaescusa-Navarro:2018bpd}
{Villaescusa-Navarro}, F., {Naess}, S., {Genel}, S., {et~al.} 2018, \apj, 867,
  137

\bibitem[{{Vlah} {et~al.}(2016){Vlah}, {Seljak}, {Yat Chu}, \&
  {Feng}}]{Vlah:2015zda}
{Vlah}, Z., {Seljak}, U., {Yat Chu}, M., \& {Feng}, Y. 2016, \jcap, 2016, 057

\bibitem[{{Wang} {et~al.}(2017){Wang}, {Yu}, {Zhu}, {Yu}, {Pan}, \&
  {Pen}}]{Wang:2017apjl}
{Wang}, X., {Yu}, H.-R., {Zhu}, H.-M., {et~al.} 2017, \apjl, 841, L29

\bibitem[{{Wang} {et~al.}(2020){Wang}, {Li}, \& {Cautun}}]{Wang:2019zuq}
{Wang}, Y., {Li}, B., \& {Cautun}, M. 2020, \mnras, 497, 3451

\bibitem[{{Wang} \& {Mathews}(2002)}]{Wang:2000js}
{Wang}, Y. \& {Mathews}, G.~J. 2002, \apj, 573, 1

\bibitem[{{Wang} {et~al.}(1999){Wang}, {Spergel}, \& {Strauss}}]{Wang:1998gb}
{Wang}, Y., {Spergel}, D.~N., \& {Strauss}, M.~A. 1999, \apj, 510, 20

\bibitem[{{Winther} {et~al.}(2017){Winther}, {Koyama}, {Manera}, {Wright}, \&
  {Zhao}}]{Winther:2017jof}
{Winther}, H.~A., {Koyama}, K., {Manera}, M., {Wright}, B.~S., \& {Zhao}, G.-B.
  2017, \jcap, 2017, 006

\bibitem[{{Wright} {et~al.}(2017){Wright}, {Winther}, \&
  {Koyama}}]{Wright:2017dkw}
{Wright}, B.~S., {Winther}, H.~A., \& {Koyama}, K. 2017, \jcap, 2017, 054

\bibitem[{{Xu} {et~al.}(2016){Xu}, {Hamann}, \& {Chen}}]{Xu:2016kwz}
{Xu}, Y., {Hamann}, J., \& {Chen}, X. 2016, \prd, 94, 123518

\bibitem[{{Zeng} {et~al.}(2019){Zeng}, {Kovetz}, {Chen}, {Gong}, {Mu{\~n}oz},
  \& {Kamionkowski}}]{Zeng:2018ufm}
{Zeng}, C., {Kovetz}, E.~D., {Chen}, X., {et~al.} 2019, \prd, 99, 043517

\bibitem[{{Zhan} {et~al.}(2006){Zhan}, {Knox}, {Tyson}, \&
  {Margoniner}}]{Zhan:2005rz}
{Zhan}, H., {Knox}, L., {Tyson}, J.~A., \& {Margoniner}, V. 2006, \apj, 640, 8

\bibitem[{{Zhu} {et~al.}(2017){Zhu}, {Yu}, {Pen}, {Chen}, \&
  {Yu}}]{Zhu:2017prd}
{Zhu}, H.-M., {Yu}, Y., {Pen}, U.-L., {Chen}, X., \& {Yu}, H.-R. 2017, \prd,
  96, 123502

\end{thebibliography}

%
\begin{appendix}

\section{Comparison with dark matter N-body simulations\label{sec:appendix1}}
\begin{figure*}
\centering
\includegraphics[width=0.4\textwidth]{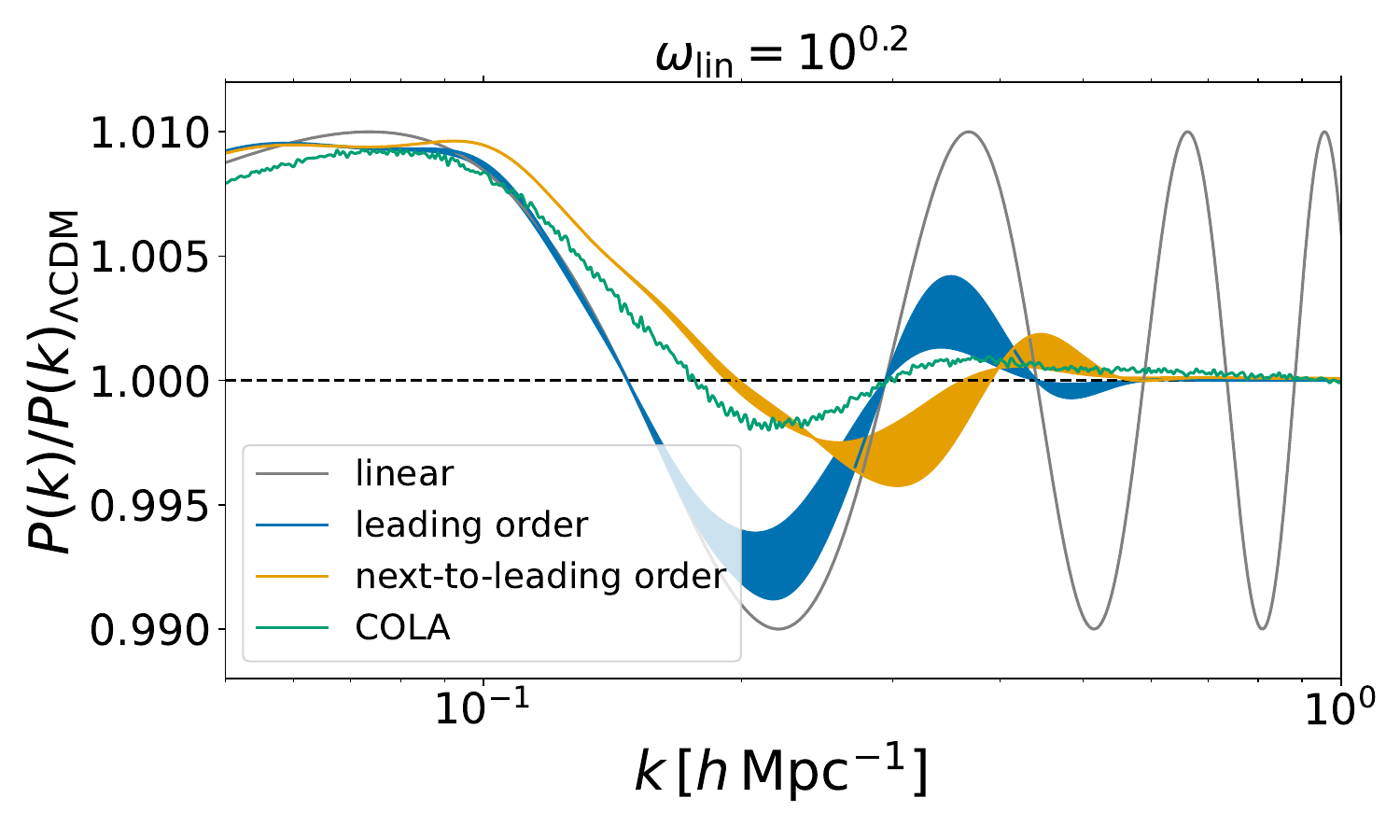}
\includegraphics[width=0.4\textwidth]{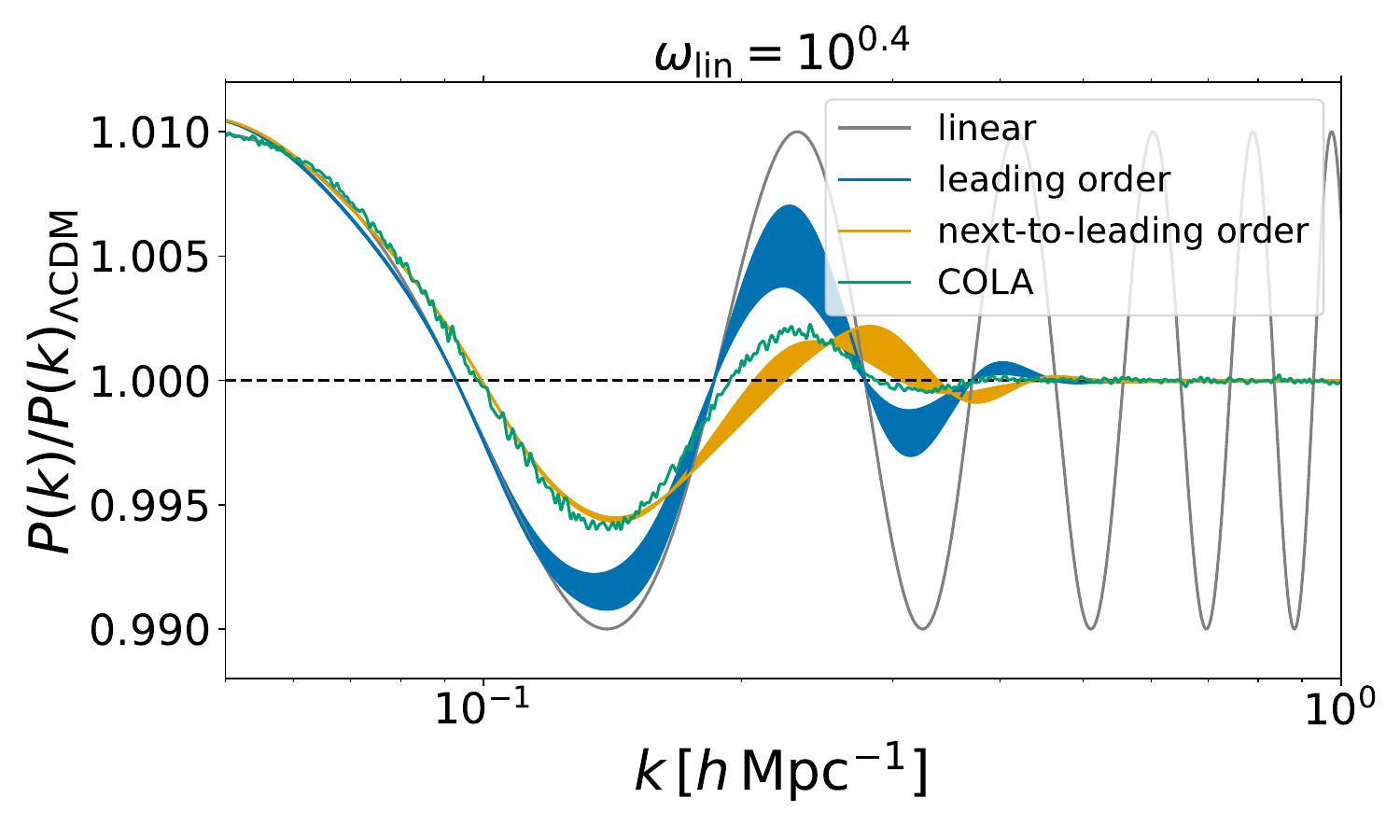}
\includegraphics[width=0.4\textwidth]{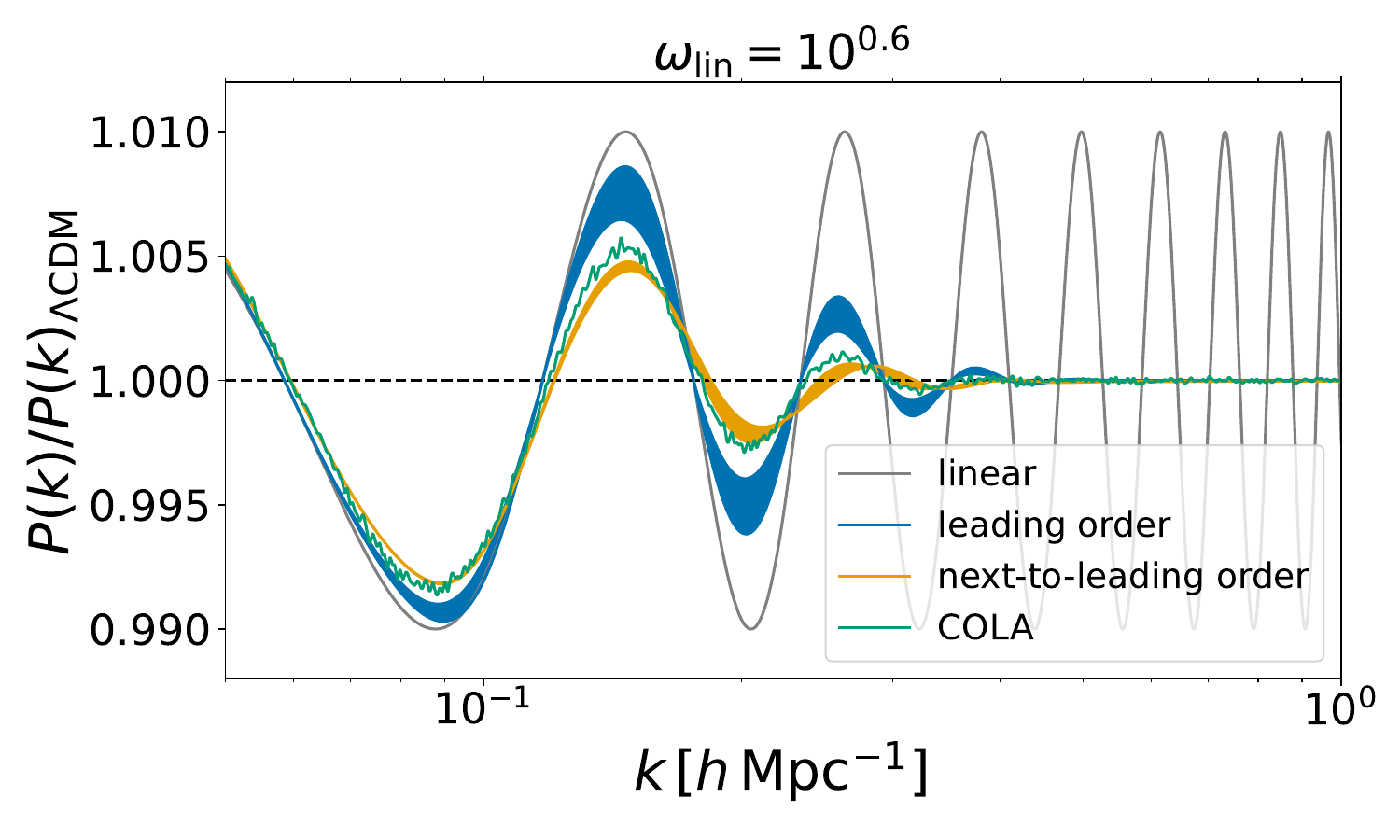}
\includegraphics[width=0.4\textwidth]{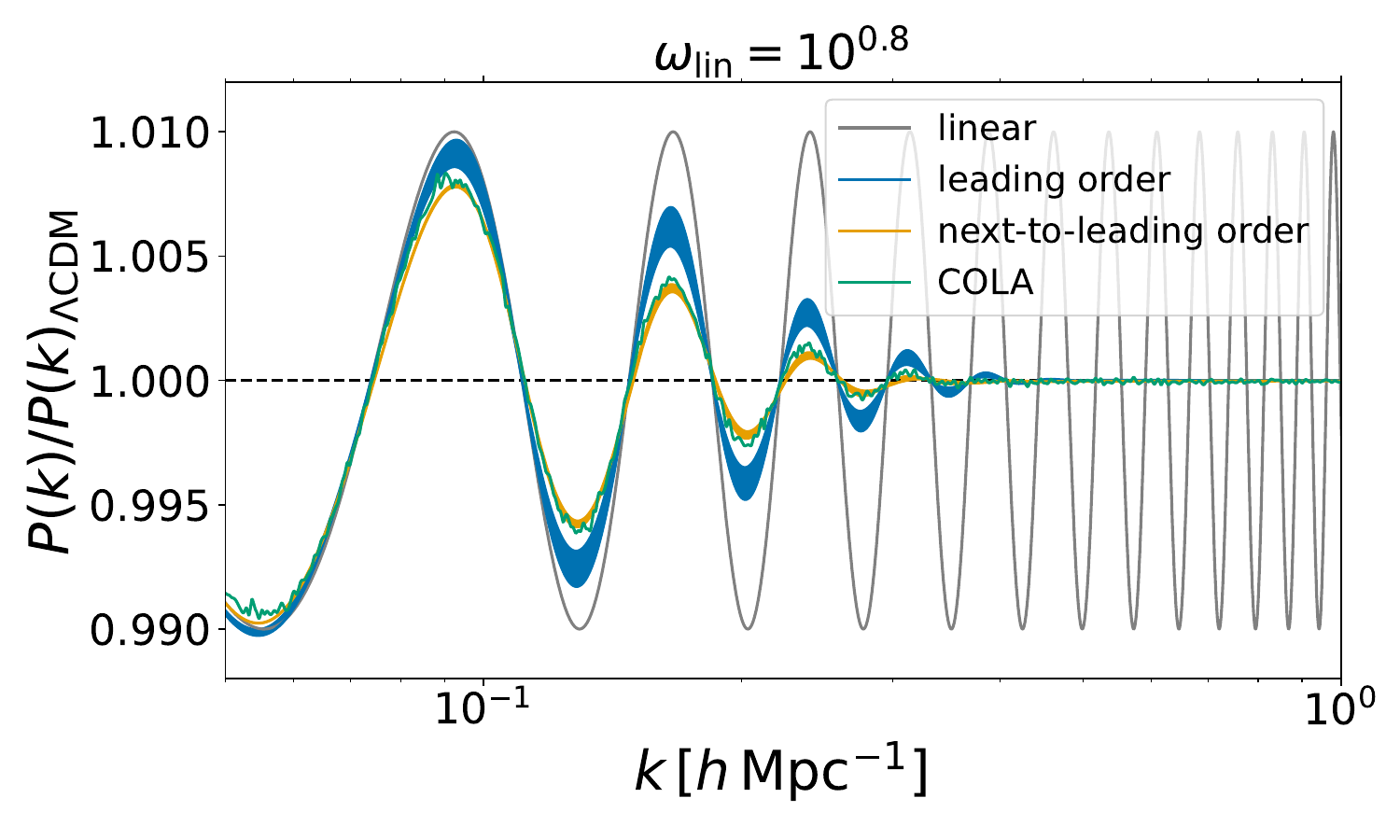}
\includegraphics[width=0.4\textwidth]{Comparison_LIN_1p0.pdf}
\includegraphics[width=0.4\textwidth]{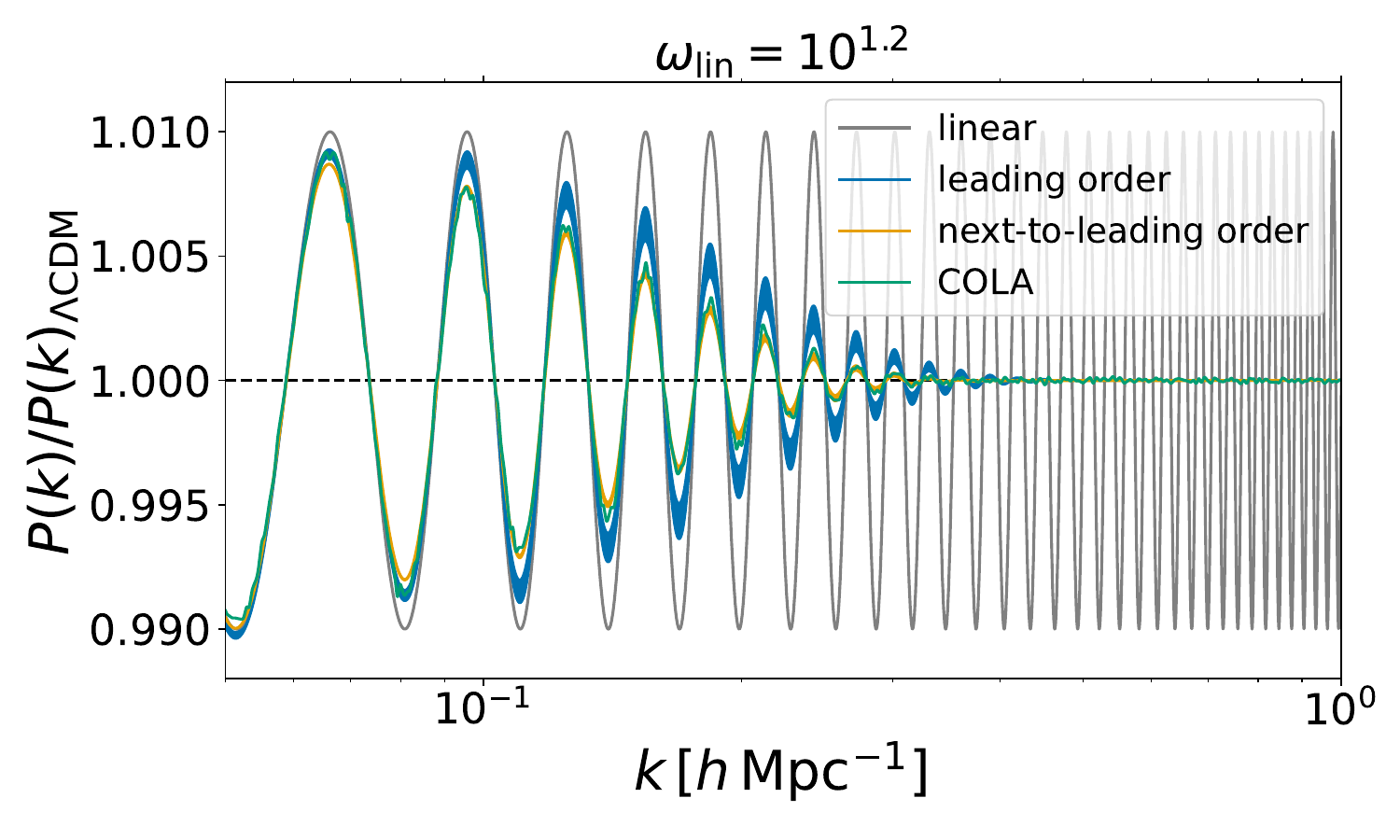}
\includegraphics[width=0.4\textwidth]{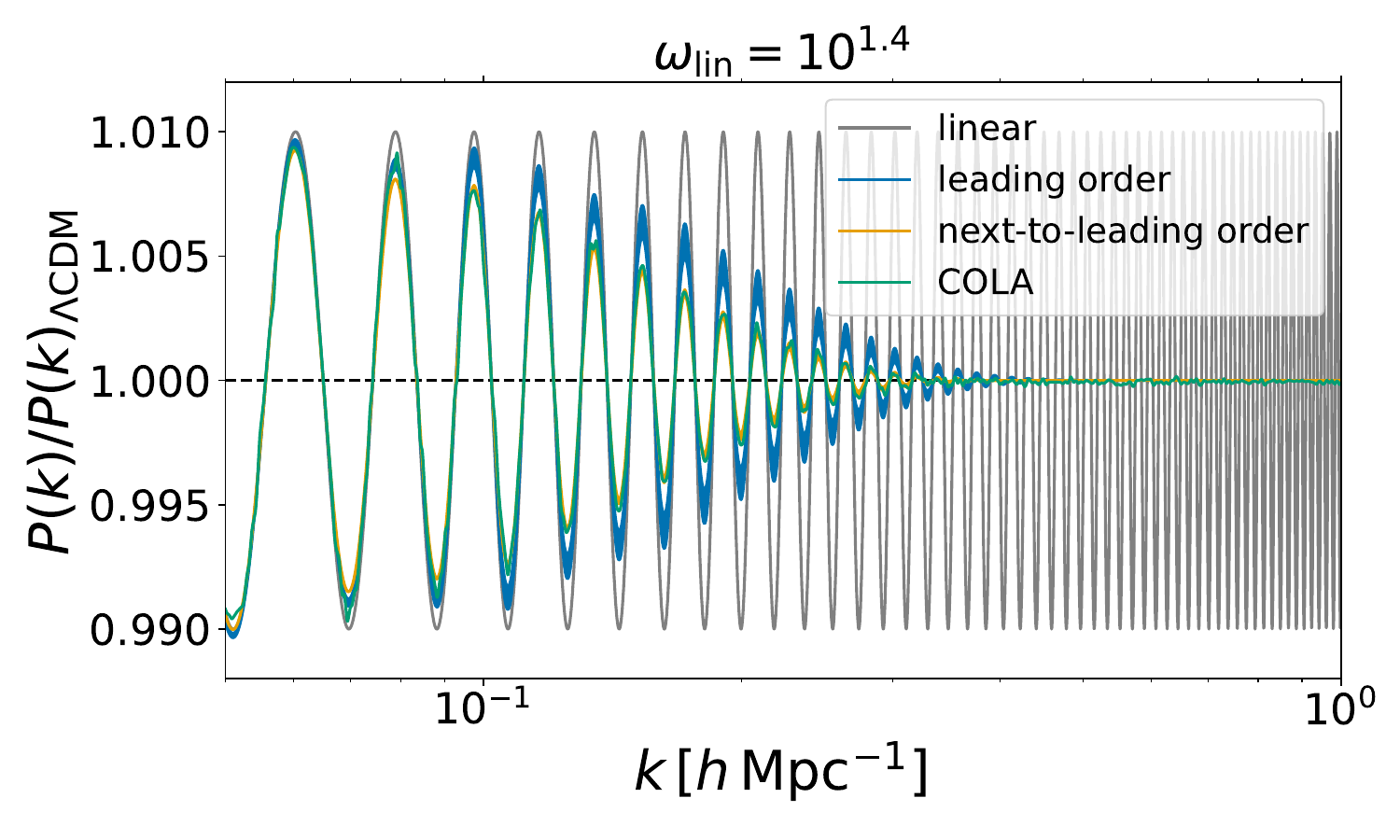}
\includegraphics[width=0.4\textwidth]{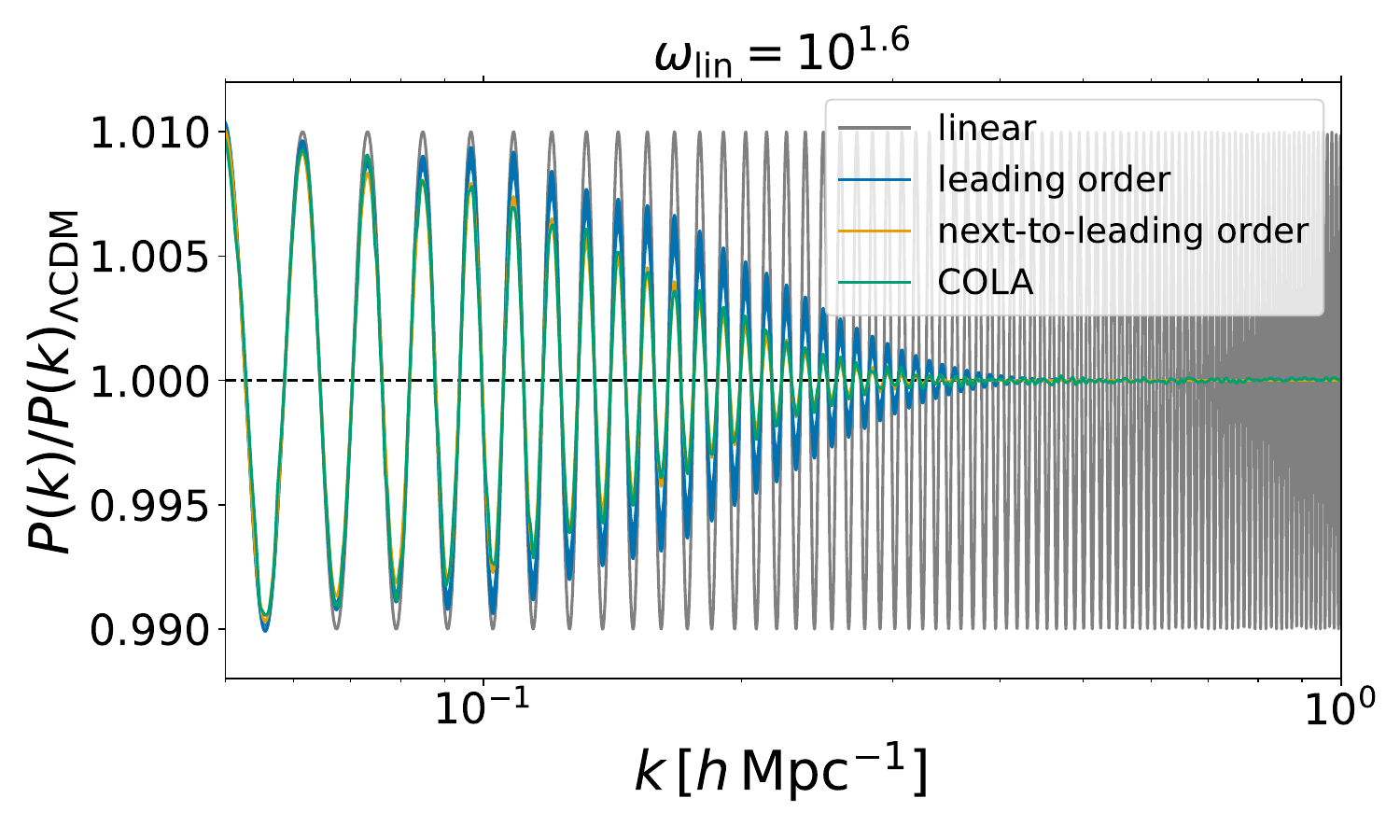}
\includegraphics[width=0.4\textwidth]{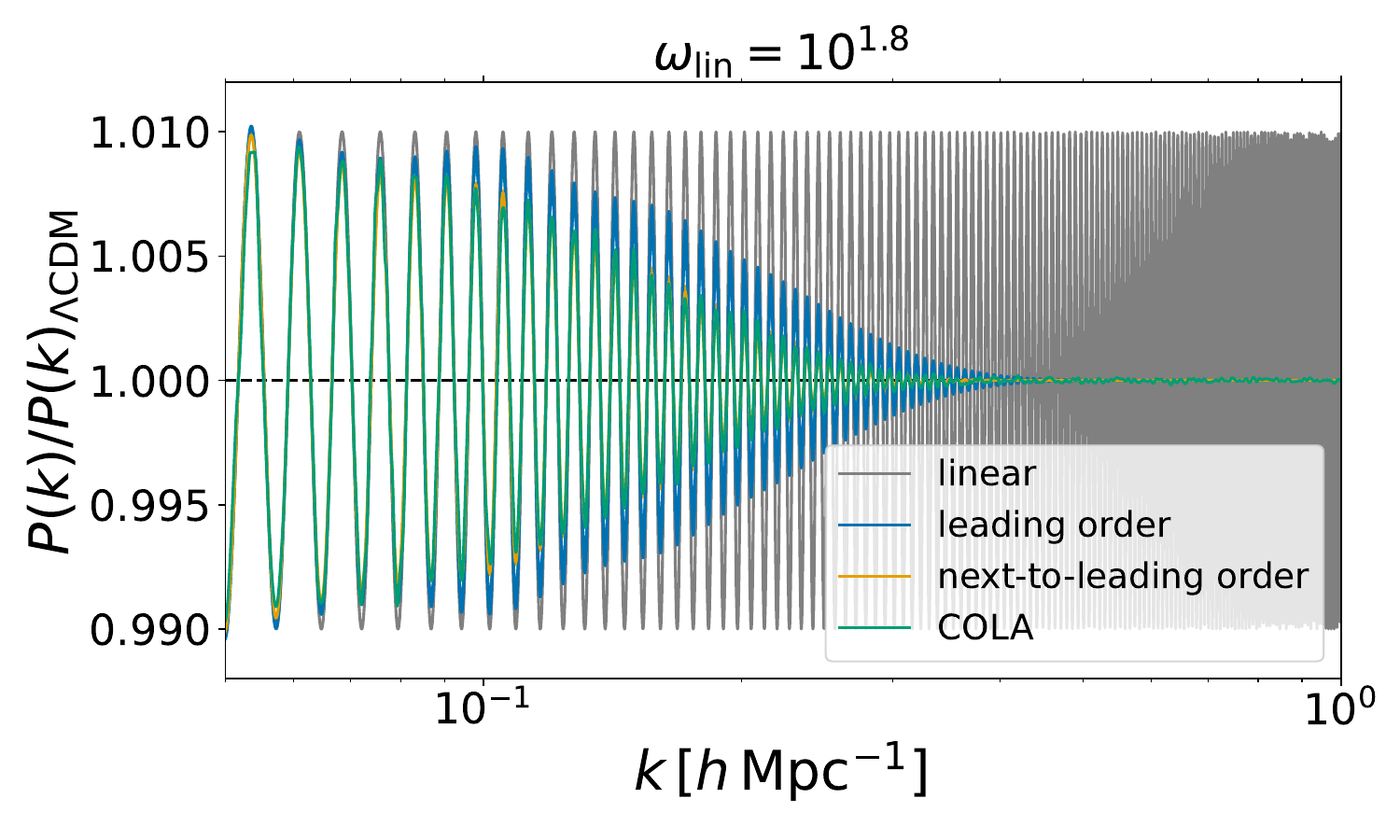}
\includegraphics[width=0.4\textwidth]{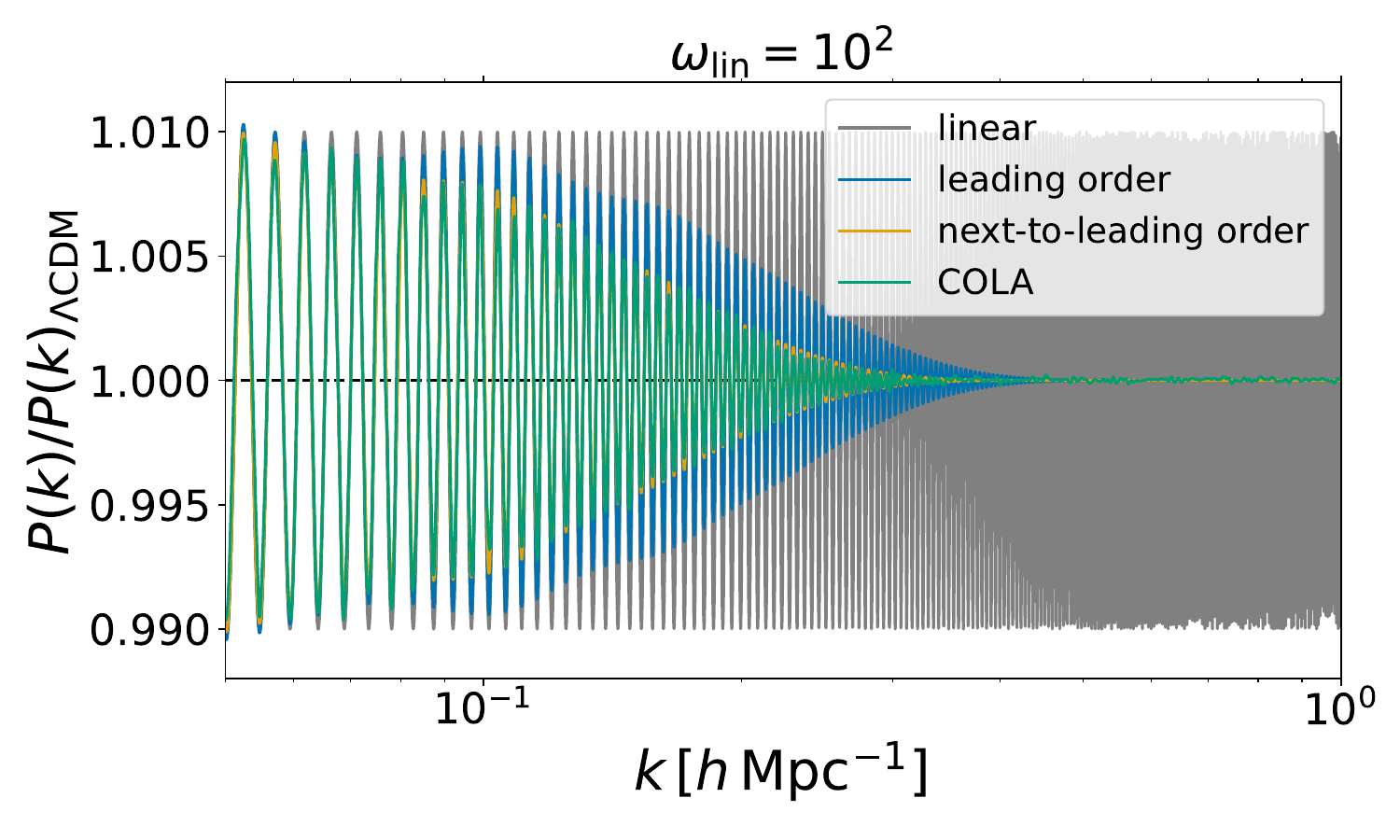}
\caption{Ratio of IR resummed matter power spectrum at LO (blue) and NLO (orange) obtained for 
the linear oscillations to the one obtained with a power-law PPS at redshift $z = 0$ 
varying the IR separation scale $k_{\rm S} = \epsilon k$ with $\epsilon \in (0.3,\,0.7)$. Also shown are  
the linear results (grey) and the ones obtained from N-body simulations (green).}
\label{fig:Comparison_lin}
\end{figure*}

\begin{figure*}
\centering
\includegraphics[width=0.4\textwidth]{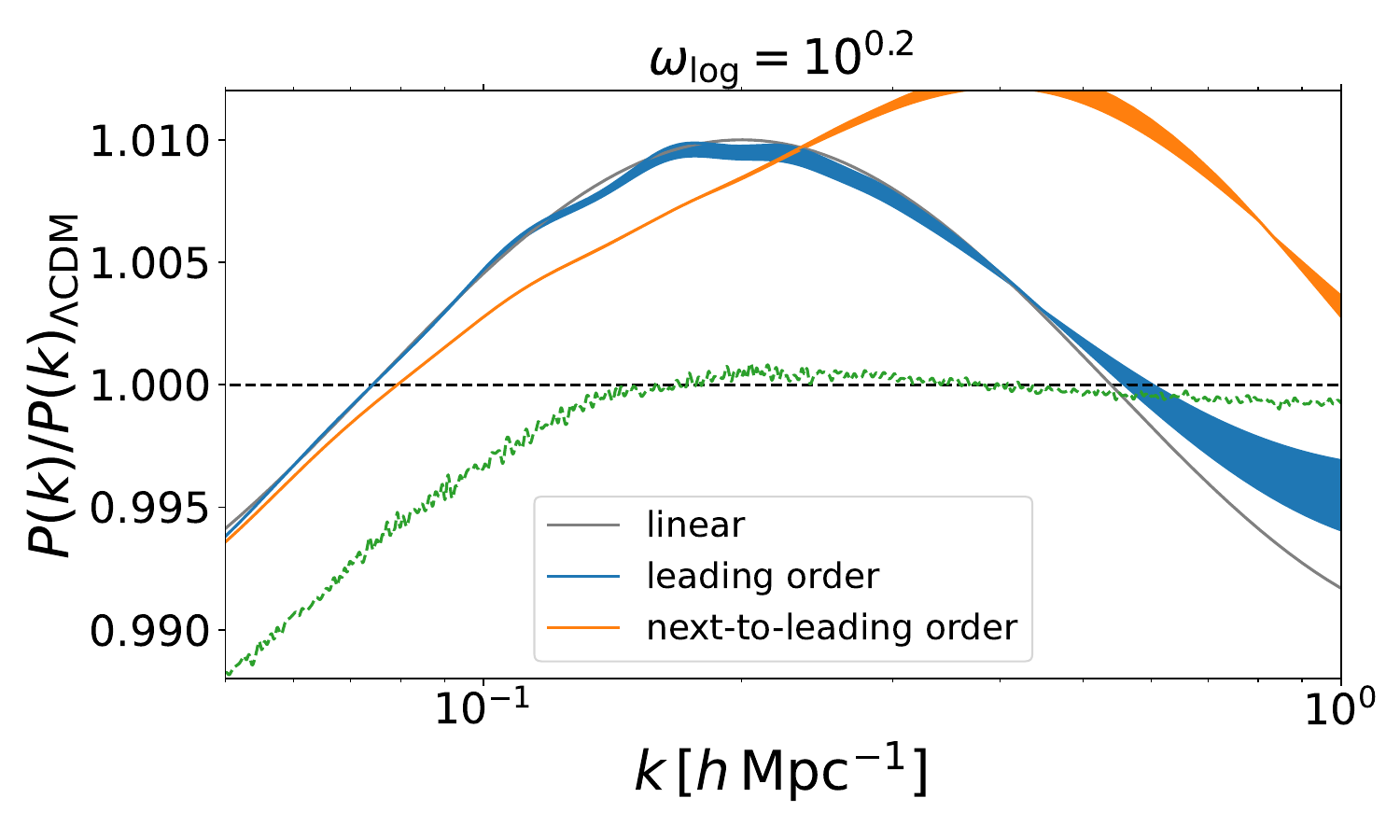}
\includegraphics[width=0.4\textwidth]{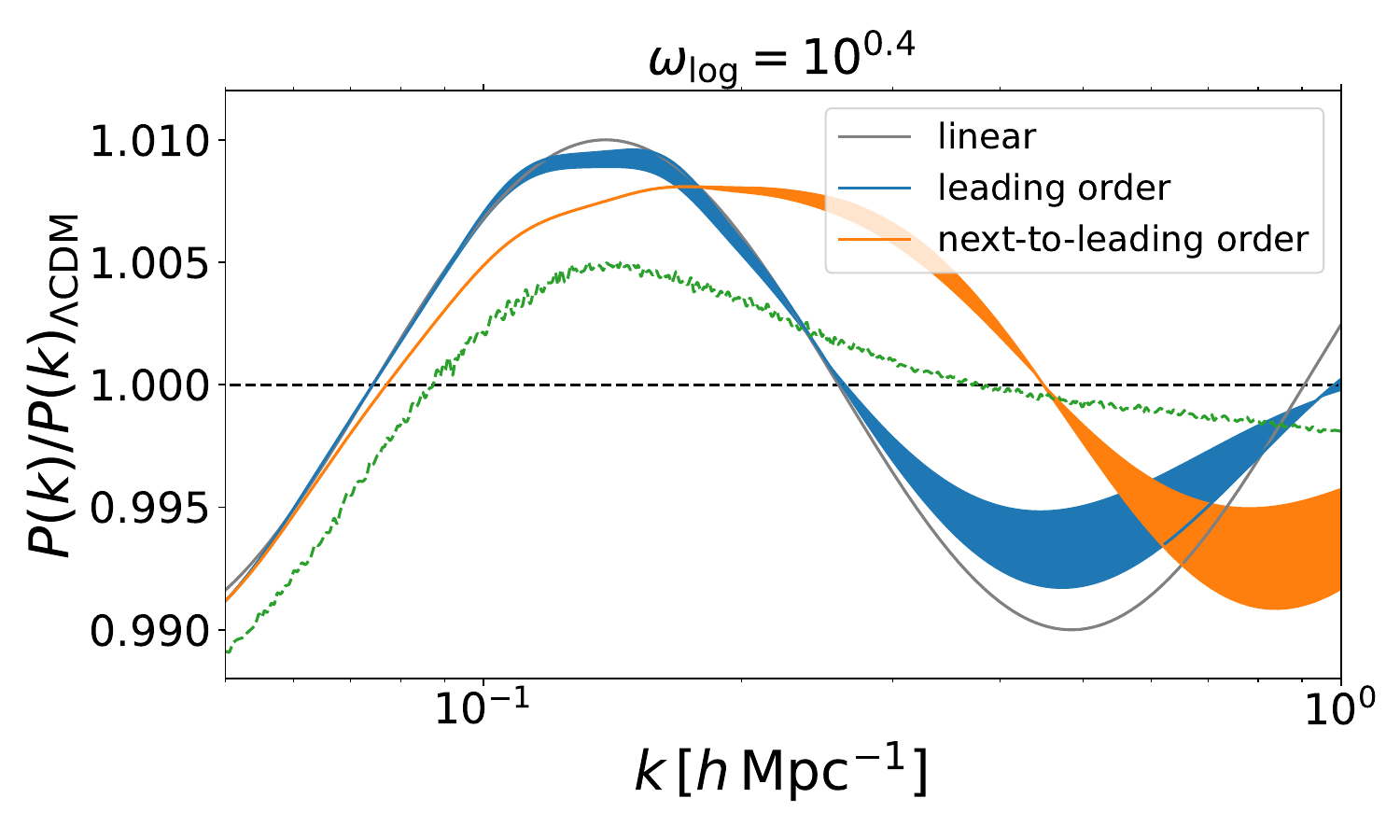}
\includegraphics[width=0.4\textwidth]{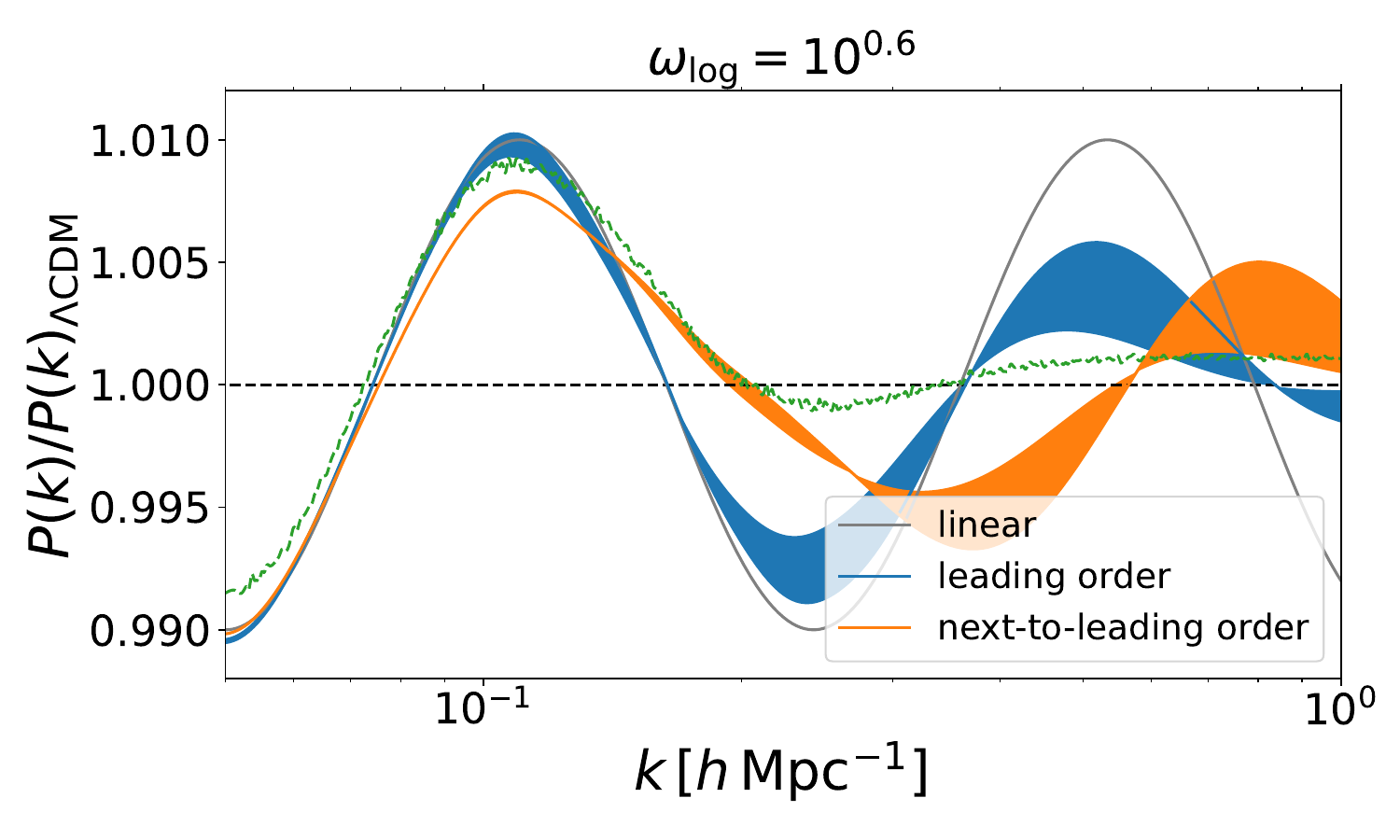}
\includegraphics[width=0.4\textwidth]{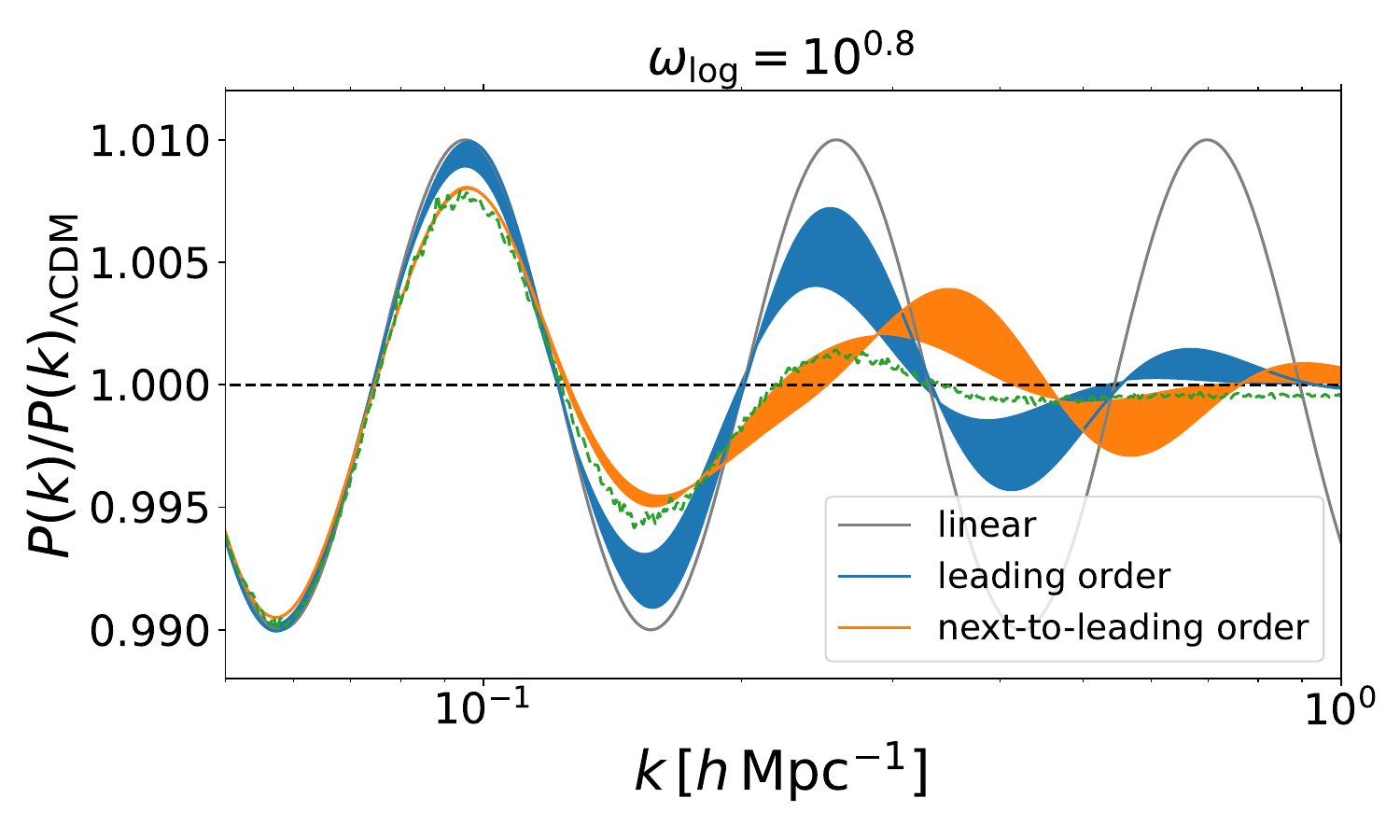}
\includegraphics[width=0.4\textwidth]{Comparison_LOG_1p0.pdf}
\includegraphics[width=0.4\textwidth]{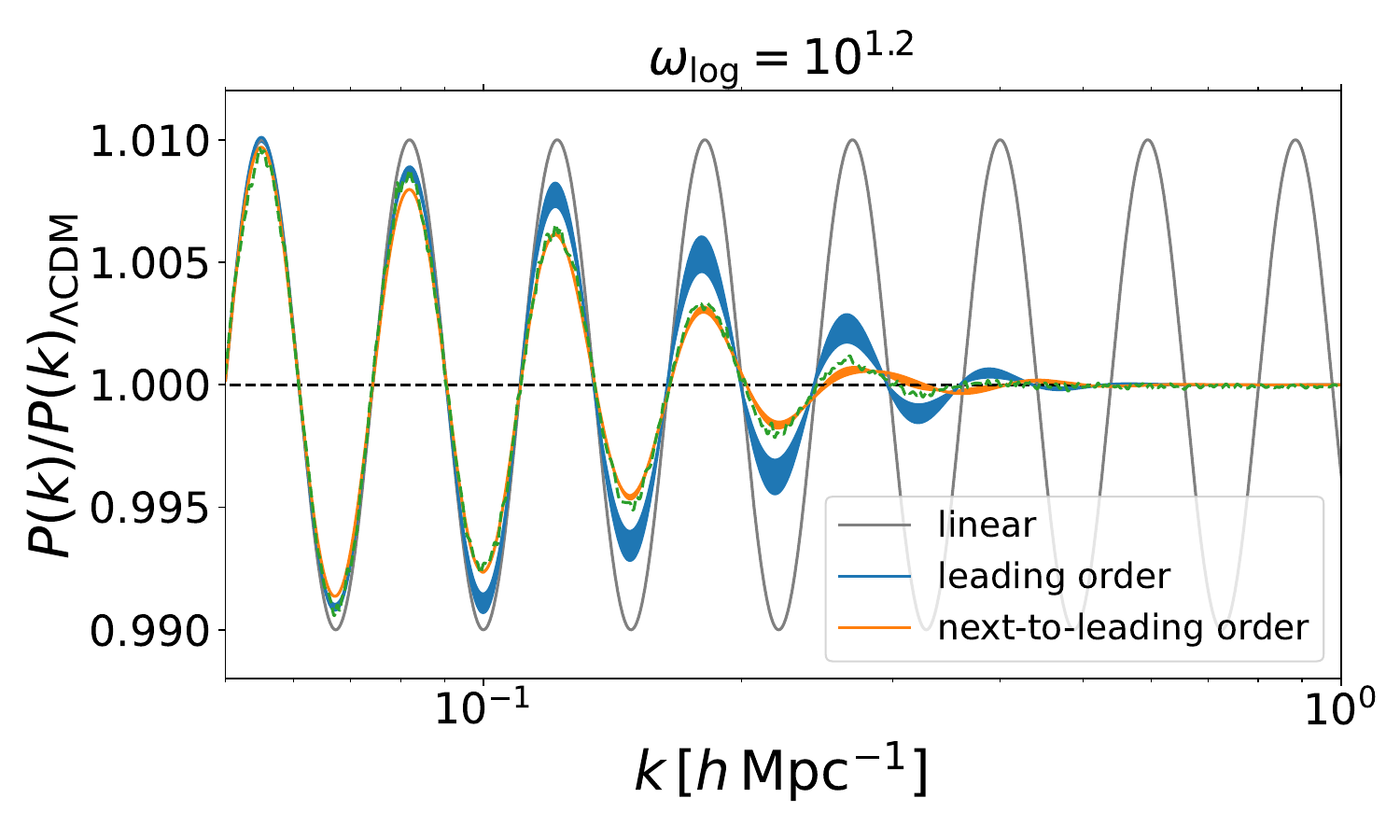}
\includegraphics[width=0.4\textwidth]{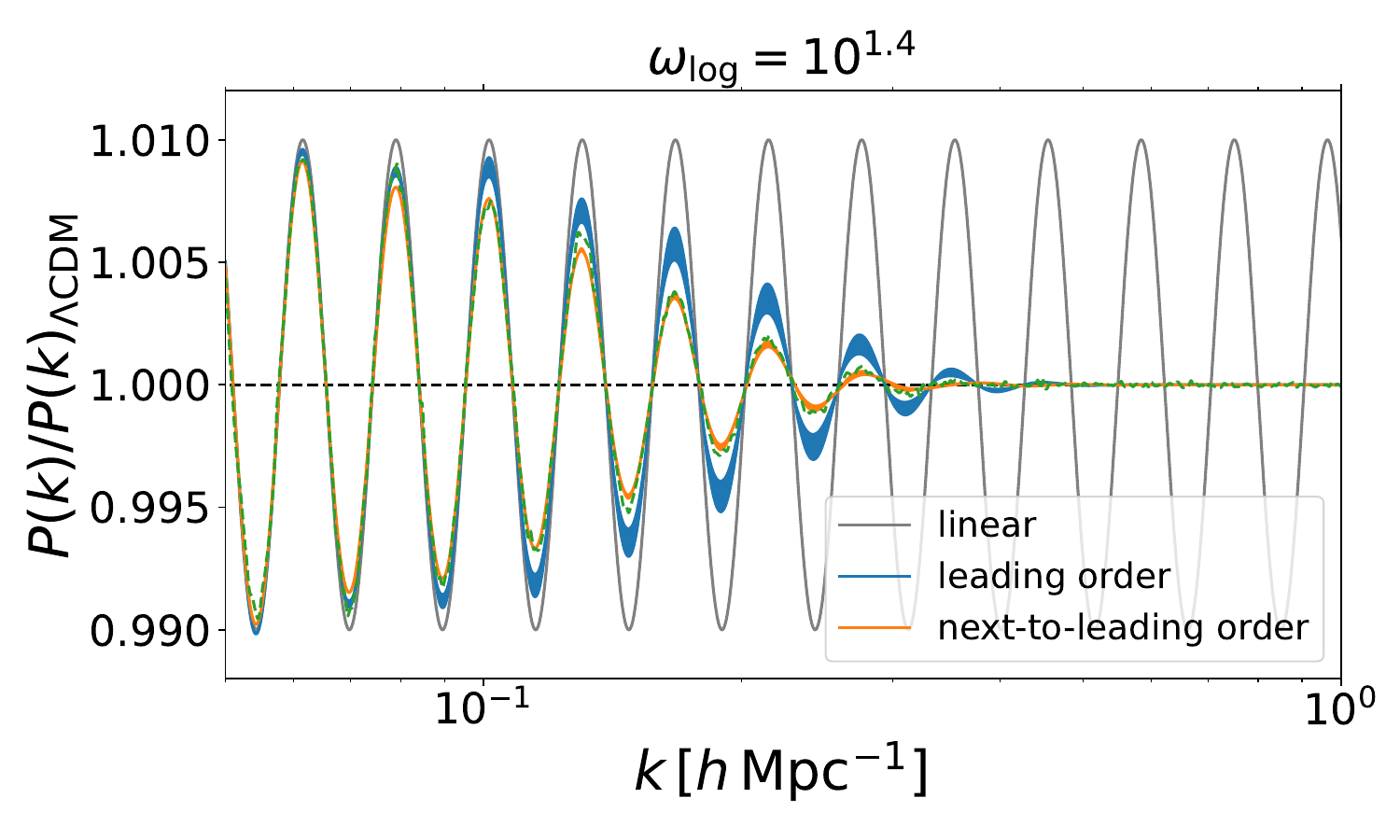}
\includegraphics[width=0.4\textwidth]{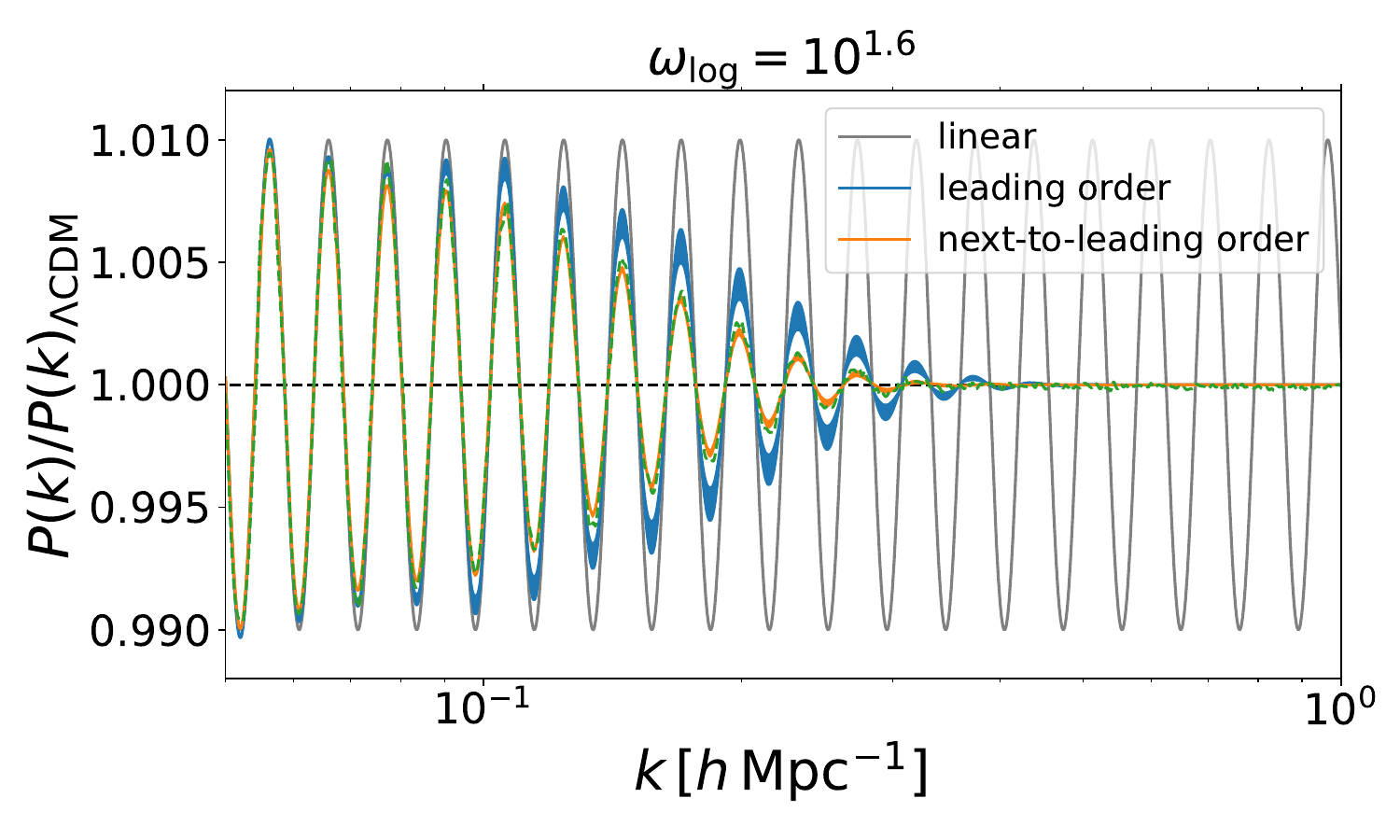}
\includegraphics[width=0.4\textwidth]{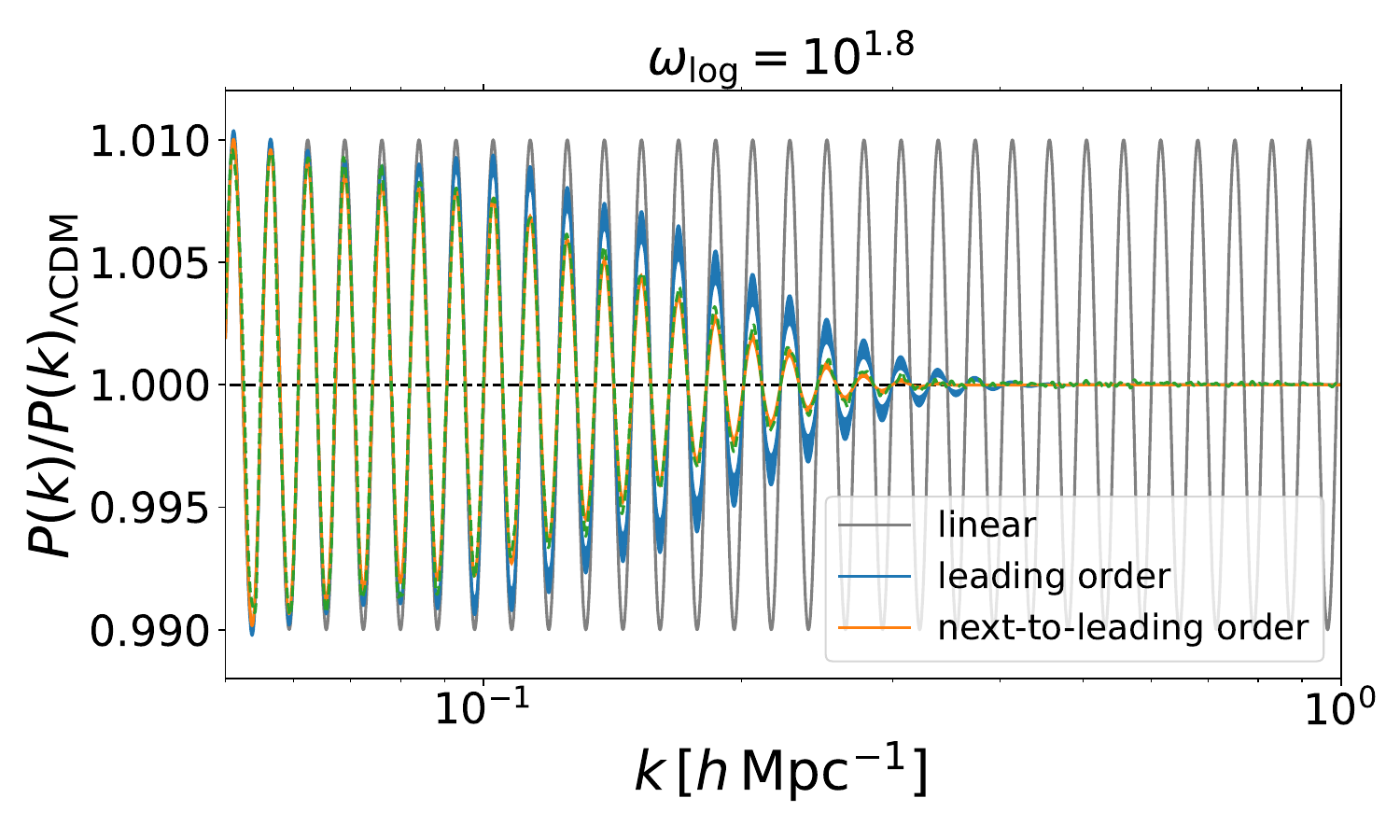}
\includegraphics[width=0.4\textwidth]{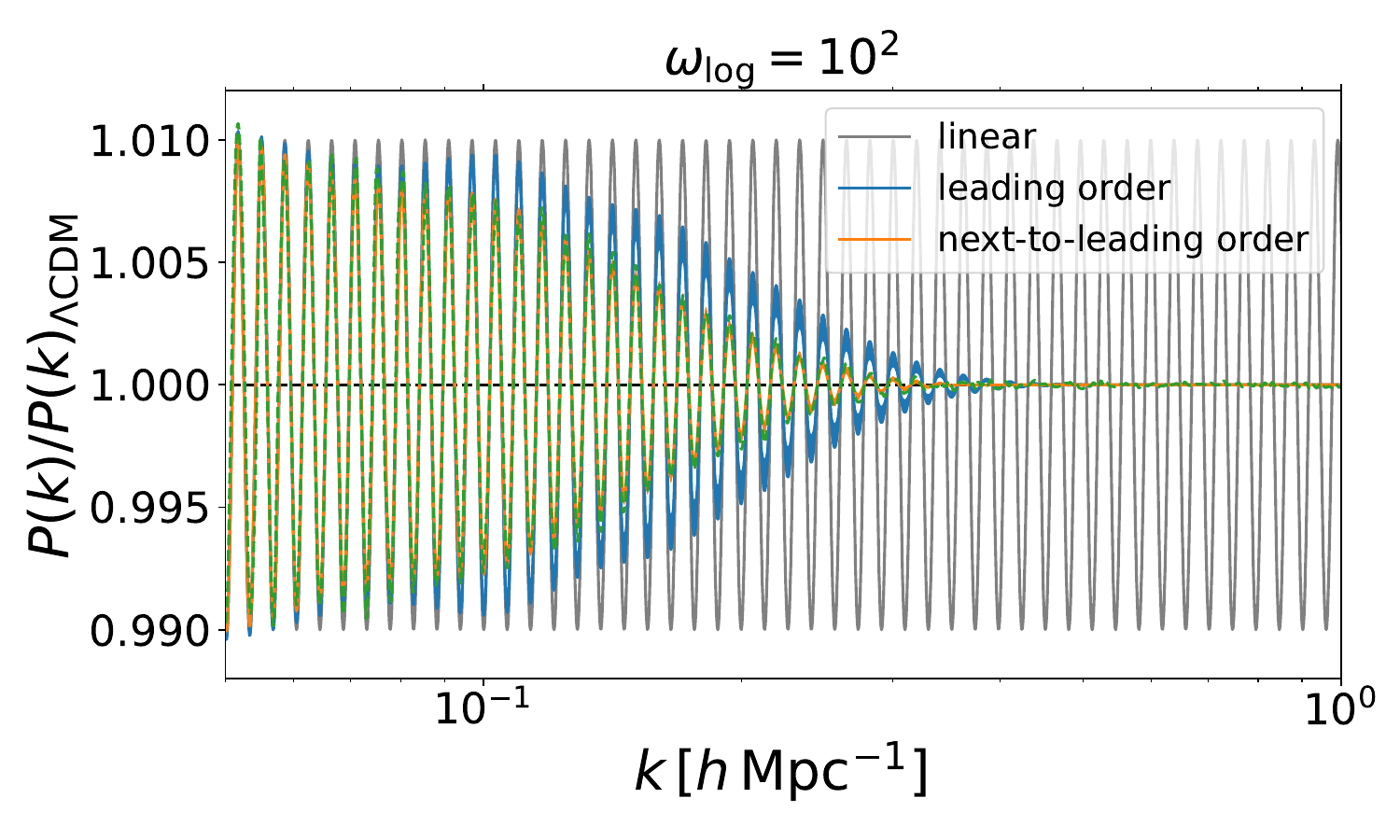}
\caption{As in Fig.~\ref{fig:Comparison_lin}, but for the logarithmic oscillations.}
\label{fig:Comparison_log}
\end{figure*}

In Figs.~\ref{fig:Comparison_lin} and \ref{fig:Comparison_log}, we compare the ratio between the matter power spectrum 
with primordial oscillations to the one obtained with a power-law PPS at redshift $z=0$.
We show analytic results obtained with PT at LO (blue) and NLO (orange) versus linear theory predictions (grey) and the nonlinear results obtained from N-body simulations (green).

\section{Nonlinear reconstruction results\label{sec:appendix2}}
\begin{figure*}
\centering
\includegraphics[width=1\textwidth]{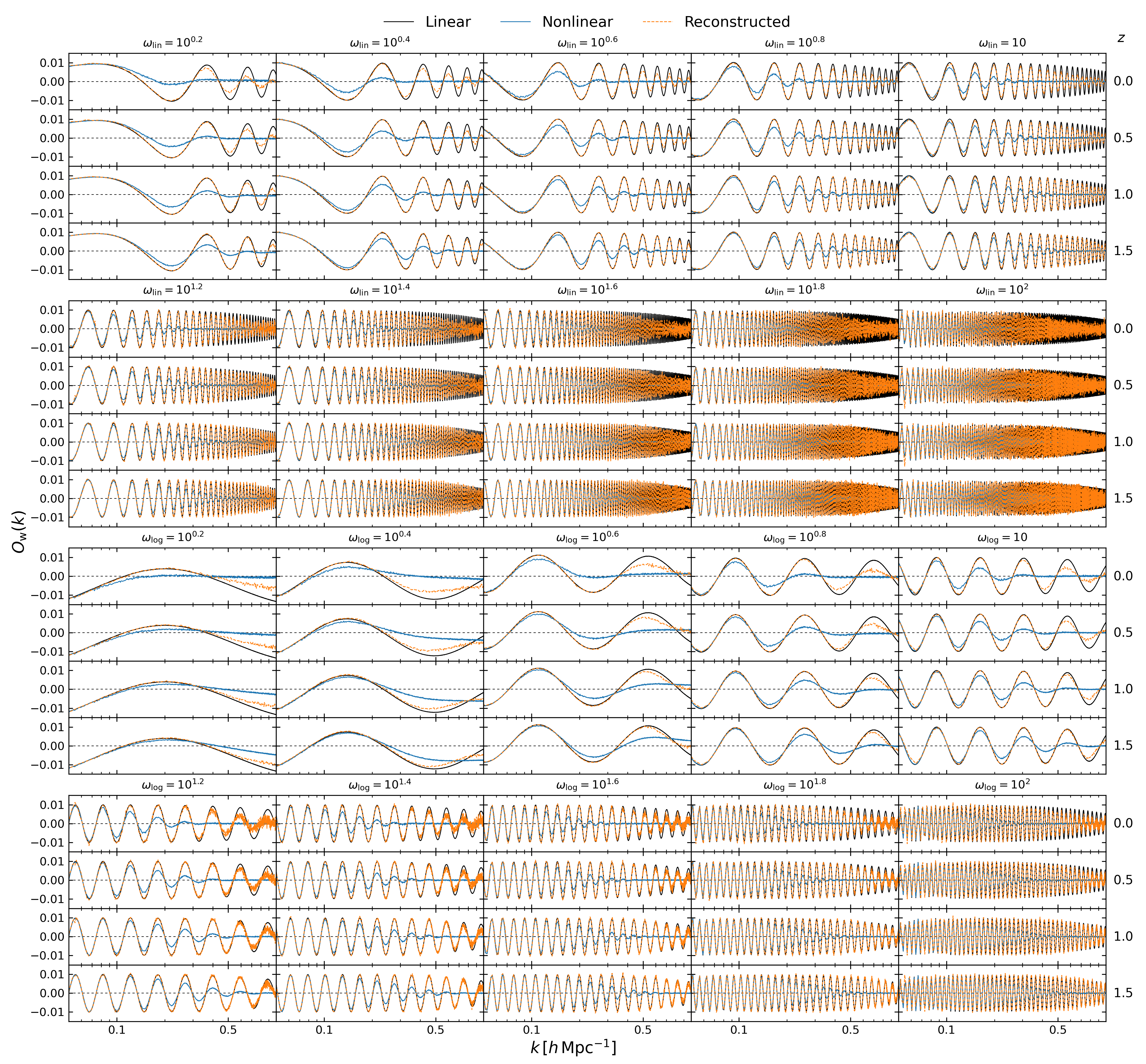}
\caption{
Ratio of the linear (black solid curves), nonlinear (blue solid curves), and 
reconstructed (orange dashed curves) density fields for the 20 selected feature models to the one obtained with a power-law PPS.
Each of the rows is respectively 
for the redshifts marked on the right. The linear $O_{\rm w}(k)$ are measured from the initial conditions at $z=9$, 
the nonlinear $O_{\rm w}(k)$ are from the output snapshots of DM, and the reconstructed $O_{\rm w}(k)$ are from the 
reconstructed density fields.} 
\label{fig:reconstruction}
\end{figure*}

In Fig.~\ref{fig:reconstruction}, we compare the wiggle $O_{\rm w}(k)$, which is the ratio between the wiggled and 
no-wiggle $P(k)$, for all 20 wiggled models: linear models in the first two rows and logarithmic models in the bottom 
two rows as shown by the legend. The four subpanels of each row, from bottom to top, are the results for $z=1.5$, $1.0$, 
$0.5$, and $0$. Linear theory predictions (which in practice are replaced by the measured $P(k)$ from the initial 
condition snapshots at $z=9$) are shown by black solid curves, while the measurements from the fully nonlinear and 
reconstructed matter density fields are shown by blue solid and orange dashed curves.

Structure formation greatly damps the wiggles, and this impact is strongest for the low-frequency wiggle models, given 
that there are fewer peaks in the $k$ range shown. The extreme example is $\omega_{\rm log}=10^{0.2}$, for which at $z<1$ 
the only peak in the entire $k$ range shown here is completely invisible. Reconstruction is expected to be useful here, 
as shown by the orange curves in the corresponding panels. Even for models with very high wiggle frequencies, such as 
$\omega_{\rm log}=10^2$, the wiggles can be well reconstructed down to $k\simeq1\,h\,{\rm Mpc}^{-1}$. 

\end{appendix}

\end{document}